\documentclass[]{emulateapj}

\usepackage{color}

\slugcomment{\\ \\ Accepted for publication in {\it{The Astrophysical Journal}}}
\shortauthors{Freedman et al.}
\shorttitle{The Carnegie Supernova Project}

\newcommand \gs{\mathrel{\raise0.35ex\hbox{$\scriptstyle >$}\kern-0.6em
                \lower0.40ex\hbox{{$\scriptstyle \sim$}}}}
\newcommand \ls{\mathrel{\raise0.35ex\hbox{$\scriptstyle <$}\kern-0.6em
                \lower0.40ex\hbox{{$\scriptstyle \sim$}}}}

\newcommand{\Nobs}{80}
\newcommand{\NTypeIa}{75}

\newcommand{\Nwithtemp}{71}
\newcommand{\Ngood}{35}
\newcommand{\Nlowz}{21}
\newcommand{\OmegaMatter}{\ensuremath{\Omega_{m} = 0.27 \pm 0.02
                 ~(\mathrm{statistical})}}
\newcommand{\OmegaLambda}{\ensuremath{\Omega_{DE} = 0.76 \pm 0.13~
                 ~(\mathrm{statistical}) \pm 0.09 ~(\mathrm{systematic})}}
\newcommand{\OmegaMatterFlat}{\ensuremath{\Omega_{m} = 0.27 \pm 0.03
                 ~(\mathrm{statistical})~ }}
\newcommand{\wsym}{\ensuremath{w}}
\newcommand{\wo}{\ensuremath{\wsym = -1.05 \pm 0.13 
                 ~(\mathrm{statistical}) \pm 0.09 ~(\mathrm{systematic})}}
\newcommand{\jo}{\ensuremath{j_{k} = 1.18 \pm 0.44 ~(\mathrm{statistical})
                                          \pm 0.27 ~(\mathrm{systematic})~}}
\newcommand{\qo}{\ensuremath{q_{o} = -0.67 \pm  0.13 ~(\mathrm{statistical})
                                           \pm 0.09 ~(\mathrm{systematic})}}
\newcommand{\NobsESSENCE}{102} 
\newcommand{\NSpec}{125}
\newcommand{\template}{host galaxy}
\newcommand{\templateh}{host-galaxy}
\newcommand{\dm}{\ensuremath{\Delta m_{\mathrm{15}}}}

\begin{document}

\title{The Carnegie Supernova Project: First Near-Infrared Hubble Diagram to
  z$\sim$0.7\footnote{This paper includes data
 gathered with the 6.5 meter Magellan Telescopes located at Las
 Campanas Observatory, Chile.}}
\author{
Wendy L.\ Freedman\altaffilmark{1},
Christopher R. Burns\altaffilmark{1},
M. M. Phillips\altaffilmark{2},
Pamela Wyatt\altaffilmark{1},
S. E. Persson\altaffilmark{1},
Barry F. Madore\altaffilmark{1},
Carlos Contreras\altaffilmark{2},
Gaston Folatelli\altaffilmark{2,3},E
Sergio Gonzalez\altaffilmark{2},
Mario Hamuy\altaffilmark{3},
Eric Hsiao\altaffilmark{4},
Daniel D. Kelson\altaffilmark{1},
Nidia Morrell\altaffilmark{2},
D. C. Murphy\altaffilmark{1},
Miguel Roth\altaffilmark{2},
Maximilian Stritzinger\altaffilmark{2},
Laura Sturch\altaffilmark{1},
Nick B. Suntzeff\altaffilmark{21} 
P. Astier\altaffilmark{6},
C. Balland\altaffilmark{6,7},
Bruce Bassett\altaffilmark{8},
Luis Boldt\altaffilmark{2},
R. G. Carlberg\altaffilmark{9},
Alexander J. Conley\altaffilmark{9},
Joshua A. Frieman\altaffilmark{10,11,12},
Peter M. Garnavich\altaffilmark{13},
J. Guy\altaffilmark{6},
D. Hardin\altaffilmark{6},
D. Andrew Howell\altaffilmark{14,15},
Richard Kessler\altaffilmark{16,11},
Hubert Lampeitl\altaffilmark{5},
John Marriner\altaffilmark{10},
R. Pain\altaffilmark{6},
Kathy Perrett\altaffilmark{9},
N. Regnault\altaffilmark{6},
Adam G. Riess\altaffilmark{17},
Masao Sako\altaffilmark{16,18},
Donald P. Schneider\altaffilmark{19},
Mark Sullivan\altaffilmark{20}, and 
Michael Wood-Vasey\altaffilmark{22}
}
\altaffiltext{1}{Carnegie Observatories, 813 Santa Barbara St, Pasadena, CA, 91101, USA}
\altaffiltext{2}{Carnegie Institution of Washington, Las Campanas Observatory, Colina El Pino, Casilla 601, Chile}
\altaffiltext{3}{Universidad de Chile, Departmento de Astronomia, Casilla 36-D, Santiago, Chile}
\altaffiltext{4}{Department of Physics and Astronomy, University of Victoria, PO Box 3055, Stn CSC, Victoria, BC V8W 3P6, Canada}
\altaffiltext{5}{Institute of Cosmology and Gravitation, University of Portsmouth, Portsmouth, P01 3FX, UK}
\altaffiltext{6}{LPNHE, CNRS-IN2P3 and Universit\'{e}s Paris VI \& VII,4 place Jussieu, 75252 Paris Cedex 05, France}
\altaffiltext{7}{APC, Coll\`{e}ge de France, 11 place Marcellin Berthelot, 75005 Paris, France}
\altaffiltext{8}{Department of Mathematics and Applied Mathematics, University of Cape Town, Rondebosch 7701, South Africa and South African Astronomical Observatory, P.O. Box 9, Observatory 7935, South Africa}
\altaffiltext{9}{Department of Astronomy, University of Toronto, 60 St. George St., Toronto, M5S 3H8, ON, Canada}
\altaffiltext{10}{Center for Particle Astrophysics, Fermi National Accelerator Laboratory, P.O. Box 500, Batavia, IL 60510, USA}
\altaffiltext{11}{Enrico Fermi Institute, University of Chicago, 5640 South Ellis Avenue, Chicago, IL 60637, USA}
\altaffiltext{12}{Department of Astronomy and Astrophysics, The University of Chicago, 5640 South Ellis Avenue, Chicago, IL 60637, USA}
\altaffiltext{13}{University of Notre Dame, 225 Nieuwland Science, Notre Dame, IN46556-5670, USA}
\altaffiltext{14}{Las Cumbres Observatory Global Telescope Network, 6740 Cortona Dr., Suite 102, Goleta, CA 93117, USA}
\altaffiltext{15}{Department of Physics, University of California, Santa Barbara, Broida Hall, Mail Code 9530, Santa Barbara, CA 93106-9530, USA}
\altaffiltext{16}{Kavli Institute for Cosmological Physics, The University of Chicago, 5640 South Ellis Avenue Chicago, IL 60637, USA}
\altaffiltext{17}{Department of Physics and Astronomy, Johns Hopkins University, Baltimore, MD and Space Telescope Science Institute, Baltimore, MD, USA}
\altaffiltext{18}{Department of Physics and Astronomy, University of Pennsylvania, 209 South 33rd Street, Philadelphia, PA 19104, USA}
\altaffiltext{19}{Department of Astronomy and Astrophysics, Pennsylvania State University, 525 Davey Laboratory, University Park, PA 16802, USA}
\altaffiltext{20}{Department of Astrophysics, University of Oxford, Keble Road, Oxford OX1 3RH, UK}
\altaffiltext{21}{Physics Department, Texas A\&M University, College Station, TX, 77843, USA}
\altaffiltext{22}{Department of Physics and Astronomy, 3941 O'Hara St, University of Pittsburgh, Pittsburgh, PA 15260, USA}

\setcounter{footnote}{23}

\begin{abstract}

The Carnegie Supernova Project (CSP) is designed to measure the
 luminosity distance for Type Ia supernovae (SNe~Ia) as a function of
 redshift, and to set observational constraints on the dark energy
 contribution to the total energy content of the Universe.  The CSP
 differs from other projects to date in its goal of providing an
 $I$-band {\it rest-frame} Hubble diagram.  Here we present the first
 results from near-infrared (NIR) observations obtained using the
 Magellan Baade telescope for SNe~Ia with 0.1 $<$ z $<$ 0.7. We
 combine these results with those from the low-redshift CSP at z $<$
 0.1 \citep{Folatelli2009}. In this paper, we describe the overall
 goals of this long-term program, the observing strategy, data
 reduction procedures, and treatment of systematic uncertainties. We
 present light curves and an $I$-band Hubble diagram for this first
 sample of \Ngood\ SNe~Ia and we compare these data to \Nlowz\ new
 SNe~Ia at low redshift. These data support the conclusion that the
 expansion of the Universe is accelerating.  When combined with
 independent results from baryon acoustic oscillations
 \citep[]{Eisenstein2005}, these data yield \OmegaMatter, and
 \OmegaLambda, for the matter and dark energy densities,
 respectively. If we parameterize the data in terms of an equation of
 state, $\wsym$ (with no time dependence), assume a flat geometry, and
 combine with baryon acoustic oscillations, we find that \wo. The
 largest source of systematic uncertainty on \wsym\ arises from
 uncertainties in the photometric calibration, signaling the
 importance of securing more accurate photometric calibrations for
 future supernova cosmology programs.
 Finally, we conclude that either the dust affecting the
 luminosities of SNe~Ia has a different extinction law ($R_V=1.8$)
 than that in the Milky Way (where $R_V=3.1$), or that there is an
 additional intrinsic color term with luminosity for SNe~Ia,
 independent of the decline rate. Understanding and disentangling
 these effects is critical for minimizing the systematic uncertainties
 in future SN~Ia cosmology studies.

\end{abstract}


\section{Introduction}

\label{sec:Intro}
Observations of high-redshift Type Ia supernovae (SNe~Ia) currently
provide the best evidence for an accelerating universe
\citep{Riess1998,Perlmutter1999,Knop2003,Astier2006,WoodVasey2007}.
Independently, the Wilkinson Microwave Anisotropy Probe (WMAP)
measurements of fluctuations in the cosmic microwave background (CMB)
\citep{Spergel2003,Spergel2007}; detections of acoustic oscillations
in the matter density spectrum \citep{Eisenstein2005}; as well as the
comparison of Hubble expansion ages from the Hubble Key Project
\citep{Freedman2001} with the ages of Milky Way globular clusters
\citep{Krauss2003}, have all led to the growing body of evidence for a
dark-energy component to the overall mass-energy density of the
Universe (see also \citet{FriemanTurner2008} for a recent review). The
above studies yield results consistent with a flat universe where the
sum of the dark energy and matter densities $\Omega_{DE}$ + $\Omega_m
$ = 1, and where $\Omega_{DE} \sim$ 0.7, with a matter density
$\Omega_m \sim$ 0.3.

At present, a physical understanding of this dark energy component
remains elusive, and there is a wide variety of possible alternatives
ranging from the cosmological constant, as originally proposed by
Einstein in 1917; a decaying scalar field; or perhaps even a
modification of general relativity.  A convenient measure is $\wsym$,
the ratio of pressure $P$ to energy density {$\rho$}, where $\wsym =
P / \rho$. In the case of a cosmological constant, $\wsym =
-1$. The time evolution of $\wsym$ is not known at present.  Given our
currently limited understanding of this critical component of the
Universe, it is clear that further observational and experimental data
are needed to constrain and characterize the properties of dark
energy.

SNe~Ia show a relation between peak brightness
and rate of decline, in addition to color (or reddening). Accounting
for these correlations yields a dispersion in the Hubble diagram of
about 7\% in distance
\citep{Phillips1993,Hamuy1995,Riess1996,Hamuy1996a,Astier2006}.  As
more and more SNe~Ia are discovered and the statistical uncertainties
in this method are decreased, the challenge becomes understanding and
controlling the remaining systematic uncertainties, as well as testing
for currently unknown systematic effects. A well-known systematic is
reddening and extinction due to dust, and there may potentially be
differences due to chemical composition and age of the progenitors
or their environment.  The requirement for
increasing measurement accuracy, the lack of a detailed theoretical
understanding of SN~Ia, the fact that most observations
have been made at rest-frame optical and ultraviolet colors (where
reddening uncertainties are large), the difficulty of obtaining
accurate K- and spectral-corrections, all point to the need to
characterize and decrease these systematic errors and uncertainties.
As other errors have been reduced over time, what were relatively
small effects previously have now become increasingly important as the
required precision in cosmology has risen.

Observations of SNe~Ia at near-infrared (NIR) wavelengths offer several
advantages in minimizing a number of systematic effects (most notably
reddening), as well as providing an additional, independent probe of
the expansion history. We make use of the $Y$ band, centered near 1.035
$\mu$m \citep{Hillenbrand2002}, which falls in between the
traditionally classified optical and infrared spectral regimes.  The $Y$
and $J$ (1.25 $\mu$m) bands provide a means of obtaining {\it
rest-frame} $I$-band magnitudes for SNe~Ia in the redshift range
0.1 to 0.7, an interval covering the time at which the influence
of dark energy  begins to dominate the expansion. Hence, NIR
observations offer an important opportunity both to minimize
systematic uncertainties and yield an independent Hubble diagram.

Aside from a few observations published by \citet{Riess2000} and
\citet{Nobili2005}, rest-frame $I$-band measurements have not yet been
routinely undertaken at higher redshifts because at z$\sim$0.25 the
$I$-band is redshifted beyond the CCD sensitivity limit at red
wavelengths.  A further challenge to NIR photometry obtained
from the ground is that the sky background is much greater than at
optical wavelengths. As a consequence, observing distant SNe~Ia in
the NIR is expensive in telescope time. However, since
current searches to find SNe~Ia now yield extensive optical
coverage of the light curves, the decline rates and time of maximum
light are already well defined, so that fewer observations are
required to derive NIR light curve parameters
accurately. With the availability of 6.5-meter class telescopes and
NIR arrays, a NIR study of SNe~Ia at z $>$ 0.1
has now become feasible.

Ongoing optical surveys aimed at discovering large samples of
SNe~Ia are signficantly decreasing the  statistical
uncertainties in SN~Ia cosmology measurements, while
simultaneously aiming to reduce the systematic errors.  The CSP
is complementary to these studies; that is, it is not designed to
rival optical surveys on numbers of objects and statistical errors,
but rather it is more focused on follow-up observations critical to testing
for and minimizing systematic uncertainties. These follow-up
observations (multiple wavelengths at low redshift, and NIR 
observations at higher redshifts) require more observing time per
individual SN~Ia.  An understanding of the systematic errors is a
critical goal for all SN~Ia programs.

To date, as part of the high-redshift CSP, we have obtained NIR
observations of \Nobs\ SNe~Ia. Subsequent \templateh\ observations have
been obtained for \Nwithtemp\ objects and we have fully analyzed
\Ngood\ objects, the sample for which photometry, light curves and a
preliminary Hubble diagram are presented in this paper.

\section{The Carnegie Supernova Project}

\label{sec:CSP}
The CSP is an umbrella name for our two coordinated SN
programs: one being carried out at the Las Campanas 1 m Swope and
2.5 m Dupont telescopes for low (0$<$z$<$0.1) redshift objects,
and the other using the Magellan Baade 6.5 m telescope at higher
(0.1$<$z$<$0.7) redshifts. Preliminary results from the low-redshift
part of this project have been reported in \citet{Hamuy2006},
\citet{Folatelli2006} and \citet{Phillips2007}; the most recent
results are presented in two companion papers \citep{Contreras2009,
Folatelli2009}.  The CSP is not a
SN {\it search} program; rather as described below, as well as
in \citet{Hamuy2006}, we are using the facilities of the Las Campanas
Observatory, in coordination with several on-going search programs
elsewhere, to provide detailed and immediate {\it follow-up}
observations of newly-discovered SNe.

\subsection{Goals}

\label{sec:goals}
Overall, the CSP is focused on obtaining observations of SNe~Ia
falling in the redshift range 0$<$z$<$0.7. As described in
\citet{Hamuy2006}, at low redshifts we are currently obtaining
10-filter ($uBVgriYJHK_s$) photometry with excellent time sampling and
optical spectroscopy to provide a database for the determination of
the Hubble constant, accurate K-corrections,
comparison with theoretical models of SNe~Ia, and a fundamental
dataset for comparison with high redshift. At high redshifts, as
described in this current paper, we are obtaining $YJ$ data near
maximum light. The NIR data, in conjunction with optical
photometry (being obtained as part of the search programs), allow the
determination of reddening corrections and a rest-frame
$I$-band Hubble diagram. One of the key goals of the CSP is to
minimize the effects of reddening in the Hubble diagram, and ensure
that the {\it rest-frame} ($BVi$) bandpasses, being observed at low
redshift, match those for a sample at higher redshift, so that
reddening corrections can be applied in a uniform way.  Ultimately the
goal is to measure accurate luminosity distances to better constrain
cosmological models. Assuming a model including dark energy, the goal
is to characterize the acceleration of the Universe and constrain the
equation of state, $\wsym$, to a  precision {\it and accuracy}
of $\pm$10\%.

\subsection{``I-band'' filter nomenclature}
There are many ``I-band'' filters in current use. We
distinguish here between the Kron-Cousins $I$-band, the SNLS MegaCam
$i_M$-band (based on Landolt standards calibrated to Vega), and the
SDSS-II $i_S$-band (based on a calibration tied to four subdwarfs,
\citet{Smith2002}). Our Las Campanas natural $i$-band calibration is
also tied to Smith et al.  The central wavelengths of these (and
other) passbands are defined in Table \ref{tab:filters}. In this
paper, we also refer to a generic `I-band' when speaking collectively
about observations in the NIR wavelength range $\sim
8000\AA$.

\subsection{I-band Light Curves SNe~Ia}

\label{sec:IbandLCs}

\begin{figure}
\centerline{
\includegraphics[angle=-90, width=5in]{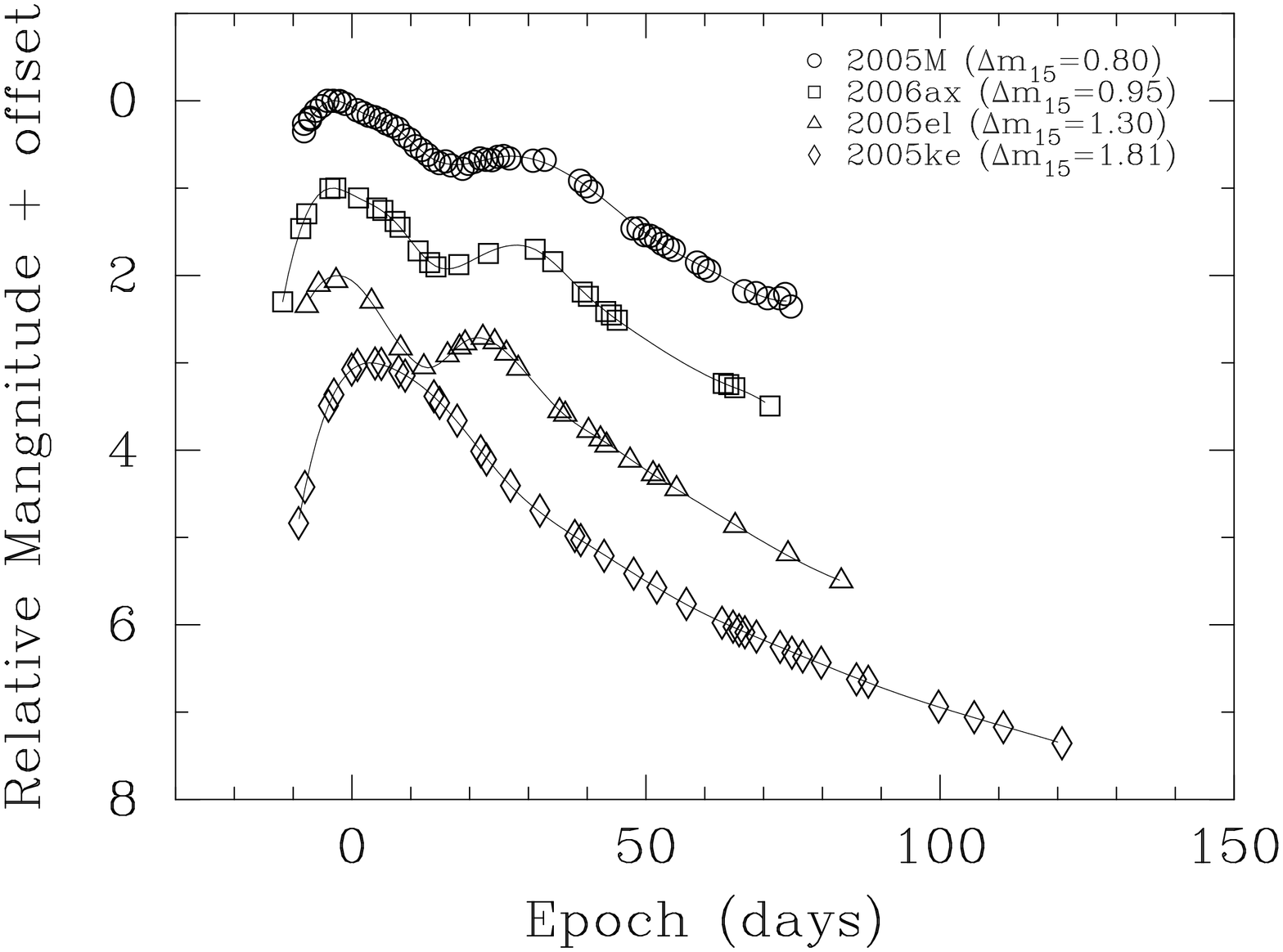}
}
\caption{A representative sample of four $i$-band lightcurves from
  the low-z CSP sample chosen to illustrate the behavior of the secondary
  maximum.  For clarity, the lightcurves have been normalized to their
  peak magnitudes and then offset by one magnitude from each 
  another. The SNe~Ia have values of \dm\ ranging from
  1.12 to 1.81. The fast-declining SN~Ia with $\dm =
  1.81$ (bottom light curve) shows no secondary maximum.
\label{fig:i_band_light_curves}}
\end{figure}

The form of NIR light curves for SNe~Ia differs from those
at optical wavelengths. At $IYJH$ and $K$, the light curves show a
minimum following maximum light, followed by an additional secondary
maximum, less luminous than the primary peak
\citep[e.g.,][]{Elias1985,Hamuy1996b,Meikle2000,Krisciunas2001,Nobili2005}.
This secondary maximum is thought to be the result of a sudden drop in
the mean opacity in the SN ejecta \citep{Pinto2000, Kasen2006}.
In Figure \ref{fig:i_band_light_curves}, we show some examples of
$i$-band light curves for four well-observed, nearby SNe~Ia from
the low-z CSP sample. Our strategy for the high-redshift CSP is to
obtain $i$ photometry covering the first peak only, and to avoid the
(fainter) secondary maximum. The optical surveys already provide
measurements of the decline rate in the $B$ band, so duplication of
these data is not required. The median start and end times of our
observations are 0 and +12 days with respect to observed $i$-band maximum,
respectively. In this paper, unless otherwise noted, we refer to
the observed (not rest) frame time.

\subsection{Targets}

\label{sec:targets}
There have been four on-going SN search programs with which the
CSP has been actively collaborating.  At low redshifts (z $<$
0.1), we are collaborating with the Lick Observatory Supernova Search
(LOSS), as described in \citet{Hamuy2006}.   For intermediate
redshifts (0.1 $<$ z $<$ 0.4), we have been collaborating with the
Sloan Digital Sky Survey II (SDSSII), while the CFHT Legacy Survey
(SNLS) and ESSENCE found SNe~Ia out to higher redshifts (0.1 $<$ z
$<$ 0.7). The three surveys relevant to this higher-redshift study are
described briefly below.

\subsubsection{Sloan Digital Sky Survey (SDSS) II }

\label{SDSS}
An extension to the Sloan Digital Sky Survey \citep{York2000}, the
SDSS-II,\footnote{http://www.sdss.org} \citep[]{Sako2008,Frieman2008}
has completed a three-year rolling search program using the 
2.5 m SDSS telescope to discover intermediate redshift (0.05 $<$ z $<$
0.35) SNe~Ia in a 300 square degree southern equatorial field,
obtaining multicolor ($ugriz$) photometry  with a cadence of $\sim$2-5 days. The data were
obtained during the months of September through November
2005-2007. 
Spectra to determine redshifts and types for the SDSSII candidates
were obtained with several telescopes with a range of
apertures. Photometry for the 130 spectroscopically confirmed SNe~Ia 
from the 2005 season have been presented in \citet{Holtzman2008}.

\subsubsection{CFHT Legacy Survey (SNLS)}

\label{sec:SNLS}
The Supernova Legacy
Survey\footnote{http://www.cfht.hawaii.edu/Science/CFHLS}, a
Canadian/French collaboration,  used the CFHT and the 1
square-degree camera (MegaCam) to obtain deep optical
($u^\prime$$g^\prime$$r^\prime$$i^\prime$$z^\prime$) images for four
fields, each of approximately one square degree around the equator
\citep[]{Astier2006}, beginning in February, 2003. The SNLS was also a
rolling search program in which each field was revisited every second
night during a 5-month campaign each semester for a total of five
years. Spectroscopic follow-up for the SNLS candidates was obtained at
the 10 m Keck telescope, the 8 m Very Large Telescope (VLT), the 
8 m Gemini North and South telescopes, as well as the Magellan telescopes.  
In their first year of
operation, they discovered 91 SNe~Ia, and published data for a sample
of 71 SNe~Ia with redshifts 0.1 $<$ z $<$ 1.1.  In total, about 450
SNe~Ia have been discovered.

\subsubsection{ESSENCE}

\label{sec:Essence}
ESSENCE\footnote{http://www.ctio.noao.edu/essence} completed its
survey using the 4 m CTIO Blanco telescope and MOSAIC II wide-field
camera covering equatorial fields at $VRI$ wavelengths over the
redshift range between 0.15 $<$ z $<$ 0.75
\citep[]{Matheson2005,Miknaitis2007}. The goal was to provide optical
light curves for about 200 Type~Ia SNe~Ia over 5 years
(2002-2007). ESSENCE was scheduled for 30 half nights during a 3-month
(October - December) campaign each year. The observing was centered on
new moon, and was scheduled every other night, for a span of 20 days,
with a gap of 10 bright nights per month.  Spectroscopic follow-up for
the ESSENCE candidates was obtained at Gemini North and South, Keck,
the VLT, MMT, and Magellan telescopes. At the time of writing, ESSENCE
has published observations for \NobsESSENCE\ SNe~Ia
\citep{WoodVasey2007}.

\subsection{CSP Magellan Follow-up Observations}

\label{MagellanFollowUp}

\begin{figure}
\centerline{
\includegraphics[angle=-90, width=5in]{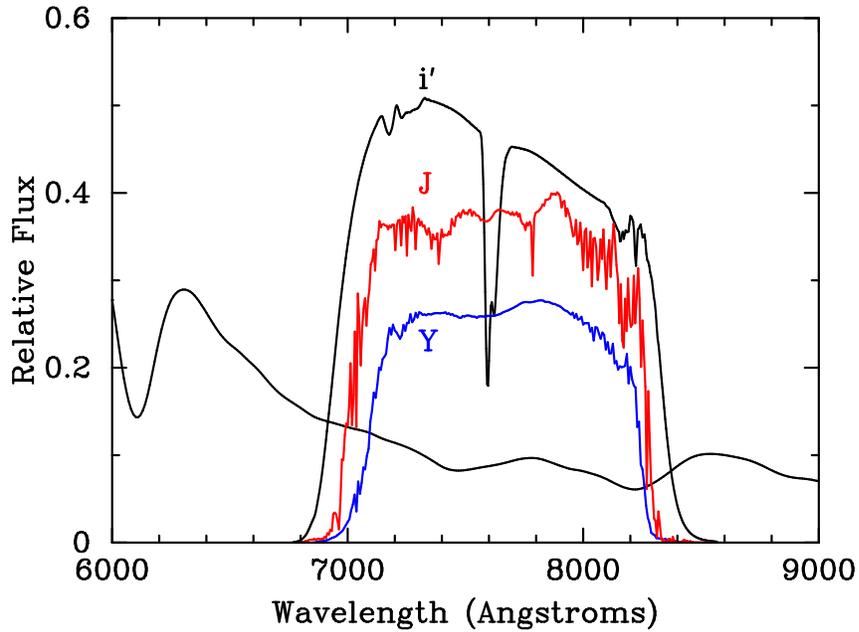}
}
\caption{Filter response functions for the $Y$ (blue) and $J$ (red)
bands blueshifted by 0.35 and 0.63, respectively.  Rest-frame
$i$-band is plotted in black. The absorption feature at
$\lambda \sim 7600\AA$ is telluric O$_2$.  Crossing the entire figure,
the SED of a typical SN~Ia at maximum \citep{Hsiao2007}
is also shown in black.  \label{fig:YJi}}
\end{figure}

Our follow-up Magellan observations consist of $Y$- and/or $J$-band
images. We chose not to observe at $H$-band because of the increased
sky background produced by atmospheric OH airglow emission.  Optical
photometry for this sample of SNe~Ia was obtained during the
course of the original surveys (SDSS-II, ESSENCE, SNLS), and was not
repeated.  The Sloan $i$ filter, which we are using at the Swope
telescope for the low redshift part of the CSP, overlaps
well with the $Y$-band for a redshift of z$\sim$0.35, and the $J$-band
at redshift z$\sim$0.63 (see Figure \ref{fig:YJi}). If targets at
these redshifts were available, they moved to the top of our observing
priority list. We aimed to obtain Magellan observations no more than a
few days after maximum light, and preferably while still on the
rise. Our criteria for selection of the SNLS, ESSENCE and SDSS-II
objects were: 1) the SNe~Ia were in the redshift range 0.2 $<$ z
$<$ 0.7, 2) the SN~Ia was estimated to be near maximum, 3)
SNe~Ia near z = 0.35 and z = 0.65 were given preference and 4)
preference was given to SNe~Ia well-separated from the host. In
practice, criteria 3) and 4) were rarely invoked since there was not
much choice after criteria 1) and 2). Our selection criteria therefore
matched fairly closely those of the parent surveys.

The SNLS, ESSENCE and SDSS-II
projects provided coordinates, finder charts and epochs of
high-probability SN~Ia candidates for the Magellan observations. In
three cases (SNLS 03D1gl, SNLS 04D2cw, ESSENCE 2004-H-280), the
spectral classifications were later revised (to types other than SNe~Ia).
In one case (SNLS 04D2al), the redshift was too large
($z=0.838$) and the SN~Ia was undetectable in our observations. In
Table \ref{tab:CSPSNe}, we list the SN~Ia name, position, number
of CSP observations, and IAU cross-identification, for the SNe~Ia.

\section{Instrumentation and Observations}

\subsection{Instrumentation}

\label{sec:instrumentation}
Our NIR imaging has been carried out using $Y$- and $J$-band
filters on the Persson Auxiliary Nasmyth Infrared Camera (PANIC)
\citep{Martini2004}.  This camera is mounted on the east Nasmyth
platform of the Magellan Baade telescope.  It contains a
1024x1024 HgCdTe Hawaii-1 array with a scale of 0.125 arcsec
pixel$^{-1}$, and covers a field of view of 2.1 $'\times $ 2.1$'$ on
the sky.

\subsection{Observations}
\label{sec:observations}

\subsubsection{Observing Strategy and Current Status}
Our observing goal was to acquire $Y$ and/or $J$ photometry with gaps in
the SN~Ia light curves no larger than 7 days, straddling the light
curve over maximum light if possible.  To avoid large gaps, PANIC was
scheduled during dark/grey time, in addition to the customary bright
time for infrared instruments. In practice, telescope time was
scheduled with gaps generally less than 5 days, although gaps of 8
days were occasionally unavoidable. A record of observations,
including finding charts of the objects, is maintained on our web site
for the project, which can be found at
http://www.ociw.edu/csp/. Generally, 3 to 5 observations per SN~Ia
were obtained.

We began this long-term project with a pilot program in November 2003
running through April 2004,  centered on the best weather
period at Las Campanas.  Our first \templateh\ images 
(once the SN~Ia had faded) were obtained in the
following year.  In the second year, we observed from October 2004
through March 2005. Unfortunately, poor weather at Mauna Kea during the
northern winters of 2003 and 2004 resulted in a very low yield of
SNe~Ia for CSP follow-up during this pilot project.  Moreover,
ESSENCE was operational only during the northern fall, and SDSS-II began
routine operations during the fall of 2005.  We therefore 
shifted the  CSP follow-up campaigns to August
through January when there was good overlap with all three SN
search programs.

\begin{figure}
\centerline{
\includegraphics[width=5in]{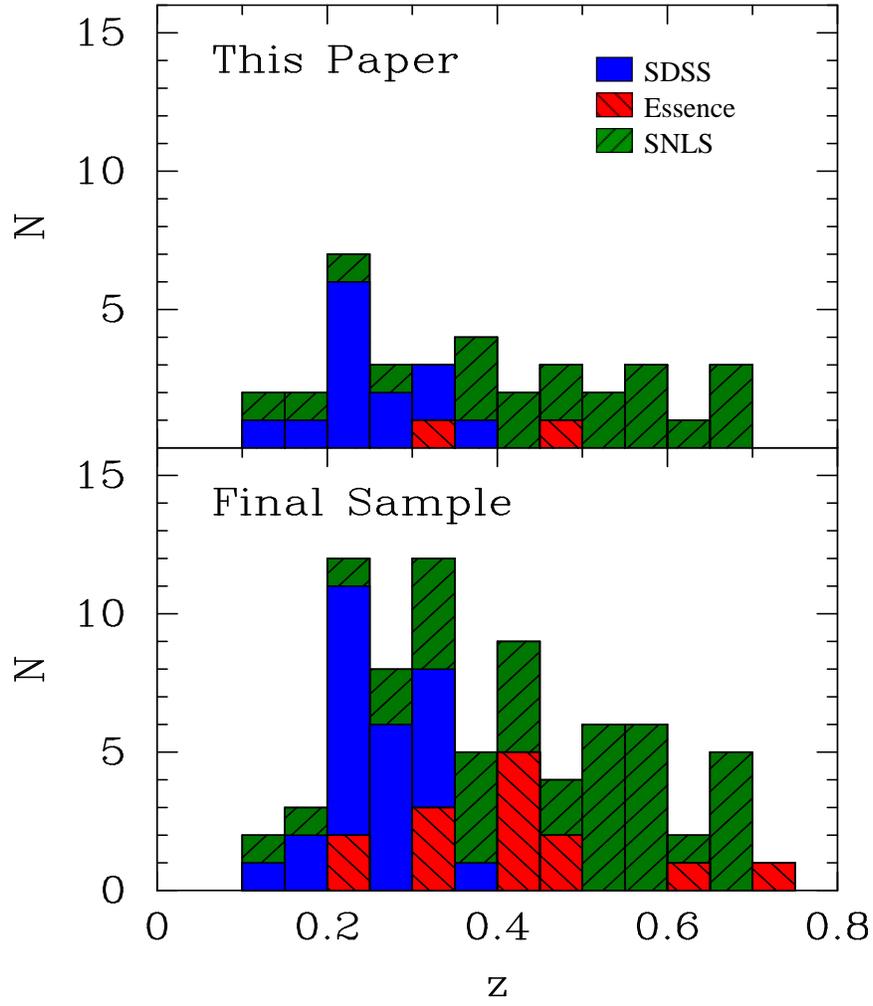}
}
\caption{Redshift distribution of objects observed as part of the CSP
  in the NIR on Magellan.  In general, the higher-redshift objects are
  found by the SNLS, and those at lower redshifts primarily by the
  SDSS-II. We have observed 14 objects discovered by ESSENCE, 37 SNLS
  objects and 24 SDSS objects.
\label{fig:zhistogram}}
\end{figure}

After four campaigns, we obtained Y and/or J photometry for a total of
\Nobs\ objects, \NTypeIa\ of which were ultimately confirmed as SNe~Ia.
Follow-up \templateh\ observations have been obtained for
\Nwithtemp\ of these SNe~Ia and we are in the process of obtaining
\templateh\ images for the remaining objects.  The redshift
distribution for our total sample is shown in Figure
\ref{fig:zhistogram}. In this paper, we report on the photometry from
the first \Ngood\ SNe~Ia.  A summary of the observations for these
\Ngood\ SNe~Ia is given in Table \ref{tab:CSPSNe}. At the current
time, data for 20, 13, and 2 SNe~Ia have been analyzed from the SNLS,
SDSS-II and ESSENCE surveys, respectively. The original goal of the
CSP was to acquire a sample of 100 objects over the redshift range 0.1
$<$ z $<$ 0.7; however, with the conclusion of the three surveys, our
expected sample will be limited to \NTypeIa\ SNe~Ia.

\subsubsection{Observing Procedures}

Twilight sky flats were taken in each filter every night in at least 5
dithered positions with 2 exposures at each position, controlling the
count level to be in the linear regime. The detector becomes nonlinear
at the 1\% level at approximately 13,000 ADU and at the 5\% level at
approximately 35,000 ADU.  Dark frames were taken in sets of 15, each
with exposure times matching those of the science images. The
SN~Ia observations were obtained using 9 dither positions with 2
exposures per position.  The exposure times used for individual images
were 120 seconds with total exposure times ranging from 2160 to 8640
sec.  Reference images for these SN~Ia fields were taken in the
same manner once the SN~Ia faded, to allow accurate background
subtraction. The signal to noise in the stacked reference
images were as high (or higher) than those in the SN~Ia images, so
as not to degrade the photometry.  The seeing for these images
ranged typically between 0.4 and 0.8 arcsec. On photometric nights, 3
to 5 stars chosen from the standard system defined by \citet{Persson1998} 
were observed throughout the night in each filter.  These
standard stars were observed in 5 dither positions with 2 exposures at
each position.  The exposure times of individual images were 3 or 4
seconds, resulting in total exposure times of 30 and 40 seconds.  On
occasion, the telescope was de-focused to ensure the counts remained
in the linear regime of the detector.

\section{Data Analysis}

\subsection{Pipeline Processing}

\label{sec:pipeline}
We have developed an automated pipeline to process the PANIC images.
The pipeline carries out (1) linearity corrections, (2) dark
combination and subtraction, (3) bad-pixel mask production, (4)
flat-field combination and division, (5) sky image computation and
subtraction and (6) combination of dithered frames into final stacked
images.  We apply a predetermined linearity correction law to every
pixel value above 8,000 ADU. The multiplicative correction ranges from
1.0 at 8000 ADU to 1.06 at 40,000 ADU.  Dome flats are created by
subtracting images of equal exposure times taken with no dome lamps
from those with the dome lamps on. The final dome flats are used only
to build bad pixel masks for each night and median-combined twilight sky
flats are used for flat-fielding.  Sky frames are subtracted from the
individual object frames using modal scale factors.  Finally, stacked
images are created by aligning and averaging the individual object
frames.

\subsection{Galaxy Template Subtraction}

\label{sec:TemplateSubtract}
To obtain accurate photometry for the SNe~Ia, the host galaxy
light must be subtracted from the images. The strategy for removing
the host galaxies from the SN~Ia data involves three steps: (1)
obtaining \templateh\ images of the SN~Ia fields in the year
following the events; (2) registering and matching the point-spread
functions (PSF) of the images; (3) subtracting the PSF-matched
SN~Ia and \templateh\ images. The algorithm developed for the
registration and non-parametric matching of the PSFs
will be described in more detail in a later paper \citep{Kelson2009};
the method is summarized briefly here. 

For a given SN~Ia image and associated \templateh\ image,
SExtractor \citep{Bertin1996} is used to identify objects down to a
threshold of 3-$\sigma$. The positions of these objects are used to
compute the coordinate transformation between each SN~Ia image and
the \templateh\ image. Matching the PSF of the \templateh\ image, $T$,
to that of the SN~Ia image, $S$, consists of solving for the
convolution kernel, $k$, that maps point-sources in $T$ to $S$. Once
$k$ is known, the entire \templateh\ image can then be convolved with
$k$ and subtracted from $S$ to isolate the SN~Ia. However, our
fields of view are small and in several instances lack suitable point
sources for determining the kernel using more traditional techniques,
for example, parameterizing the kernel with a Gaussian (Alard \&
Lupton 1998, Alard 2000). Instead, we use a non-parametric technique
that utilizes all objects in the field to constrain the kernel. The
kernel consists of a $\left(2M+1\right)\times\left(2M+1\right)$ matrix
indexed by $u$ and $v$, and is determined by minimizing

\begin{equation}
\chi^{2}=\sum_{i}^{N}\biggl|{\frac{S\left(x_{i},y_{i}\right)-\sum_{u}\sum_{v}k\left(u,v\right)[T(x_{i}-u,y_{i}-v)]}{\sigma(x_{i},y_{i})}}\biggr|^{2}
\nonumber 
\end{equation}

\noindent
where the sum is over $N$ pixels in the source and template images.

The matrix of $T(x_{i}-u,y_{i}-v)$ is decomposed using singular value
decomposition, and its constituent eigenvectors contain the kernel's
natural set of orthogonal basis functions.  The eigenvalues represent
each eigenvector's sensitivity to noise.  We eliminate those basis
functions that do not contribute to reducing the $\chi^{2}$ per degree
of freedom. Because image re-binning is mathematically equivalent to a
convolution, errors in the registration are fully accounted for by the
convolution kernel that minimizes $\chi^{2}$.

\begin{figure}
\centerline{
\includegraphics[width=4in]{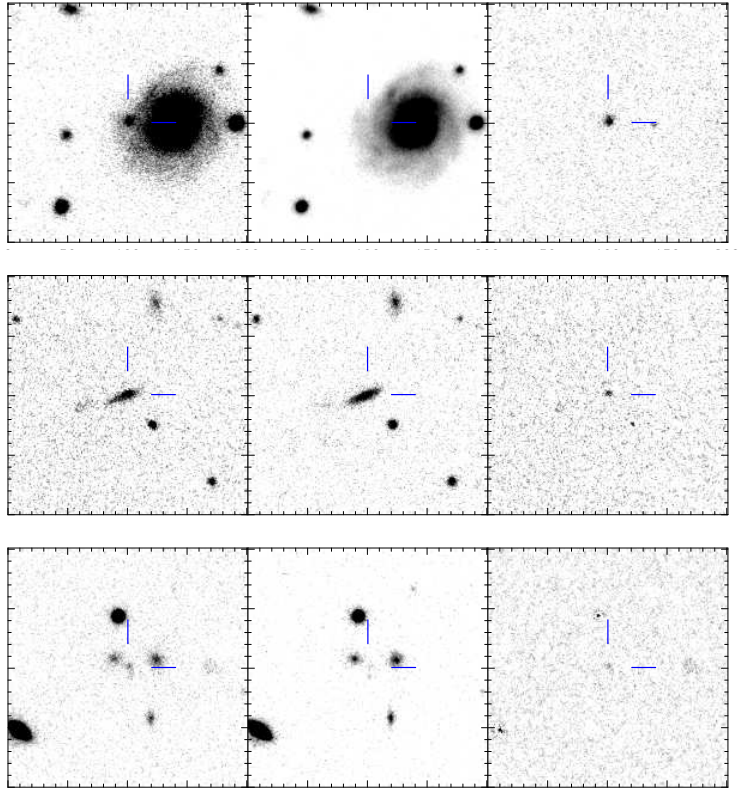}
}
\caption{Sample of SNe~Ia, post-SNe~Ia and difference frames
for three SNe~Ia: SN~3241 (discovered by SDSS-II, top panel),
04D1rh (discovered by SNLS, middle panel), and 05D2bt
(discovered by SNLS, bottom panel).  SN~3241 ($z=0.25$) was observed
in the Y-band, 04D1rh ($z = 0.435$) was observed in both Y and J
(J is shown),
and 05D2bt ($z=0.679$) was
observed in the J-band. The scale of these images is 200 pixels = 25
arc-seconds on a side.
\label{fig:temp_subtract} } 
\end{figure}

Ideally, one wishes to convolve (i.e., degrade) the \templateh\ images
so as not to decrease the signal-to-noise of the SN~Ia. However,
approximately 8\% of our SN~Ia images were taken under
exceptional seeing conditions (less than 0.35 arcsec) which, to date,
have not been matched in our \templateh\ observations. As a result,
this subset the SN~Ia images have to be degraded to match the
poorer image quality of the \templateh\ images. Fortunately, these
SNe~Ia have lower redshifts ($z\leq0.3$) and are relatively
bright, so that the signal to noise remains high. The SN~Ia images
in our sample to date have a range of image quality, from 0.3 arcsec to
1.4 arcsec; the \templateh\ images have a range of image quality, from
0.3 arcsec to 1.3 arcsec. We find that our image subtraction
technique works well for removing the host galaxies for a wide range
of seeing conditions, position of the SN~Ia relative to the host
galaxy, and redshift. Three examples of a \templateh-subtracted image
are shown in Figure \ref{fig:temp_subtract}, for objects at redshifts
of 0.25, 0.30, and 0.68, respectively.

\subsection{Photometry}

\label{sec:photometry}
Observing SNe~Ia in the NIR is more challenging than in
the optical owing to higher sky background. The
contrast of SN~Ia to galaxy is also less in the red, and overcoming
these effects require longer integrations than in the optical. We
measure the flux with a two-step approach.  The first step is to
measure the magnitudes of several stars in each SN~Ia field, which
we shall refer to as tertiary standards. The primary standards are
those that establish the JHK system \citep{Elias1982}, on which the
secondary standards \citep{Persson1998} that we observe are based.  In
choosing the tertiary standards, we require that: 1) the star is no
closer than 20 arcsec to the edge of the PANIC field of view 2) the star
appears in all observations of the SN~Ia and the \template; and 3)
there are no significant residuals for the star after \templateh\
subtraction.  We use DAOPHOT and DAOGROW \citep{Stetson1990} to measure
and fit a growth curve (flux versus aperture size) for each star. We
then compute an aperture correction using these fits to get the flux
measured through a 10-arcsec-diameter aperture, matching the aperture
used on the standards. These fluxes are therefore calibrated with
respect to the secondary standards observed that evening, and are then
averaged over all photometric nights.

The second step is to measure the flux ratios between the SN~Ia
and the tertiary standards. At the highest redshifts, we are working
at the detection limit of the telescope where DAOPHOT and DAOGROW are
no longer robust.  Instead, we use the optimized extraction algorithm
of \citep{Naylor1996}. In brief, the brighter stars in the field are used
to estimate the PSF, $P_{i,j}$, modeled as a
superposition of Gaussian and Moffat profiles. This estimated PSF is
then used as a weight mask in summing the flux from the SN~Ia
pixels: $F=\sum_{i,j}w_{i,j}\left(D_{i,j}-S_{i,j}\right)$ where
$w_{i,j}=\frac{P_{i,j}}{\sum P_{i,j}^{2}}$, $P$ is the model PSF, $D$
is the measured counts, and $S$ is the sky flux. Following 
\citet{Naylor1996}, we
have assumed that the variance in the weights is dominated by the sky
($V_{ij}\simeq V_{s}$ in his equation (10)). This yields an optimized
estimate of the flux for the SN~Ia. We then use the same weight
mask to measure fluxes and compute flux ratios between the SN~Ia
and each tertiary standard. These flux ratios, together with the
calibrated fluxes from step 1, provide estimates of the calibrated
flux of the SN~Ia.  These estimates are averaged to yield the
final calibrated flux for the SN~Ia. We have done a number of
tests in which fake point sources are inserted into the frames and then
recovered, and find that the method works extremely well. These
results will be reported in \citet{Kelson2009}.

We have also taken great care in computing the variances of our
fluxes.  In the process of correcting for image distortion in the
PANIC pipeline, rectifying the images, convolving with a kernel,
and then subtracting a \template\ image, the pixels in each image
have become correlated.  We produce variance maps for each
observed field early in the PANIC pipeline. A simple propagation
of errors is done at each step of the pipeline, updating the
variance in each pixel in the maps.  The final variance maps are
used to compute the noise rather than image statistics, ensuring
that we can properly estimate the variance of the tertiary
standards and the SN~Ia. 

\subsection{Absolute Calibration}

\label{sec:calibration}

The absolute calibration of the Magellan photometry is based on $JHK$
standard stars from \citet{Persson1998}, where the zero point is tied
to Vega.  Currently, we are applying the $Y$-band absolute calibration
described in \citet{Hamuy2006}, using Kurucz model spectra and
\citet{Hillenbrand2002} standard star measurements.
\citet{Contreras2009} have recently obtained new observations
confirming this calibration to an accuracy of $\pm$0.01 mag. We have
observed common standard stars and adopted the same procedures for the
reduction of the standard stars so that differences in the calibration
of low- and high-z samples are minimized.  Adopting atmospheric
extinction coefficients of k$_Y$ = 0.10 and k$_J$ = 0.12 mag/airmass
following Hamuy et al., we solve for the nightly zero points in each
filter.  Photometry for the standard stars is obtained with apertures
of diameter 10 arcsec. The statistical errors in the zero point are
determined from the scatter of individual measurements and range from
$\pm$ 0.01 to $\pm$ 0.05 mag. A minimum of three photometric nights
determines the absolute flux for each SN~Ia (see \S
\ref{sec:photometry}). The average scatter in the zero points based on
observations of standard stars is 0.023 mag at $J$ and 0.013 mag at
$Y$.

In this (and other CSP) papers, we are presenting our photometry in
our own natural system. The advantage of using the natural system is
that it avoids the uncertainties resulting from the broad features
present in the spectra of SNe~Ia, which present challenges for the
transformation onto the standard system. As outlined in
\citet{Contreras2009}, the natural photometry is obtained by first
computing color terms that transform local standard sequences of stars
to the system appropriate for each filter (\citet{Landolt1992}
standards for $BV$, \citet{Smith2002} standards for $ug ri$, and
\citet{Persson1998} for $YJ$).  These color terms are then used in
reverse to transform the standard magnitudes to our natural system and
it is these magnitudes that are used to calibrate the SN~Ia
photometry.  The natural magnitudes are therefore equal to the
standard magnitudes at zero color.  Our natural system magnitudes can
be straightforwardly transformed to other systems. Transmission curves
for our NIR and optical filters are given in 
\citet{Hamuy2006} and \citet{Contreras2009}, and updated versions 
are available online at
the CSP web site. In the SN~Ia target fields, we identified
several isolated stars to serve as tertiary standards to determine
absolute flux for those nights that were not photometric. Photometry
for these secondary standards is also available on line. In Table
\ref{tab:filters}, we list the filters, effective wavelengths, and
published references to the calibrations relevant for the CSP. A
comparison of our own NIR $YJ$ photometry (K-corrected to
the $i$ band) to that of the optical surveys (SNLS, ESSENCE and
SDSS-II) requires that we adopt a standard with a measured spectral
energy distribution (SED)
spanning the optical and NIR. For this
purpose, we have chosen the \citet{Bohlin2004} model for Vega, as
updated in \citet{Bohlin2007}.

\section{Systematic Effects}

\label{sec:systematics}

As the number of objects has increased, SN~Ia cosmology has reached
the stage where the systematic uncertainties are becoming the dominant
source of error.  The CSP has been designed to minimize known
systematic effects, particularly those due to K-corrections,
reddening, and cross calibrations to different photometric
systems. Below we discuss our current approach to dealing with
K-corrections and reddening.

\subsection{K-corrections}

\label{sec:K-correct}
The observed SEDs of SNe~Ia are shifted and
stretched with redshift due to the expansion of the Universe. Accurate
corrections for these effects (K-corrections) for SNe~Ia remain an
observational challenge, and much effort has been put into creating
libraries of SN~Ia spectra with which to assemble SEDs
that can be used to estimate the K-corrections and their
variance \citep{Nugent2002,Hsiao2007}.  The K-corrections are computed
in a manner similar to \citet{Kim1996} and \citet{Nugent2002}, using
the following formula:

\noindent
\begin{eqnarray}
K_{AB}\left(t\right) & = & 2.5\log\left(1+z\right) + \\
                     & & 2.5\log\left[
    \frac{ \int R_B\left(\lambda\right) \Phi^\prime \left(\lambda ;t\right) \lambda d\lambda}
         { \int R_A\left(\lambda\right) \Phi^\prime \left(\lambda\left(1+z\right) ;t\right)
         \lambda d\lambda}\right] + \mathcal{Z}_A - \mathcal{Z}_B
         \nonumber \label{eq:Kcorr}\end{eqnarray}

\noindent
where $A$ represents the observed filter and $B$ represents the
rest-frame filter to which we are transforming, $R_A$ and $R_B$ are
the corresponding observed and rest-frame filter response curves, and
$\mathcal{Z}_A$ and $\mathcal{Z}_B$ are the photometric zero-points.
The K-corrections are therefore applied in the following sense: $
m_{B} = m_{A} - K_{AB}$.  The spectral energy distribution (SED)
$\Phi^\prime\left(\lambda ;t\right)$
at epoch $t$ is obtained by ``color-matching'' the corresponding SED
template from \citet{Hsiao2007}.  Initially, K-corrections are
computed using unmodified \citet{Hsiao2007} SED templates and applied
to the photometry.  Light-curve templates are then fit to the $N$
filters for which there is optical and NIR photometry
yielding $N-1$ colors as a function of epoch.  We then construct a
smooth function $S\left(\lambda\right)$ which, when multiplied by the
template SEDs, yield synthetic colors equal to the observed colors;
i.e., we model the template rather than the (more noisy) observed
colors.  Finally, these improved SEDs are used in Equation
\ref{eq:Kcorr} to compute the final K-corrections.  This
color-matching simultaneously accounts for the intrinsic color
variations from SN~Ia to SN~Ia as well as reddening
corrections due to dust.

\citet{Hsiao2007} demonstrate that while K-corrections are determined
mainly by broad-band colors, accounting for differences in spectral
features is also necessary. These corrections are based on a larger
library of spectra, with a greater number of epochs and wavelength
coverage than previously available. They include 67 spectra from the
CSP (34 of which cover the I-band), obtained at the DuPont telescope,
and made available to Hsiao et al.  for this purpose.  Overall, this
new library contains many more spectra with red wavelength coverage,
and the telluric features at $\lambda \simeq 6880$\AA, $7200$\AA,
$8200$\AA, and $9400$\AA ~have been identified and removed. In the
$I$-band, the Hsiao et al. template includes approximately 250
spectra. Determining accurate K-corrections in the $i$-band requires
careful attention to the broad Ca II triplet P-Cygni absorption
feature (8498, 8542 and 8662\AA).  Fortunately, for the purposes of
the CSP, it is only well after maximum light has occurred that the Ca
II feature changes dramatically \citep[see Figure 5]{Hsiao2007}.
Currently, in the spectral region of the $i$-band, \NSpec\ spectra
have been used for the K-correction template. We are continuing to
acquire additional spectroscopy of our low-redshift candidates,
specifically to improve the K-corrections as a function of both epoch
and decline rate of the SNe~Ia.

\begin{figure}
\centerline{
\includegraphics[angle=-90, width=5in]{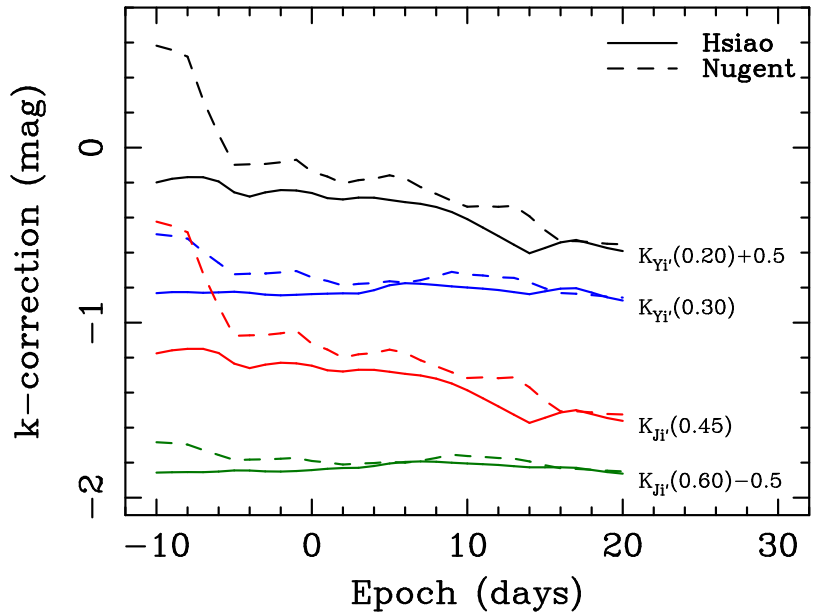}
}
\caption{Cross-band K-corrections (K$_{iJ}$ and K$_{i
Y}$), based on SED templates of \citet{Hsiao2007} and
\citet{Nugent2002}, shown as a function of epoch for redshifts z =
0.20, 0.30, 0.45 and 0.60. The solid lines are the Hsiao et
al. corrections and the dashed lines are Nugent et al.  The largest
differences occur at about 10 days before maximum. Given the
significantly larger template library available to Hsiao et al., and
the corrections for the telluric features, we have adopted the Hsiao
et al. corrections.
\label{fig:K-corr}}
\end{figure}

We show in Figure \ref{fig:K-corr} typical cross-band K$_{iY}$ and
K$_{iJ}$ corrections based on the \citet{Hsiao2007} library, plotted
as a function of time since maximum light at $B$, for redshifts 0.2,
0.3, 0.45, and 0.6. For comparison, we also show the K-corrections
from \citet{Nugent2002}. Given the currently larger number of
available $I$-band spectra, and correction for the presence of
telluric features, we have adopted the newer K-corrections from Hsaio
et al. for the purposes of this study. The greatest differences occur
at early times, 10 days before peak magnitude.  The corrections are
greatest at late times in the SN~Ia evolution, as well as very early
times for z=0.6. For our observations around peak magnitude, the
K-corrections at these redshifts range from $-0.5$ to $-0.3$
magnitudes in Y-band and from $-1.1$ to $-0.9$ magnitudes in J-band.
Table \ref{tab:Photo} contains our currently adopted values for the
K$_{iY}$ and K$_{iJ}$ -corrections.  In order to estimate the
uncertainties in the K-corrections, we apply the same procedure to the
SNe~Ia whose spectra were used to generate the template SEDs
themselves.  For those epochs with spectra, we can compute the
K-corrections using the color-matched SED and the observed library
spectrum, and compare the results.  For each redshift, we compute
synthetic photometry and K-corrections based on the library spectrum
and appropriately redshifted filter functions, simulating what would
be observed if this SN~Ia were at redshift $z$.  The SED template is
then color-matched to these synthetic observed colors and the
K-correction computed again.  This is repeated for each library
spectrum and the $rms$ difference between the library and template
K-corrections is computed. Since ESSENCE, SNLS, and SDSS-II have three
different filter sets, we have separately computed the uncertainties
in each case.

\begin{figure}
\centerline{
\includegraphics[angle=-90, width=5in]{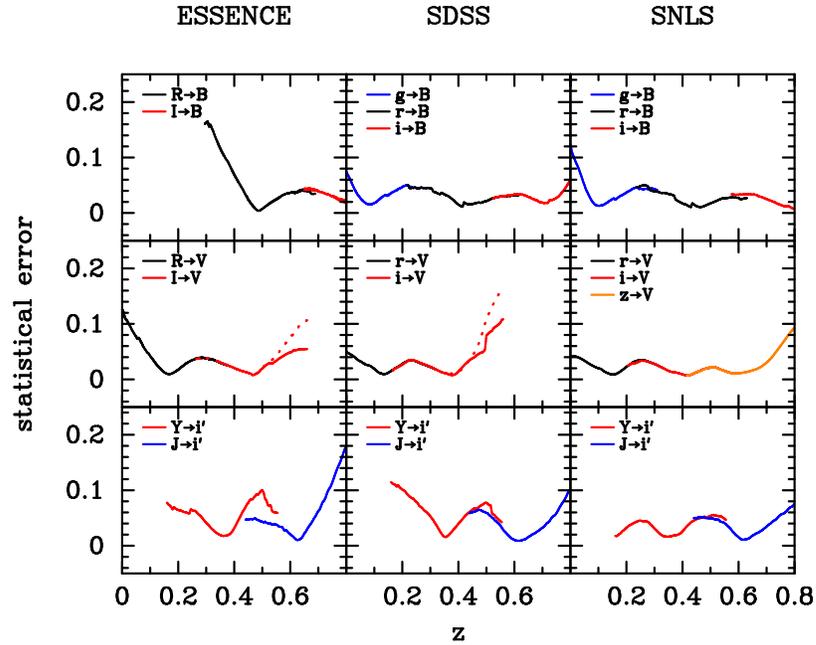}
}
\caption{Estimated dispersion in the K-corrections as a function of
redshift for different photometric filter sets.  Solid lines represent
the dispersion when full optical and NIR filter sets are
used to color-correct the template SED, whereas dashed lines show the
dispersions when no NIR data are included.  The dispersion
in the $K_{RB}$ for ESSENCE increases dramatically at low redshift
because $R_{4m}$ is redder than rest-frame $B$ and there are therefore
no observations to anchor the SED template on the blue side to which
we are transforming.  Our lowest redshift ESSENCE object in this
analysis is at z = 0.34.  At the high-redshift end, the uncertainty
for $K_{IV}$ is higher since $I_{4m}$ is shifted to the blue side of
rest $i$, and there is no anchor on the red side.  Adding the
NIR photometry helps in the color-matching and decreases the
dispersion.  There is little effect on the SNLS dispersions because
our CSP sample includes only objects with redshifts less than 0.8 and
the $z_m$ band serves as a red anchor over this entire range.
\label{fig:k-disp}}
\end{figure}

In Figure \ref{fig:k-disp} we plot the statistical dispersion in the
K-corrections as a function of redshift for the filters used in
ESSENCE, SDSS-II and SNLS.  The dispersion is sensitive to the filter
set used, which determines the accuracy with which we can
color-correct the template SED.  As expected, the dispersions are
lowest at the redshifts where the observed and rest-frame filters
overlap and increase at higher and lower redshifts. The largest
uncertainties arise when there are no filters to anchor the blue side
of the SED at low redshifts or, alternatively, the red side at high
redshifts. For comparison, we plot the dispersions with (solid) and
without (dashed) including the NIR photometry; in most
cases, the curves overlap and are indistinguishable.  In general the
statistical uncertainties in the K-corrections range from $\pm$0.005
to $\pm$0.05 mag.  The largest uncertainty is for d149 at a redshift
of 0.34, where the transformation to the $B$-band reaches an uncertainty
of $\pm$0.1. Similarly, for four SDSS objects at z$<$0.25 (SN~3331,
5549, 7243, and 7512), the uncertainties also reach $\pm$0.1 mag. For
the SNLS objects, the statistical uncertainties in the K-corrections
are generally less than $\pm$0.04 mag. The errors for individual
epochs for a given object are correlated, and a combined, weighted
uncertainty is applied to the distance modulus (or peak
magnitude). These uncertainties are included in the total statistical
uncertainties computed for each individual SN~Ia.  We carry these
uncertainties in the analysis of cosmological parameters described
throughout this paper.  Fortunately, progress is continuing to be made
in improving K-corrections for SNe~Ia. The spectral library will
continue to grow in the next few years with additional spectra from
numerous groups, including the low-redshift CSP.

\subsection{Extinction Corrections}

\label{sec:extinction}
Correcting accurately for extinction remains a challenge for SN~Ia
cosmology. The issue is complicated by many factors:

1)  there are at least four separate potential sources of dust: a)
foreground (Milky Way Galaxy) dust, b) dust within the host galaxy, c)
dust associated with the circumstellar material of the SN progenitor and
d) dust in the intergalactic medium.

2) there is no {\it a priori} knowledge of the dust properties in
the latter three of these environments and the reddening laws
could, in principle, be different in all three environments,
perhaps being a function of metallicity, atmospheric environment of
the SN~Ia, or even evolving as a function of time. 

3) even if the extinction law(s) is known perfectly, correcting for
reddening requires knowledge of the intrinsic colors (and their
dispersion) to derive a reddening curve. The observed color of a
SN~Ia is determined by the intrinsic SED
of the SN~Ia, reddening due to dust, and the expansion of the
Universe (the K-correction). Hence, there is an inherent circularity
in the problem, and generally some assumptions are made (e.g., that
the intrinsic colors of SNe~Ia are known, and that they are known
as a function of redshift and environment, and/or that the reddening
law is universal). At present, for SNe~Ia, it is still not possible to
distinguish unambiguously between a different reddening law and
differing intrinsic colors.

The reddening law can be characterized by a ratio of
total-to-selective absorption, $R_\lambda$, which generally increases
toward shorter wavelengths: $R_\lambda =  A_\lambda /  E(B -
V)$, where A$_\lambda$ is the total absorption at each wavelength,
$\lambda$. On average, the ratio of total-to-selective absorption, $R_\lambda$
decreases from 4.9 at $U$, to 4.1 at $B$, 3.1 at $V$ and 1.7 for the
$I$-band in the Galaxy \citep[e.g.,][]{Cardelli1989}.  For a reddening of, say,
$E(B-V) = 0.02$ mag, the corrections to the rest-frame $U$-band
magnitude would be $\sim$0.10 mag and $\sim$0.03 mag at $I$. Thus,
longer-wavelength observations offer a significant advantage in
minimizing systematic effects due to reddening.

In general, it has been concluded that the dust properties in other
host galaxies appear to be similar to those in the Milky Way
\citep[see, for
example][]{Riess1996,Phillips1999,Knop2003,Riess2004,Jha2007}.  To
date, no empirical evidence for grey dust has been found
\citep[e.g.,][]{Knop2003,Riess2004,Riess2007}; i.e., larger dust
grains with wavelength-neutral effects.  However, some studies have
indicated that the reddening law for SNe~Ia is consistent with
$R_V$ $\sim$ 2.5, lower than the Galactic reddening law
\citep[e.g.,][]{Wang2006}.  Correcting for SN~Ia reddening has
generally made use of one or more of the following approaches: 1)
using a sample of SNe~Ia where the reddening is expected to be
negligible in order to define a zero-extinction fiducial sample
\citep[e.g., SNe~Ia in elliptical
galaxies:][]{Hamuy1996c,Phillips1999} and/or 2) making use of a
discovery by \citet{Lira1995} that after $\sim$30 days, the $(B-V)$ colors
of SNe~Ia show a very small dispersion
\citep[e.g.,][]{Phillips1999,Prieto2006,Jha2007}, or 3) not correcting
directly for reddening, but solving for a general $(B-V)$ color term
that treats the differences in SN~Ia intrinsic colors and
reddening as indistinguishable \citep[e.g.,][]{Tripp1998,Astier2006}.

We have dealt with the extinction using two different methods.  First,
we use a ``reddening-free'' magnitude, as described below. For
comparison, we also have solved for the reddening explicitly following
\citet{Phillips1999}, in which an intrinsic color is assumed for a
given value of \dm\ and the observed color (after
K-corrections and Milky-Way reddening corrections are applied) yield
the reddening due to the host galaxy. We differ from Phillips et. al
in that we allow for negative reddenings in the models.  While a
negative reddening is not {\it physical}, measuring a negative value
for the reddening parameter is certainly statistically possible given
the uncertainties in the measured photometry, and our uncertainty in
the intrinsic dispersion of SN~Ia colors.  For the whole sample of
SNe~Ia, the results are statistically unbiased. We list the
reddenings, corrected for Galactic foreground reddening, for our
individual SNe~Ia in Table \ref{tab:lcparams}.  The mean reddening
for the low redshift sample is $<E(B-V)>$ = 0.06 with a standard
deviation of 0.09, consistent with that for the high redshift sample,
with $<E(B-V)>$ = 0.05 and a standard deviation of 0.10.

We proceed to compute a reddening-free magnitude, $w$
\citep{Madore1982, Freedman2001},  defined here as:

\begin{equation}
w^{i}_{BV} = i - R^{i}_{BV} (B-V) = i_0 - R^{i}_{BV} (B-V)_0
\nonumber 
\end{equation}

\noindent
where the subscript, $0$, refers to intrinsic (unreddened) magnitudes
and magnitudes without subscripts are observed magnitudes. The
reddening coefficient $R^A_{BC}$ is defined as
$$ R^A_{BC} \equiv \frac{ A_A} { E(B - C)} = \frac{R_A}{R_B - R_C}$$

\noindent
where $R^I_{BV} = 1.9$ and $R^I_{BI}$ = 0.8 for the case of a standard
reddening law (with $R_V = 3.1$), while $R^I_{BV} = 1.1$ and
$R^I_{BI}$ = 0.5 for $R_V=2.0 $\citep{Cardelli1989}.  The advantage of
reddening-free magnitudes is that no knowledge of either the intrinsic
colors of SNe~Ia, nor a sample of unreddened SNe~Ia is needed:
$w^{i}_{BV}$ is defined such that the observed and intrinsic
combinations of these magnitudes and colors are numerically
equivalent. This method is in wide use for Cepheid variables. However,
(just as for other methods), the same color coefficient is applied to
both the nearby and distant samples.  It should be noted that this is
equivalent to using the color term, $\beta$, of \citet{Astier2006} if
the fiducial color of a SN~Ia is 0.  Indeed, it would simply be a
reddening correction if all SNe~Ia had zero colors.  An advantage of
this method is in the case where the reddening is solely due to dust
extinction when this approach will correct for it without the need to
isolate an unreddened sample. Furthermore, the $\beta$ coefficient
used by other authors will be the reddening coefficient in such a
case.

Given the definition of $w^{i}_{BV}$, and because we are adopting a
single reddening coefficient, it makes no difference to the final
results whether we deal with reddening-free magnitudes or
reddening-corrected magnitudes; the results are mathematically
equivalent. For both methods, we used the high- and low-redshift data
to determine the best value of $R_V$ by minimizing the scatter in the
Hubble diagram, while simultaneously solving for the best-fit
cosmology.

\subsection{Other Systematic Uncertainties:  Evolution, Metallicity
  and Weak Lensing}
\label{sec:systematics2}

In addition to the uncertainties in K-corrections and extinction
discussed above, there are other potential uncertainties on the
luminosities of SNe~Ia (e.g., evolution, metallicity, and weak
lensing). For completeness we briefly summarize the observational
situation with respect to these effects.  It is observed that spiral
galaxies host slower decliners, and hence, more luminous SNe~Ia
\citep{Hamuy1996c,Riess1998}. In general brighter SNe~Ia occur in
bluer, lower-luminosity galaxies \citep{Hamuy2000}.  In addition, the
scatter in the Hubble diagram is observed to be a function of
morphological type of the host galaxy \citep{Sullivan2003}.
\citet{Gallagher2008} have noted a correlation for nearby E/S0
galaxies such that SNe~Ia in older galaxies are fainter than those in
galaxies with younger global ages, as estimated from stellar
population models. Further, they find that residuals in the Hubble
diagram correlate with the host-galaxy metallicity.  Constraints on
differing individual SN~Ia properties are now coming from detailed
comparisons of SN~Ia spectra \citep{Hook2005,Balland2006,
Blondin2006,Riess2007}. To date, these studies have revealed no
evidence for significant evolution or metallicity
differences. However, there appear to be systematic differences at
shorter wavelengths, particularly in the restframe ultraviolet
\citep{Ellis2008,Foley2008}. Although empirically such effects appear
to be small relative to the cosmological effect being
measured \citep[e.g.,][]{Hamuy1995,Riess2004,Knop2003,Astier2006},
understanding at what level these factors affect the observed
properties of SNe~Ia is critical to SNe~Ia cosmology.  We will return
to these questions as more data become available and we complete the
CSP. Finally, we note that the net magnification due to weak lensing
is not predicted to be significant at redshifts z$<$0.7
\citep{Holz1998}, the redshift interval of the CSP sample.

\section{Light Curves and Decline Rates}
\label{sec:lightcurves}
\begin{figure}
\centerline{
\includegraphics[width=6in]{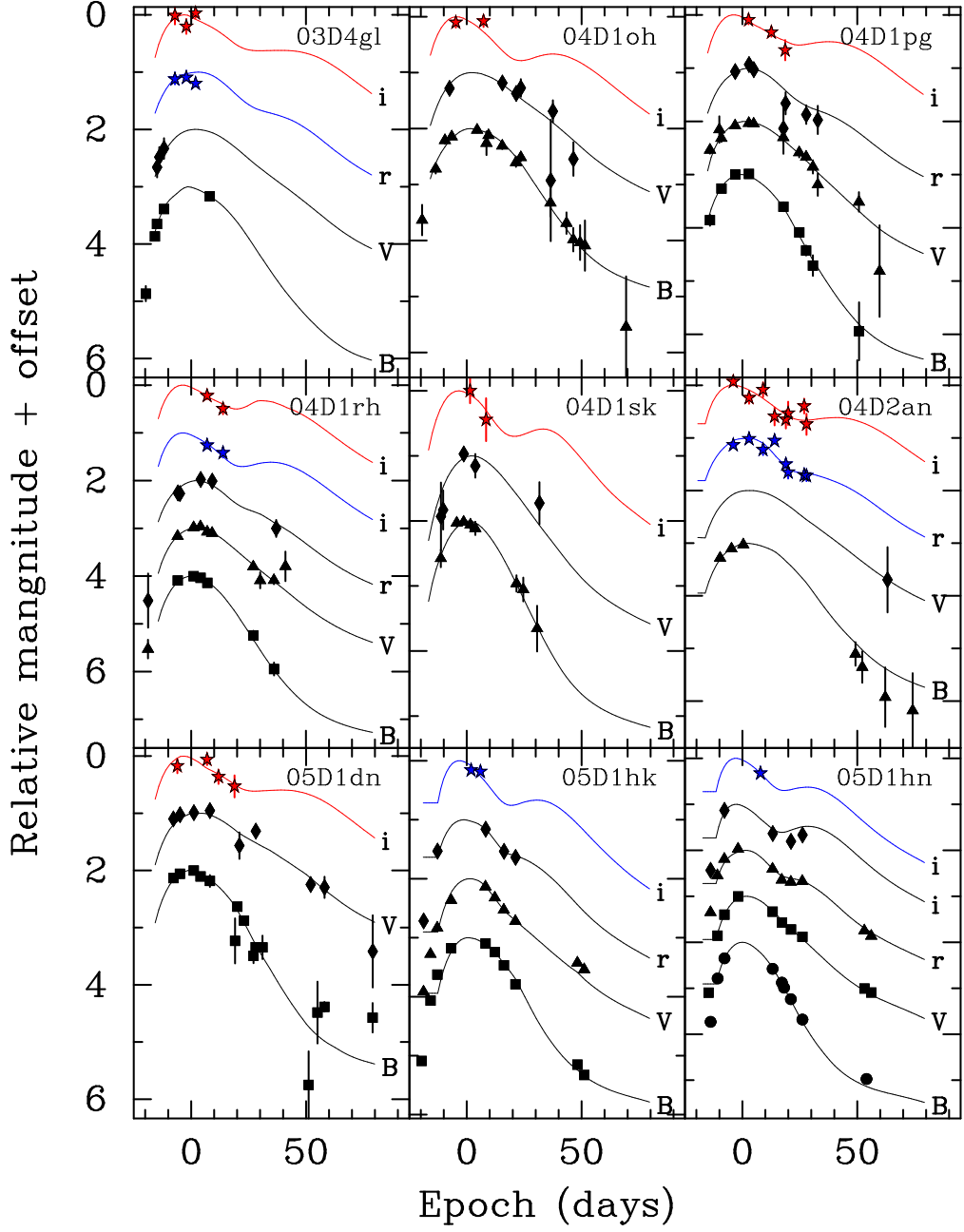}
}
\caption{\small Optical and $YJ$-band light curves for SNe~Ia
  discovered as part of the SNLS, ESSENCE and SDSS-II projects,
  followed up using PANIC on Magellan. Light curves and templates are
  shown in the observer (not rest) frame. The template $BVri$ light
  curves (solid lines) are  based on the low-redshift CSP data
  \citep{Folatelli2009}, and are expanded by $(1+z)$ and K-corrected
  to fit the observed light curves. The curves are labelled $BVri$ to
  indicate which restframe template has been used to fit the observed data.  An
  offset of 1 mag is applied so that the curves do not intersect.  In
  some cases, both the observed $Y$ and $J$ data are de-redshifted to
  the $i$ band.  The maximum-light magnitudes for each filter are
  presented in Table \ref{tab:lcparams}.  The red and blue stars
  correspond to the CSP's J- and Y-band data, respectively.  The black
  circles correspond to g$_m$ (SNLS) or g$_s$ (SDSSII); the black
  squares correspond to r$_m$ (SNLS), r$_s$ (SDSSII), or R$_{4m}$
  (ESSENCE); the black triangles correspond to i$_m$ (SNLS), i$_s$
  (SDSSII), or I$_{4m}$ (ESSENCE); and the black diamonds correspond
  to z$_m$ (SNLS).  \label{fig:lightcurves}} \end{figure}

\begin{figure}
\centerline{
\includegraphics[width=6in]{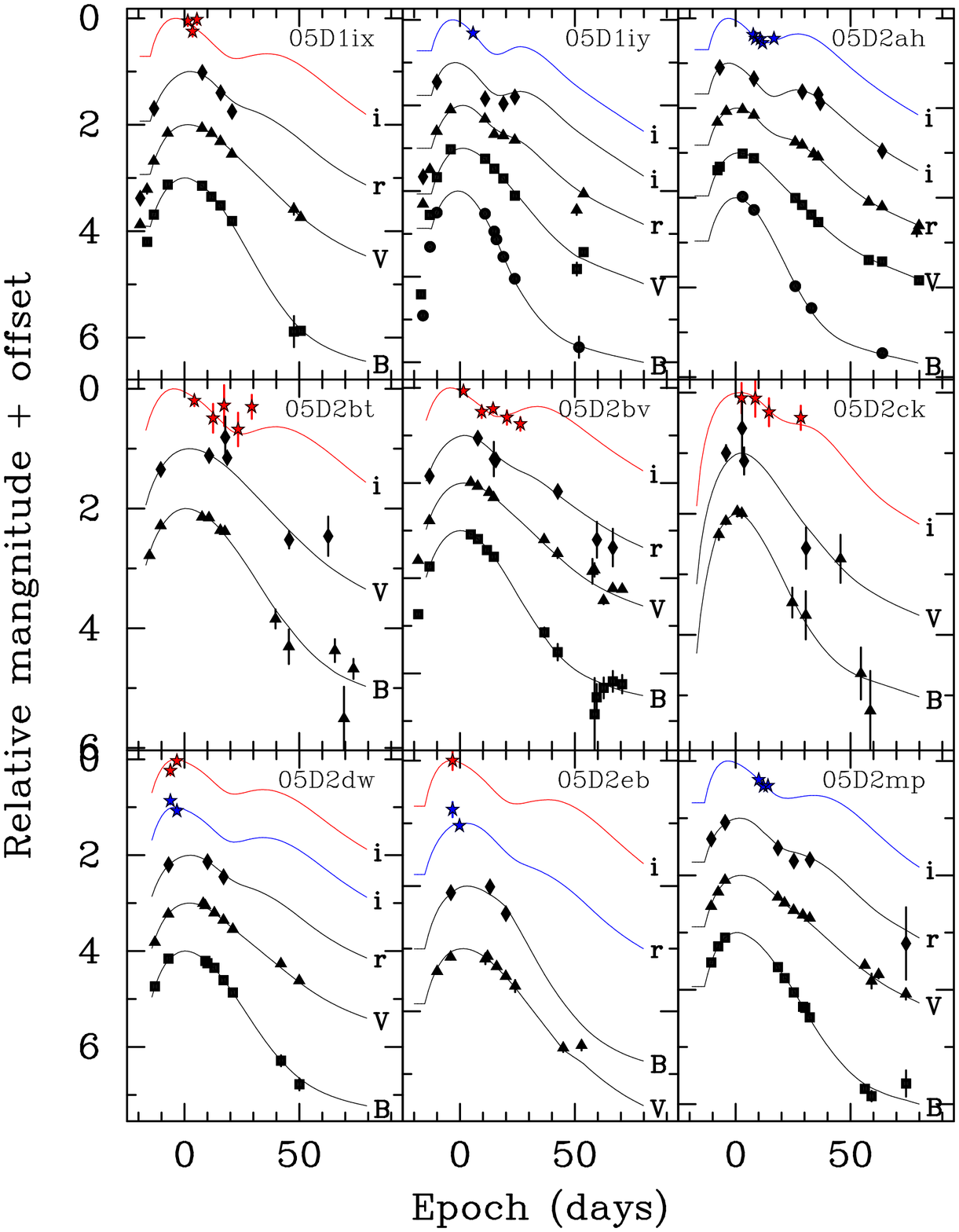}
}
\caption{\small Same as Figure \ref{fig:lightcurves}.
\label{fig:lightcurves_b}}
\end{figure}

\begin{figure}
\centerline{
\includegraphics[width=6in]{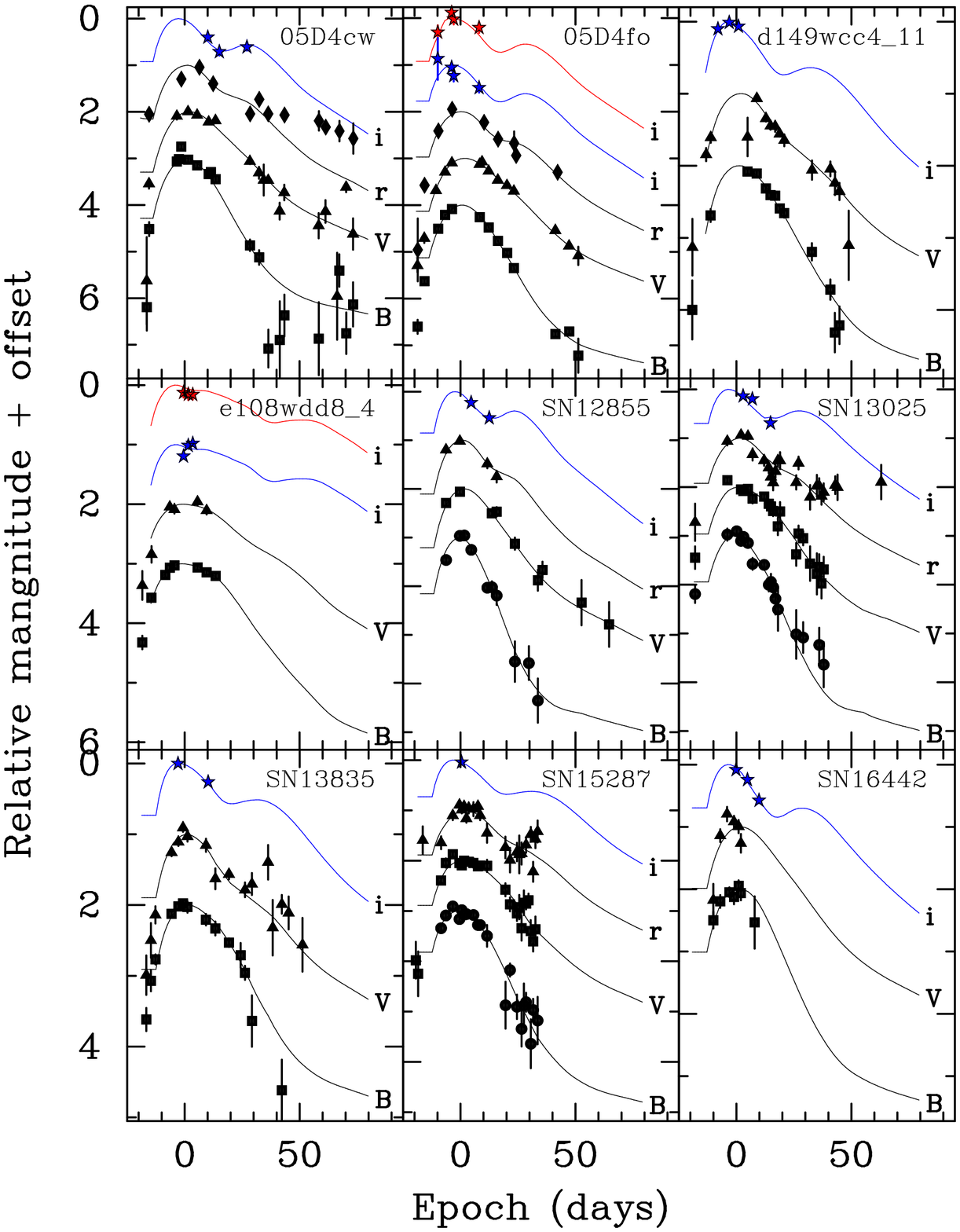}
}
\caption{\small Same as Figure \ref{fig:lightcurves}.
\label{fig:lightcurves_c}}
\end{figure}

\begin{figure}
\centerline{
\includegraphics[width=6in]{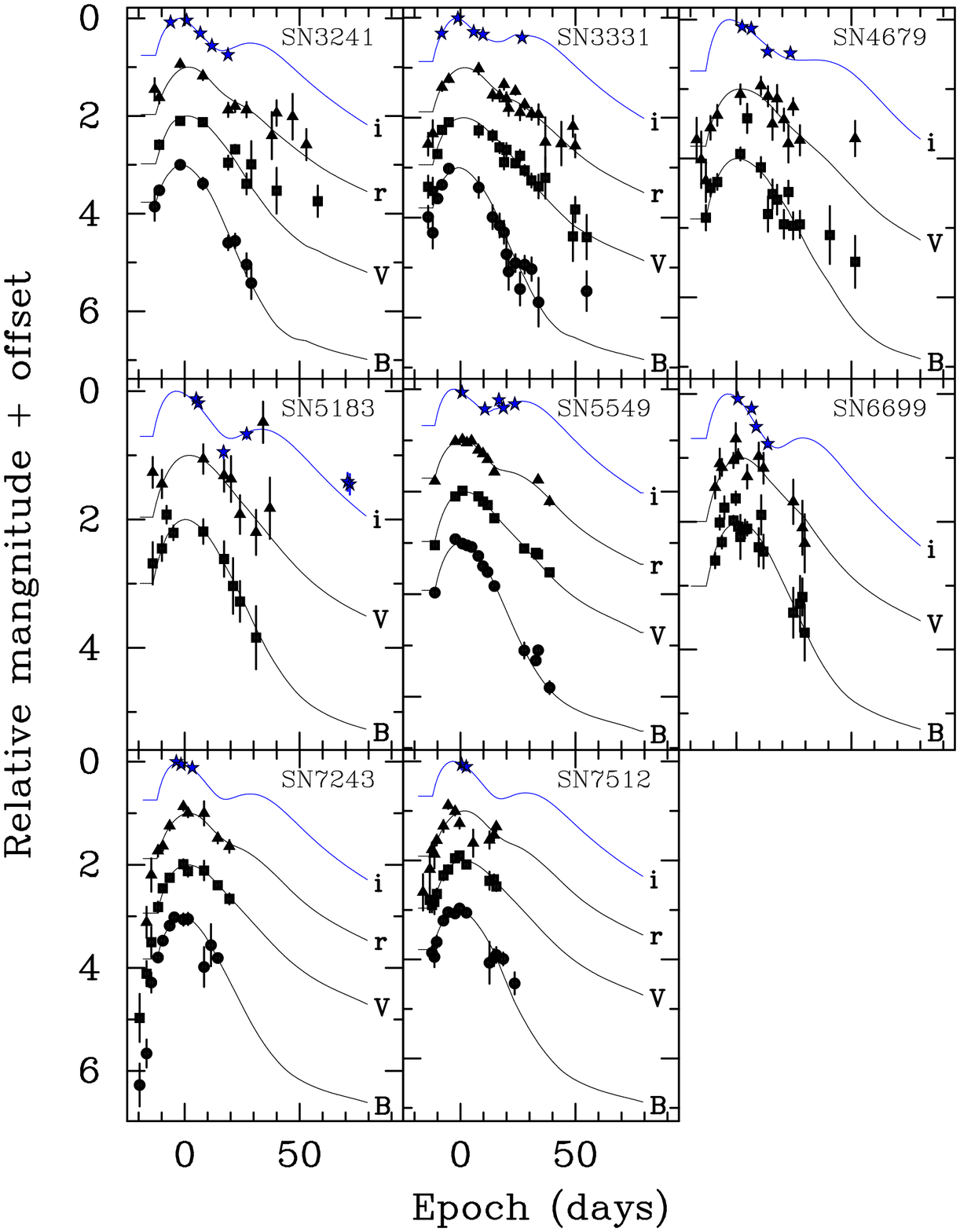}
}
\caption{\small Same as Figure \ref{fig:lightcurves}.
\label{fig:lightcurves_d}}
\end{figure}

In Figures \ref{fig:lightcurves} to \ref{fig:lightcurves_d}, we
present optical and $YJ$-band light curves for the \Ngood\ SNe~Ia with
\template\ subtractions. The $YJ$-band data and uncertainties are
given in Table \ref{tab:Photo}. We also show the optical light-curve data
from the SNLS, ESSENCE, and SDSS-II surveys.
The solid lines are template light curves,
generated as described below.  The quality of the light curves in
general is quite good, and there is an excellent correspondence
between the optical and the NIR data. For the CSP
photometry, the average $Y$-band uncertainties are $\pm$0.03 mag,
$\pm$0.06 mag ($\pm$0.10 mag at $J$), rising to $\pm$0.08 mag
($\pm$0.19 mag at $J$) for the redshift intervals 0.1 $<$ z $<$ 0.3;
0.3 $<$ z $<$ 0.5; and 0.5 $<$ z $<$ 0.7, respectively.

We make use of our own CSP low-redshift, optical data
\citep{Contreras2009,Folatelli2009} for comparison with our
NIR sample at higher redshifts. The low-redshift sample is
currently comprised of \Nlowz\ well-observed SNe~Ia (those labeled
Best Observed in Table 1 of \citet{Folatelli2009}), with redshifts
z $>$ 0.01, and with $E(B-V) <$ 0.5
mag. As described in more detail in \citet{Burns2009}, we construct a
set of $BVgri$ light-curve templates based on this sample of \Nlowz\
CSP SNe~Ia.  We use a technique similar to that of \citet{Prieto2006}
to generate the light-curve templates shown in Figures
\ref{fig:lightcurves} to \ref{fig:lightcurves_d}, and determine the
\dm\ decline-rate values and time of $B$ maximum using
$\chi^{2}$ minimization. The fits are done in flux space
simultaneously for the optical and NIR data. This technique
will be refined as additional data are obtained during the
low-redshift part of the CSP. As shown in Figure
\ref{fig:i_band_light_curves}, the $i$-band light curves usually
exhibit a second maximum, which can vary in strength from event to event.
The
variations of the $i$-band template are included in the error budget
when fitting the $i$-band templates; these become a statistical error
when the $i$-band distance moduli are plotted in the Hubble diagram.
As described in \S \ref{sec:K-correct}, K-corrections were computed by
first color-matching the SED from
\citet{Hsiao2007} to the light curves at each epoch.  Independent
co-authors (CB, GF, and MP) performed a double-blind check of the
light-curve parameters (\dm), the K-corrections
and the reddenings.  Our derived light-curve parameters are given in
Table \ref{tab:lcparams}. As discussed earlier, some of the derived
$E(B-V)$ values have negative values. Given the measurement
uncertainties and color fluctuations of $\pm$0.06 mag, however, these
negative values are not significant.

\section{A Comparison of Properties of the CSP Low- and High-Redshift SNe~Ia}
\label{sec:V-Icolor}

\begin{figure}
\centerline{
\includegraphics[width=4in]{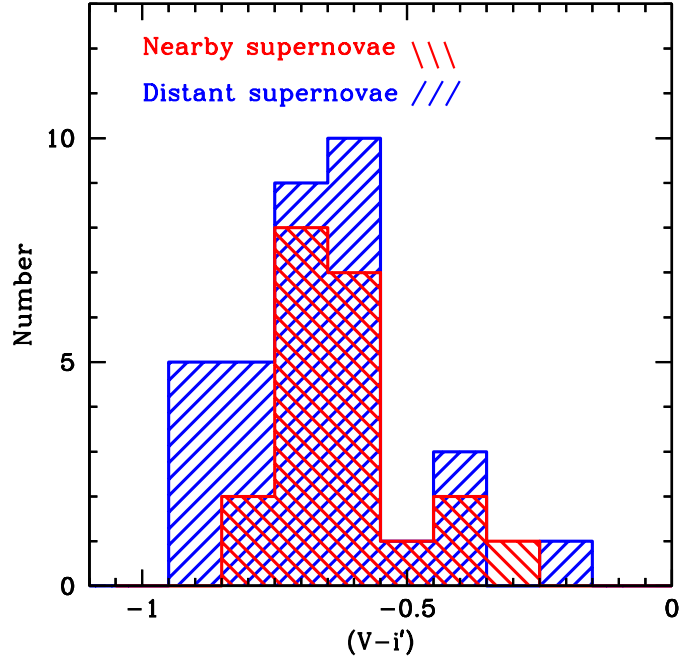}
}
\caption{Histograms of the ($V-i$) color distributions at low and high
redshifts for the SNe~Ia labeled best observed in Table 1 of
\citet{Folatelli2009}), with redshifts z $>$ 0.01, and with $E(B-V) <$
0.5 mag. Note that the ($V-i$) colors on our natural system are all
negative, and bluer than ($V-I$). The object SN~04D2an is not included
in this plot as it has no rest-frame $V$
observation.\label{fig:VIhistogram}}
\end{figure}

\begin{figure}
\centerline{
\includegraphics[width=4in]{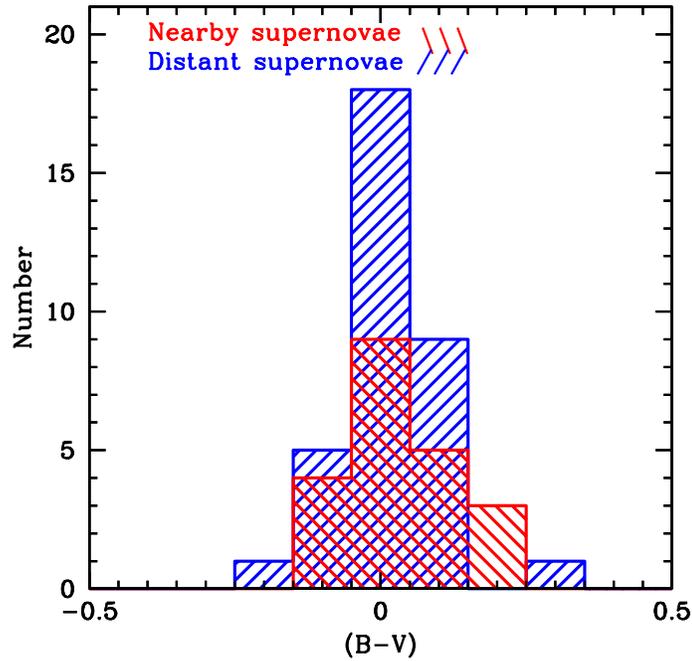}
}
\caption{Histograms of the $(B-V)$ color distributions at low and
high redshifts. \label{fig:BVhistogram}}
\end{figure}

\begin{figure}
\centerline{
\includegraphics[width=4in]{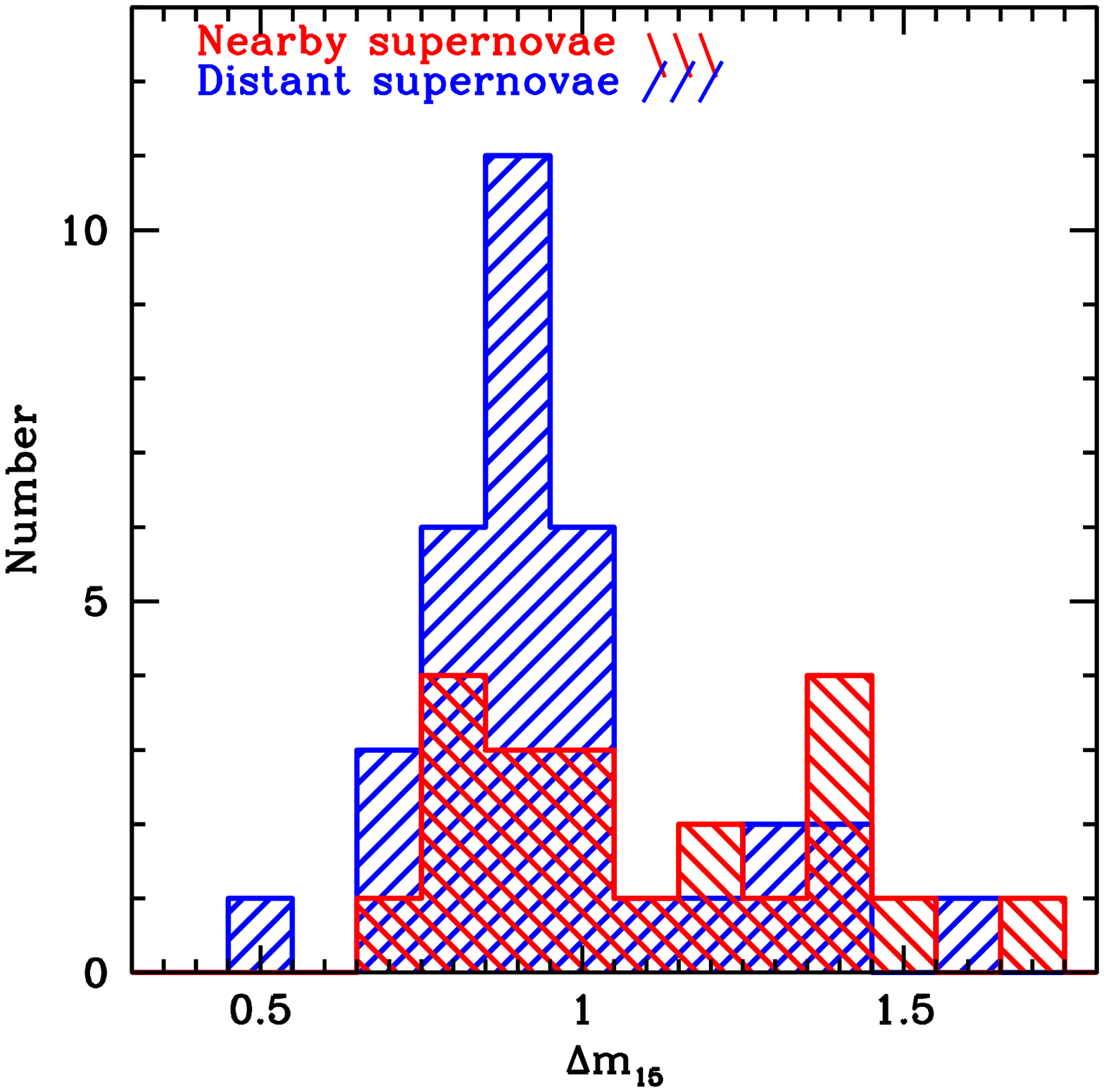}
}
\caption{Histograms of the \dm\ distributions at low and
high redshifts. \label{fig:dm15histogram}}
\end{figure}

We compare here the restframe $(B-V)$ and $(V-i$) colors for the CSP
nearby and high-redshift samples analyzed in this paper. These colors
are computed based on the peak magnitudes in each band. In Figures
\ref{fig:VIhistogram} and \ref{fig:BVhistogram} we show histograms for
the \Nlowz\ nearby SNe~Ia ($z < 0.1$) and \Ngood\ more distant CSP
SNe~Ia with ($0.12 < z < 0.70$).  The mean colors and 1-$\sigma$
dispersions for our sample are ($V-i$) = -0.63 $\pm$ 0.12 mag and
($V-i$) = -0.67 $\pm$ 0.16 mag at low and high redshifts,
respectively. For $(B-V)$, the mean colors are $(B-V)$ = 0.04 $\pm$
0.09 mag and $(B-V)$ = 0.02 $\pm$ 0.10 mag, respectively. The colors
are consistent to within the uncertainties. However, the very red
CSP objects SN~2005A and SN~2006X have not been included in this analysis
since they do not fit our redshift or reddening criteria, as defined
above.  In Figure \ref{fig:dm15histogram} we show a comparison of the
distribution of values for \dm\ for the low- and high-redshift
samples. The mean values for \dm\ are 1.11 $\pm$ 0.28 and 1.18 $\pm$
0.33 for the nearby and more distant samples, respectively.  There is
good overlap in the two distributions, although the distant sample is
more peaked. Given the different selection effects for the different
samples and the possibility of SN~Ia evolution, we do not necessarily
expect the low- and high-redshift distributions to agree. For this
small sample, quantitative comparison is limited; however, we conclude
that the low-redshift sample spans the parameter space of
high-redshift color and \dm.

\section{Distance Moduli, Reddenings and Errors}
\label{sec:distmod}

We now turn to the determination of the distance moduli, reddenings
and the errors associated with these quantities. 
We undertake a simultaneous fit for the distance modulus
as well as three parameters used to define the absolute magnitudes. 
We include both the low- and high-z SNe~Ia in this
analysis.  For each independent choice of three filters used to define
the reddening-free magnitude (generally $BVi$), we use the
following three parameters to define the absolute luminosity of a SN~Ia.
We treat these three quantities as nuisance parameters in this
analysis.

\noindent
1) $W_0$, the absolute reddening-free magnitude  of a
SN~Ia with $\dm = 1.1$ 

\noindent
2) $b$, the slope of the $W$-\dm relation  and

\noindent
3) $R_V$, the ratio of total-to-selective absorption.

\noindent
The absolute (reddening-free) magnitude for a SN is then $W_{BV}^{i} =
W_0 + b\left(\dm - 1.1\right)$.  To compute the distance
moduli for the entire sample, we then measure an apparent
reddening-free magnitude at maximum light,
$w^{i}_{BV}\left(t_{max}\right)$, for the
restframe $BVi$ photometric bands. The distance modulus is then, by
definition, $ \mu_0 = w^{i}_{BV} - W^{i}_{BV}$. Distance moduli are
listed in Table \ref{tab:lcparams}. (Here, a value of $H_0$ = 72
km/sec/Mpc is adopted. However, the value of $H_0$ cancels out for the
determination of other cosmological parameters in \S \ref{sec:Hubble}.)

Using a reduced $\chi^2$= 1 approach
\citep{Tremaine2002}, we then compute $\chi^2$ by
comparing the measured distance modulus, $\mu_0$, with a theoretical
distance modulus $\mu_T\left(\mathcal{C}\right)$ where $\mathcal{C}$
represents the set of cosmological parameters
(e.g., $\Omega_m,\Omega_{DE},\wsym$) we consider:

\begin{equation}
   \chi^2 = \sum_j \frac{
      \left[
         \mu_j - 
         \mu_T\left(\mathcal{C}, z_j\right)
      \right]^2
   }
      {\sigma_{j}^2 + \sigma_{SN}^2
   }
   \label{eq:chi_cosmo}
\end{equation}

\noindent
where $\sigma_{SN}$ is the intrinsic dispersion of SNe~Ia, chosen
such that reduced-$\chi^2$ = 1 (for which
we find a value of $\sigma_{SN}$ = 0.09), and $\sigma_{j}^2$ is the total variance for
observation $j$ and is given by

\begin{eqnarray}
   \sigma^2_{j} & = & \sigma^2(i_{max}) + R^{i2}_{BV}\sigma^2(B_{max}-V_{max})
      + b^2\sigma^2(\dm) \nonumber\\
      & & - 2R^{i}_{BV}\sigma(i_{max},B_{max}-V_{max})
      - 2b\sigma(i_{max},\dm) \nonumber\\
      & & +2R^{i}_{BV}b\sigma(B_{max}-V_{max},\dm)    
\label{eq:sigmasq}
\end{eqnarray}

\noindent
where $\sigma^2(x)$ is the variance in parameter $x$ and $\sigma(x,y)$
is the covariance between parameters $x$ and $y$. A peculiar velocity
term of $\pm$300 km/sec is also included.  By minimizing this $\chi^2$
(i.e., the residuals in the Hubble diagram), we simultaneously
determine the cosmology and the three nuisance parameters $W_0$, $b$,
and $R_V$. In this sense, our method is similar conceptually to that
described by \citet{TrippBranch1999,Astier2006,Conley2008}.  We defer
a discussion of the results for cosmology until \S \ref{sec:Hubble},
and discuss first the results for the decline-rate and reddening
parameters from this method.

Using the $i$-band data at maximum, corrected for reddening based on
the $BV$ photometry, the results for $W^{i}_{BV}$ are $W_0 = -18.45
\pm 0.05$ (statistical) $\pm 0.01$ (systematic), $b = 0.38 \pm 0.08$
(statistical) $\pm 0.01$ (systematic), and $R_V = 1.74 \pm 0.27$
(statistical) $\pm 0.1$ (systematic).\footnote{ Unless otherwise
noted, all reported uncertainties are statistical and correspond to
1-$\sigma$ errors (68\% confidence).}  Since we have data for three
filters, we can independently correct for reddening using different
combinations of bandpasses (e.g., $B-V$, $V-i$, or $B-i$). However, we
prefer to solve for the reddening using simultaneous data from a
single photometric/telescope system ($B-V$), which is available for
all of the SNe~Ia, rather than a hybrid optical-NIR combination.  If
instead we use the $B$-band data and correct for reddening based on
the $BV$ photometry, $W^{B}_{BV}$, we find $W_0 = -19.11 \pm 0.07$, $b
= 0.65 \pm 0.13$, and $R_V = 1.66 \pm 0.27$.  Encouragingly, $R_V$ is
consistent for the two filter combinations.
The final results for the
cosmology are consistent to within the uncertainties, with all
combinations of bandpasses.

A value of $R_V$ = 1.74 is significantly lower than a Milky-Way value
of $R_V$ = 3.1. A lower value of $R_V$ is also found by
\citet{Tripp1998}, \citet{TrippBranch1999}, and \citet{Wang2006}.
\citet{Conley2008} and \citet{Astier2006} would find an even lower value than ours,
consistent with $R_V$ (= $R_B -1$) $\sim$ 1 (their $\beta = 2$ would
correspond to $R_B$), if their color term was associated with
the extinction law.

As a check on our reddening corrections, two independent analysis
methods were used by two of this paper's co-authors (CB and GF) to
determine the nuisance parameters: the reddening-free method outlined
above, and the extinction method outlined in \citet{Phillips1999}.  We
initially discovered a large discrepancy between the two derived
values of $R_V$.  The reason for this discrepancy can be understood
from equations \ref{eq:chi_cosmo} and \ref{eq:sigmasq}.  The
denominator of $\chi^2$ includes terms that depend on both $R_V$ and
$b$.  Depending on the magnitude of the variances and covariances,
these terms have leverage on the final solution.  We have found that a
significant issue affecting the derived reddening law is what is
assumed for the variance in SNe~Ia colors, $\sigma^2\left(B-V\right)$,
with the value of $R_V$ increasing with increasing
$\sigma^2\left(B-V\right)$ (see Appendix \ref{sec:app_systematics}).
The two analysis methods agree well if consistent errors are adopted.
It should therefore be stressed that accurate estimates of the
variances in the data are crucial for the determination of $R_V$.  As
a further test of these different approaches, \citet{Burns2009} have
begun to explore this issue with an unbiased estimator using a Monte
Carlo Markov Chain (MCMC).  This work goes beyond the scope of the
present paper, but we note that the preliminary results agree well
with those presented here, with $R_V$ = 1.8. Our current analysis and
that of \citet{Folatelli2009} are consistent with a picture in which,
in addition to corrections for Milky-Way-like dust, there is an
intrinsic dispersion in the colors of SNe~Ia that is correlated with
luminosity, but not the decline rate.

Before concluding this section, it is worth emphasizing that
disentangling a possible additional intrinsic color term from
extinction leaves open a challenge to SN~Ia cosmology.  Potential
differences in either the intrinsic or observed sample populations of
SNe~Ia with redshift and/or environment, or differences in the
reddening law may limit the ultimate accuracy for cosmology, and for
reaching systematic uncertainties significantly below the $\sim\pm
10\%$ level. Having highlighted the remaining challenges, we note,
however, that the level of the effect we are discussing amounts to
less than $\pm 10\%$, and reaching a systematic uncertainty at this
level for SNe~Ia is already excellent progress.  Few other methods in
use for dark energy measurements currently offer this kind of
accuracy. We now turn to a discussion of the Hubble diagram, and then
to implications of different values of $R_V$ on the cosmology.

\section{Hubble Diagram and Cosmological Results}
\label{sec:Hubble}

\subsection{The First $I$-band Hubble Diagram for z $>$ 0.1}
\label{sec:ibandHubble}

\begin{figure}
\centerline{
\includegraphics[angle=-90,width=6in]{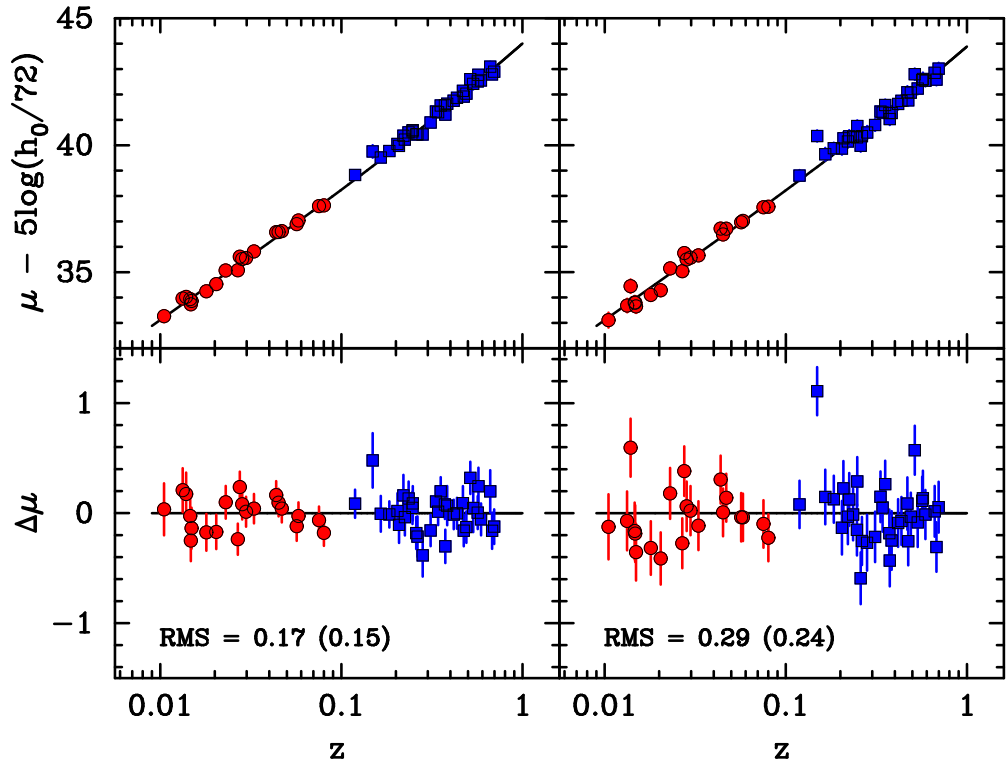}
}
\caption{ Top panel: $i$ and $B$-band Hubble diagrams for
\Nlowz\ low-redshift and \Ngood\ high-redshift SNe~Ia from the CSP,
{\it uncorrected for reddening}. Bottom panel: The residuals about the
best-fit to these data. The values for $rms$ scatter about the best fit
to these data are labelled. The $rms$ value in brackets excludes the most
discrepant (highly reddened) SNLS 05D1hn.
\label{fig:rawHubble}}
\end{figure}

In Figure \ref{fig:rawHubble}, we show the $i$ and $B$-band
Hubble diagrams, uncorrected for reddening. The residuals about the
best-fit to these data are shown in the panels below.  We can
immediately see the advantage of observing SNe~Ia at red wavelengths
relative to the optical. The $rms$ dispersions in these plots amount to
$\pm$0.17 and $\pm$0.29 mag, respectively.

\begin{figure}
\centerline{
\includegraphics[width=6in]{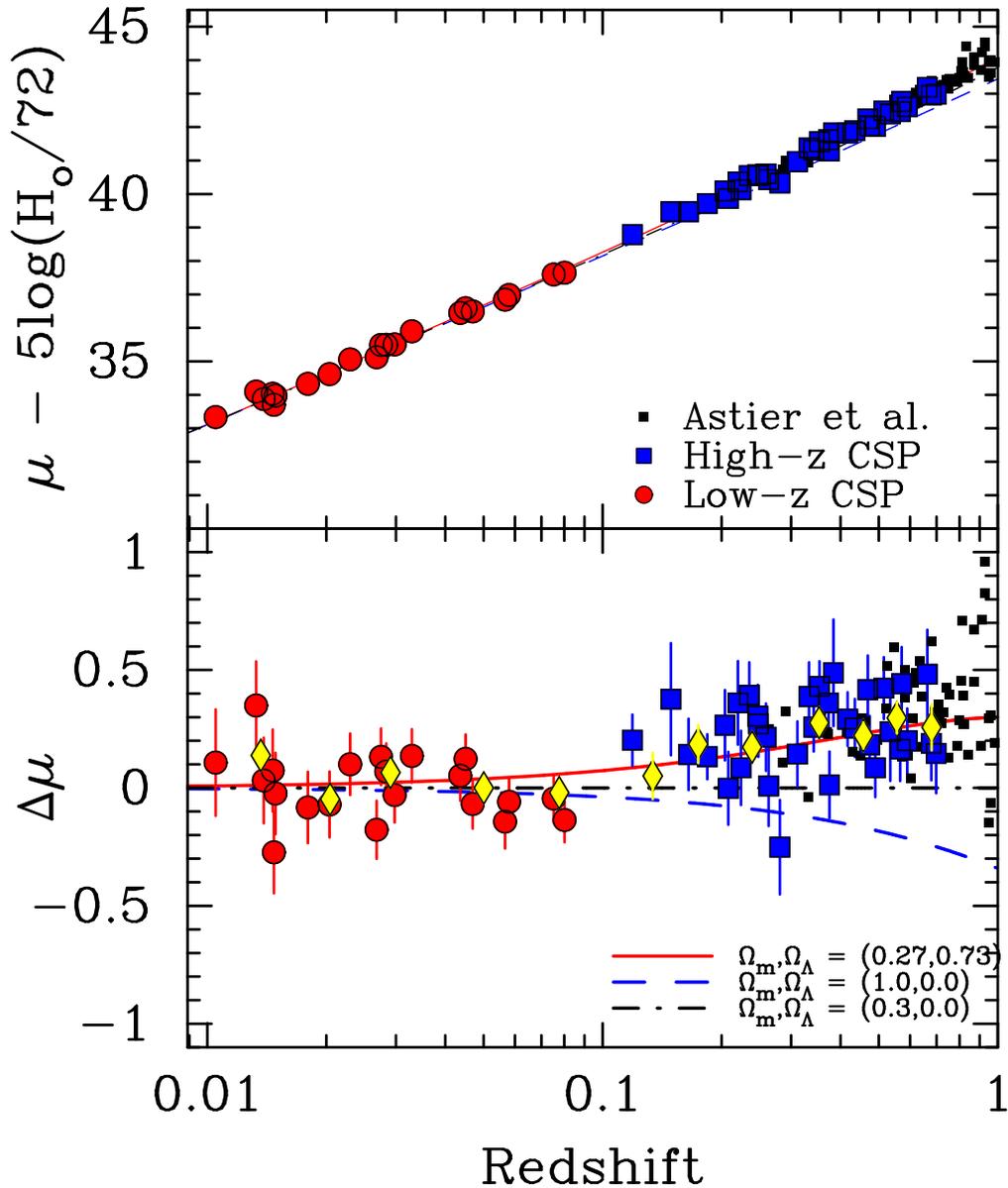}
}
\caption{ $i$-band rest-frame Hubble diagram for a total of 56 CSP
 SNe~Ia, \Ngood\ SNe~Ia from the Magellan CSP sample (blue squares),
 and \Nlowz\ low-redshift data (red solid circles) from
 \citet{Folatelli2009}.  For comparison, distances determined by
 \citet{Astier2006} are shown as black squares, but are not included
 in the fits.  Error bars shown are 1-$\sigma$. A value of H$_0$ = 72
 km s$^{-1}$ Mpc$^{-1}$ has been adopted for the plot. The solid
 (red), dot-dashed (black), and dashed (blue) lines represent
 $\Omega_m$ = 0.3, $\Omega_{DE}$ = 0.7 ; $\Omega_m$ = 0.3,
 $\Omega_{DE}$ = 0 ; and $\Omega_m$ = 1 models, respectively.  The
 data are consistent with the standard (accelerating) cosmological
 model. To minimize the effects of peculiar velocities, the fit to the
 low-redshift sample is restricted to z$>$0.010. In the bottom panel,
 the data are shown relative to the standard model, shown as the
 solid line.  The yellow diamonds are the result of binning the
 data (see table \ref{tab:binned_hubble}).
\label{fig:Hubble}}
\end{figure}

In Figure \ref{fig:Hubble}, we present the first $i$-band Hubble
diagram for the CSP sample of \Ngood\ SNe~Ia based on Magellan data
(solid blue squares) using the reddening-free magnitude technique
described in \S \ref{sec:lightcurves}, adopting a value of $R_V = 1.74
\pm 0.27$. The error bars shown in the lower panel represent the
formal 1-$\sigma$ uncertainties in the distance modulus and include
the combined errors in the photometry, the dispersion in the
light-curve templates, the estimated dispersion in K-corrections, the
error in the foreground Galactic reddening, and the co-variances
between the light-curve parameters. We also include a
peculiar-velocity component of $\pm$300 km/sec.  The top curve
corresponds to an $\Omega_m$ = 0.27 , $\Omega_{DE}$ = 0.73 cosmology.
For comparison, the small black squares indicate distance moduli from \citet{Astier2006}.  The
current CSP sample of \Nlowz\ nearby SNe~Ia from
\citet{Folatelli2009} (solid red circles) is also shown in Figure
\ref{fig:Hubble}.  The low-redshift sample is restricted to nearby SNe~Ia
in the Hubble flow, having redshifts greater than z $>$ 0.010, so
that the scatter due to random peculiar velocities is minimized. (The results
remain consistent, to within the quoted uncertainties, if we further
restrict the sample to z $>$ 0.015 or z $>$ 0.02.)

The immediate conclusion we can draw from this Hubble diagram is that
the CSP data alone provide independent evidence for a (standard)
cosmological model with $\Omega_m$ = 0.3 and $\Omega_{DE}$ = 0.7. The
CSP constraints can be further improved by combining them with other
independent measurements, for example, baryonic acoustic oscillations
\citep[e.g.,][]{Eisenstein2005}.  A weighted fit to our \Ngood\ data
points in combination with baryon acoustic oscillations (assuming
$\wsym = -1$) yields the solution: \OmegaMatter, \OmegaLambda.  The
systematic uncertainties for SNe~Ia do not have much impact on the
determination of $\Omega_m$ since this parameter is determined largely
by the matter power spectrum. We quote only statistical uncertainties
for $\Omega_m$ here.  The statistical uncertainties are determined by
marginalizing over all other parameters and fitting the 1D probability
distribution to a Gaussian.   The systematic uncertainties are treated below in \S
\ref{sec:cosmo_errors}.

\begin{figure}
\centerline{
\includegraphics[angle=-90,width=5in]{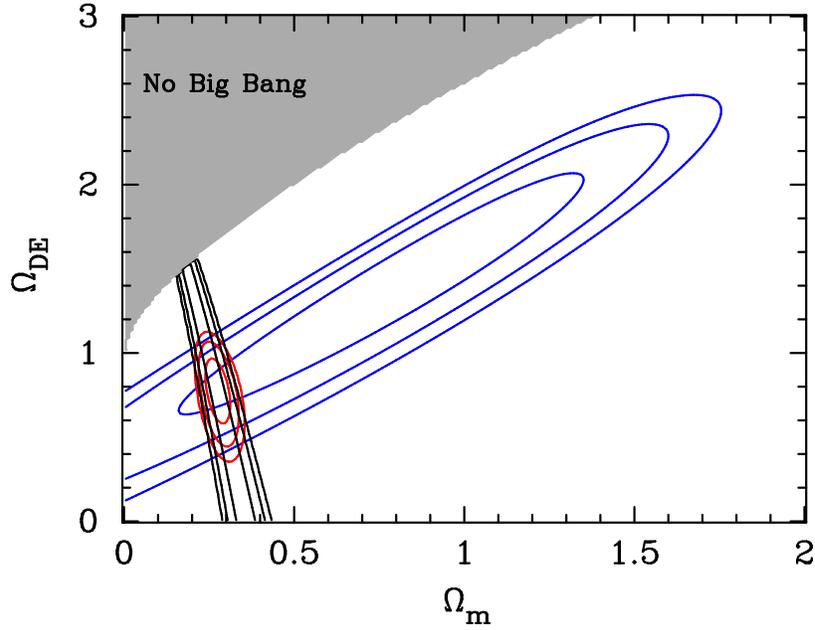}
}
\caption{Our best-fit cosmological model in the
$\Omega_m$--$\Omega_{DE}$ plane assuming a constant equation of state
parameter $\wsym=-1$.  Our 68\%, 95\%, and 99\% confidence intervals are
shown as solid blue (diagonal) contours.  The constraints from baryon acoustic
oscillations \citep[]{Eisenstein2005} are shown as solid black
(nearly-vertical) contours and the combined confidence intervals are
shown as red contours.  
\label{fig:CSP+BAO_OM_OL}}
\end{figure}

Based on the fit to the Hubble diagram above, we show, in Figure
\ref{fig:CSP+BAO_OM_OL}, our error ellipses in the
$\Omega_m$-$\Omega_{DE}$ plane. Consistent with previous SN~Ia
studies, we find that based on the SN~Ia data alone (and the
assumption that $\wsym=-1$), a value of $\Omega_{DE} >$ 0 is required
at greater than the 99\% confidence level.

\begin{figure}
\centerline{
\includegraphics[width=4in,angle=-90]{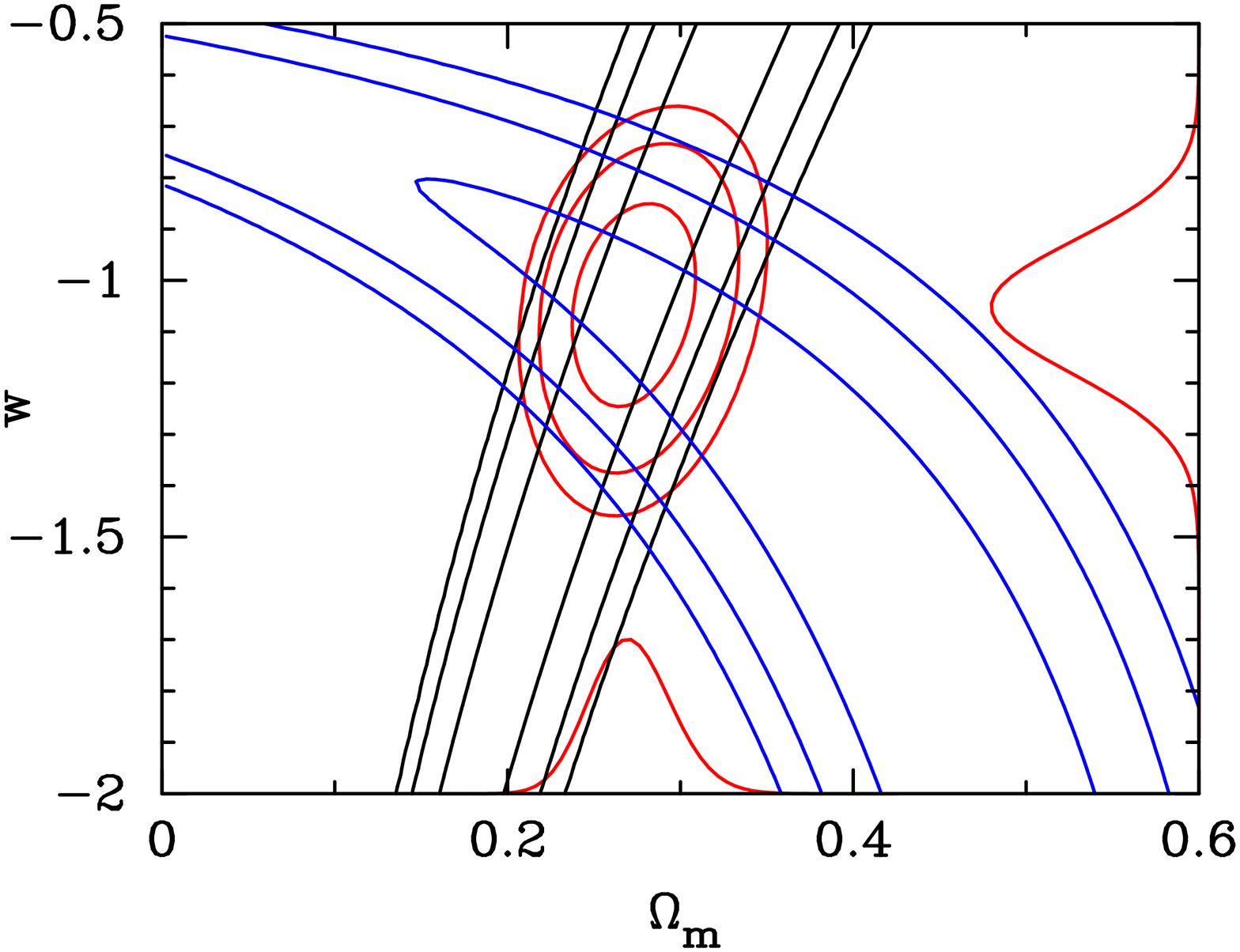}
}
\caption{Combining the CSP constraints with baryonic acoustic
 oscillations \citep[]{Eisenstein2005} and assuming
 $\Omega_k = 0$.  The CSP and BAO data combined
 are consistent with a value of \wo\ and \OmegaMatter . Our 68\%,
 95\%, and 99\% confidence intervals are shown as solid blue
 (banana-shaped) contours.  The constraints from baryon acoustic
 oscillations \citep[]{Eisenstein2005} are shown as solid black
 contours and the combined confidence intervals are
 shown as red contours.  The 1-D marginalized probabilities for
 each parameter are plotted as red lines on the axes.
\label{fig:CSP+BAO_OM_w0}}
\end{figure}

As an alternative to making assumptions about $\wsym$, we can use the
CSP data to calculate $\wsym$ under the assumption of flatness
($\Omega_k$ = 0).  Here we again combine the CSP results with
independent measurements of baryonic acoustic oscillations
\citep[e.g.,][]{Eisenstein2005}, as shown in Figure
\ref{fig:CSP+BAO_OM_w0}.  Assuming a flat cosmology, these joint
constraints yield a value of \OmegaMatterFlat and \wo.  For the
purposes of this calculation, we are assuming that $\wsym$ is a
constant (i.e., $\wsym_a$ = 0).  These results are in excellent
agreement with other joint constraints from SN~Ia studies and
baryon acoustic oscillations \citep{Astier2006, WoodVasey2007,
Riess2007}, which also yield values of $\wsym = -1$ and $\Omega_m$ =
0.3, to within the quoted measurement uncertainties. We have also
combined our CSP results with the two-dimensional probability contours
from the 2dF galaxy redshift survey \citep{Hawkins2003}. We find
$\Omega_m$ = 0.27 $\pm$ 0.09 (statistical) and $\wsym_0$ = -1.03 $\pm$
0.23 (statistical), in good agreement with the SDSS data, but with
larger uncertainties on the value of $\wsym_0$.

\begin{figure}[t]
\centerline{
\includegraphics[width=4in]{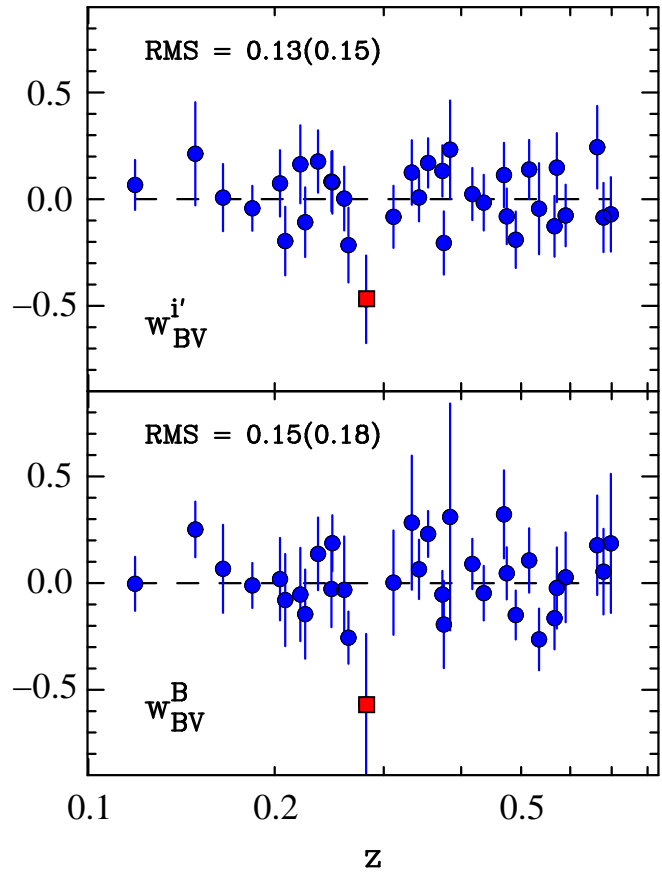}
}
\caption{ Residuals in the $i$ and $B$ Hubble diagrams.  The top panel
shows reddening-free $i$ magnitudes and the bottom panel shows the
reddening-free $B$ magnitudes. Both are computed using the $(B-V)$
color.  The $rms$ dispersion is shown in the upper left.  SN~04D2an is
not shown as it has no $V$-band observation. If one includes  SN~16442 
(red square), the dispersion in the Hubble diagram
increases to $\pm$0.15 at $i$, and $\pm$0.18 at $B$ as indicated in
parentheses.  The dashed line correponds to the best-fit cosmological
model.
\label{fig:BViW}}
\end{figure}

In Figure \ref{fig:BViW}, we show a comparison of the residuals in the
Hubble diagram for the high-z data ($z>0.10$) for the $i$- and $B$-bands
relative to the best-fit flat, constant \wsym\ cosmological model, again
for a value of $R_V = 1.74$.  The $rms$ scatter in the $i$-band Hubble
diagram amounts to $\pm0.13$ mag about the fit ; the $B$-band scatter
amounts to $\pm$0.15 mag.\footnote{The $rms$ values excludes SN~04D2an
at a redshift of z = 0.62, which has no reliable rest-frame $V$-band
observation and therefore no color correction is possible using only
the optical data.}  The resulting cosmological
parameters are in good
agreement, with differences in \wsym\ and $\Omega_m$ amounting to only
0.03 and 0.01, respectively, well within the measurement
uncertainties. The $i$-band measurements have smaller systematic
uncertainties due to smaller reddening corrections, and also have
somewhat smaller observed scatter.

As discussed in \S \ref{sec:lightcurves}, we find that the smallest
scatter in the Hubble diagram at low (as well as high) redshift is
found for a value of the (color coefficient or) ratio of
total-to-selective absorption $R_V = 1.74 \pm 0.27$.  The standard
Galactic reddening law is $R_V$ = 3.1.  In Table
\ref{tab:cosmologies}, we show the fits to $\Omega_m$ and
$\Omega_{DE}$ and the $rms$ scatter about the cosmological fit for two
values of $R_V$, 1.74 and 3.1, for both the $i$ and $B$-band data. For
an adopted value of $R_V = 1.74$, the resulting values of $\Omega_m$
and $\wsym$ are in excellent agreement between both bands.  However,
adopting instead a value $R_V$ = 3.1 results in a value of $\wsym =
-1.20 \pm 0.13$ for the $i$ data and $\wsym = -1.24 \pm 0.16$ for the
$B$ data.  The solutions for these different adopted values of the
reddening are still mutually consistent, but the scatter in the Hubble
diagram increases by about 50\% for $R_V$ = 3.1.  Given the increase
in the Hubble diagram scatter at larger redshifts for values of $R_V =
3.1$, and the excellent fit to the low redshift CSP sample for $R_V =
1.74$, we conclude that currently the CSP data are consistent with a
lower value of a color coefficient and/or reddening law.

It is interesting to ask what happens if reddening effects are ignored
altogether.  In this case we find (based on the data shown in Figure
\ref{fig:rawHubble}), that for $i$, $\Omega_m$ = 0.29 $\pm$
0.03 (statistical) and $\wsym_0$ = -0.90 $\pm$ 0.14 (statistical),
consistent to within the systematic uncertainties with the results
from the reddening-free and reddening-corrected data discussed above.
For the $B$-band data, however, $\Omega_m$ = 0.31 $\pm$ 0.03
(statistical), $\wsym_0$ = -0.70 $\pm$ 0.21 (statistical), and the
results for the equation of state are much more sensitive to the
treatment of reddening. These results illustrate quantitatively the
advantage of the $i$ band in minimizing systematic effects for
SNe~Ia cosmology. We have also shown that the 
$i$-band can be calibrated as well as the $B$-band; in fact, there is
a further advantage to the $i$-band, where the
luminosity-\dm\ relation has a smaller slope. 

It is beyond the scope of this paper to ascertain unambiguously 
the reddening properties for SNe~Ia; however, we  make the following
observations. The simplest assumption, the application of a
standard reddening law coefficient, $R_V=3.1$ alone, does not provide
the best fit to either the low redshift or high redshift CSP data as
defined by reduced scatter in the Hubble diagram. As noted previously,
this in and of itself does not imply a different reddening law within
the host galaxies. For example, it could be indicating an intrinsic
color term or perhaps an unusual kind of reddening resulting from
dust in the vicinity of the SN~Ia. In fact, it would be a somewhat
unusual circumstance for the dust within the interstellar medium of
all SN~Ia host galaxies to have different properties from the dust
in nearby galaxies;  e.g., LMC, SMC and the Milky Way, which are the only
galaxies for which the value of $R_V$ can be measured to high accuracy
and where on average it is found to be equal to 3.1 \citep{Draine2003}.
\citet{NobiliGoobar2008} have determined the average reddening law for
a sample of 80 SNe~Ia and find that it agrees with
\citet{Cardelli1989} for $R_V$ 1.75 $\pm$ 0.27.  It has also recently been
shown that normal dust which is distributed locally around the SN can
produce a reddening law with lower value of $R_V$ \citep{Goobar2008}.
In other words, this abnormally low value of $R_V$ may simply be due
to the geometry of local dust around the SN, which dominates any
extinction produced by the host galaxy.

We plan a detailed investigation of the reddening law for SNe~Ia based
on our low redshift CSP sample, which was designed to provide
independent determinations of reddening at multiple wavelengths. For
the present discussion, we adopt a general color coefficient term  
consistent with a reddening law (or color term), $R_V = 1.74 \pm 0.27$.

\subsection{Dark Energy and Other Parameterizations of the Expansion }

How do we best compare the observational SN~Ia data with
cosmological models?  While a simple question, this issue is
non-trivial. As discussed in \S \ref{sec:Intro}, the expansion of the
Universe can be parameterized in terms of the equation of state
parameter, $\wsym$. Lacking a physical explanation for the dark energy,
this is a helpful parameterization, but it is still just that
-- a parameterization. It is important to keep in mind what the actual
observables are: observations yield magnitudes and redshifts -- not
accelerations, equations of state or second derivatives of scale
factors (the deceleration parameter).  In the case of a flat universe,
the luminosity distance is given by:

\begin{equation}
d_L\left(z\right) = r\left(z\right) \left(1+z\right) = \int _0^z{ \frac{z^\prime}
{H\left(z^\prime\right) }},
\nonumber 
\end{equation}

\noindent
where $r(z)$ is the comoving distance.  Because the luminosity distance
relates (inversely) to the {\it integral} of $H(z)$, it does not provide
information on how the expansion rate may have changed from higher
redshifts to today.  A model is required to infer this evolution.

We now turn to a discussion of the equation-of-state parameter,
$\wsym$, which is commonly used in current cosmological models.  A number of ways to parameterize $\wsym$ have been suggested
in the literature \citep[e.g.,][]{Linder2006,Albrecht2006}. The
functional forms for the evolution in $\wsym$ have included simple
terms that are either linear functions of redshift, $z$ (or scale
factor, $a$), but range also to more exotic functional forms. It must
be emphasized that while SN~Ia data alone provide a very strong
case for a non-zero value for $\Omega_{DE}$ (i.e., an additional
component to matter and radiation), SN~Ia data {\it alone} do not
provide a sensitive means of constraining either the value of $\wsym$
at the current epoch, or its time evolution.

As discussed in \S \ref{sec:ibandHubble}, under the assumption of
flatness ($\Omega_k$ = 0), SNe~Ia can provide a strong constraint on
$\wsym$ when combined with another experiment like baryon acoustic
oscillations.  Given current evidence, flatness is not an unreasonable
assumption; for example, we note that either recent $H_0$ or SNe~Ia
results, in combination with the WMAP observations, yields $\Omega_k =
-0.01 \pm 0.01$ \citep{Spergel2007}.  However, additional methods
(e.g., baryon acoustic oscillations, cluster growth, weak lensing)
must be combined with the SN~Ia data to provide meaningful constraints
on $\wsym$. At the current time, the accuracy required is not yet
sufficient for any of these methods alone, and is the goal of future
studies.

The cosmological parameterization $(\Omega_m,\Omega_{DE},\wsym)$ rests
on assumptions about the matter/energy content of the Universe and the
functional form for the dark energy equation of state. Here we also
investigate another parameterization that is independent of this
theoretical framework and involves a purely kinematic model described
by the 3 parameters: $(q_0, j_0, \Omega_k)$. Here
$q_0=-\ddot{a}\dot{a}^{-2}a$ is the cosmic deceleration;
$j_0=\stackrel{...}{a}\dot{a}^{-3}a^2$ is the third derivative of the
scale factor, the so-called cosmic jerk; and $\Omega_k$ is the
curvature parameter.  The only assumption that enters into this
parameterization is that the Robertson-Walker metric accurately
describes the geometry of the Universe.  It is therefore free of
assumptions about the energy content of the Universe or even that the
Einstein field equations are the correct description of gravity; it is
a purely kinematic model.  Although currently somewhat out of fashion,
it still remains the closest parameterization of the data, since it is
acceleration that is actually being measured. In the standard
cosmological model $(\Omega_m,\Omega_{DE},\wsym) = (0.3,0.7,-1)$, $j =
1$, and $q_0 = -0.67$. As noted by \citet{Blandford2004}, lacking an
understanding of the dynamics of the Universe, a purely kinematic
description remains a well-motivated family of models to explore.

The scale factor is expanded as a Taylor series:
\begin{equation}
\frac{a(t)}{a(t=t_0)} \simeq 1 + H_0(t-t_0) - \frac{1}{2}q_0H_0^2(t-t_0)^2 +
   \frac{1}{6}j_0 H_0^3(t - t_0)^3 + \ldots
\end{equation}
from which one can then derive a luminosity distance.
\citet{CaldwellKamionkowski2004} provide a convenient expansion of the
luminosity distance to third order in $z$ and show that to within the
precision of this truncation of the Taylor series, the cosmic jerk and
curvature can be combined into one parameter ($j_k = j + \Omega_k$):
\begin{eqnarray}
d_L\left( z\right) & \simeq & \frac{cz}{H_0}\left\{
    1 + \frac{1}{2}\left( 1-q_0\right) z - \right.\nonumber \\
    & & \left.\frac{1}{6}\left( 1 - q_0 - 3q_0^2 + j_0 + \Omega_k\right) z^2 + 
    \mathcal{O}\left( z^3\right) \right\}\nonumber .
\end{eqnarray}

Using the
definitions of these parameters and the Friedmann equation, one can
derive the transformation equations between the two parameterizations:
\begin{equation}
\begin{array}{ccc}
 q_{0} & = & \frac{1}{2}\left(\Omega_{m}+\Omega_{DE}\left(1+3w\right)\right)\\
 j_{0} & = & \Omega_{m}+\frac{\Omega_{DE}}{2}\left(2+9w\left(1+\wsym \right)\right)\\
 \Omega_{k} & = & \Omega_{m}+\Omega_{DE}-1\end{array}
 \nonumber
 \end{equation}

\begin{figure}
\centerline{
\includegraphics[width=4in,angle=-90]{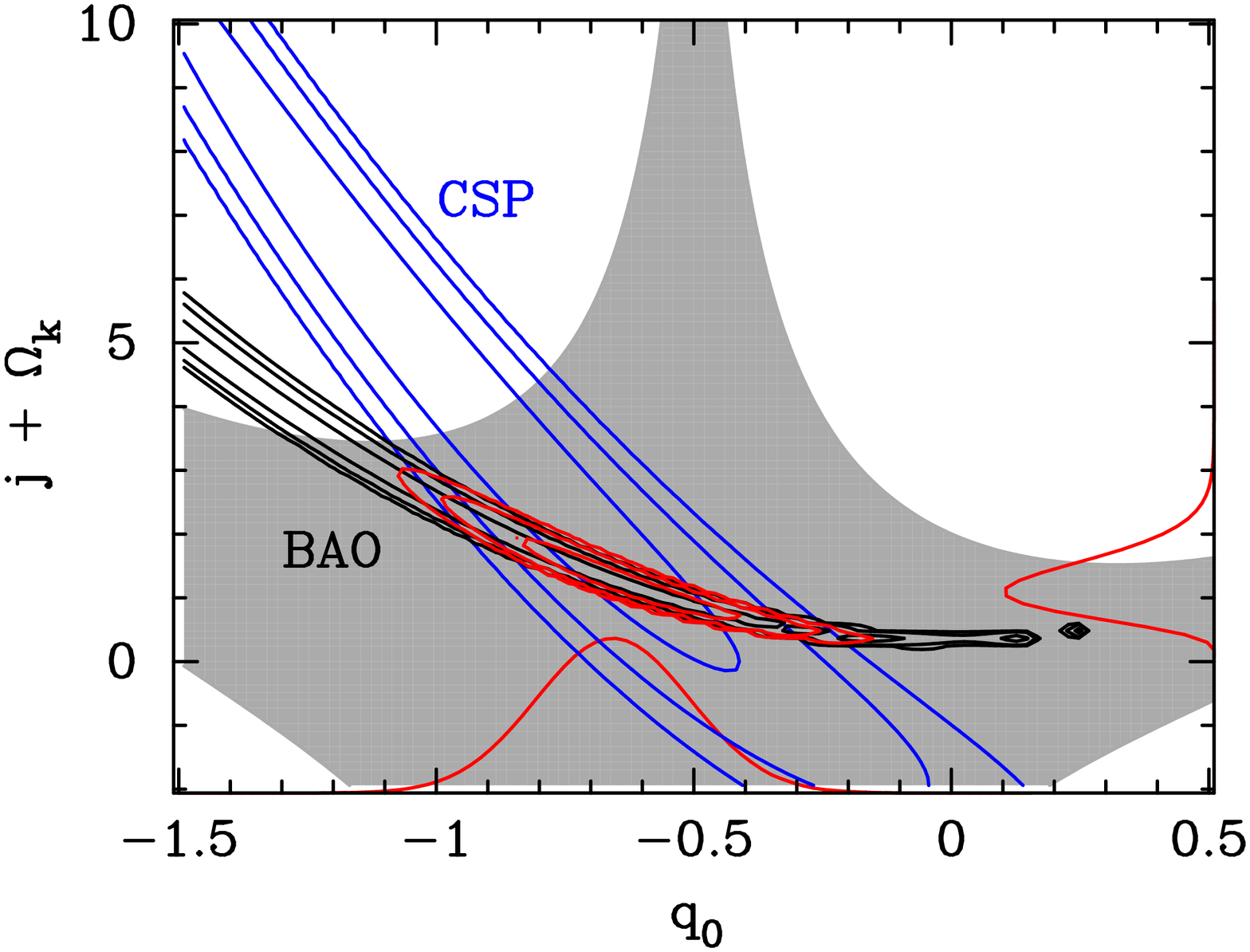}
}
\caption{The sum ($j_k$) of the jerk ($j$) and curvature parameter
  ($\Omega_k$) as a function of the deceleration parameter (q$_0$).
  The grey shading indicates the region where the luminosity distance
  expansion is valid, as described in the text. The best-fit values
  including both baryon acoustic oscillation and the CSP data are \jo\
  and \qo\ at the 95\% confidence level.  The 1-D marginalized probabilities 
  for each parameter are plotted as red lines on the axes.
\label{fig:j0q0}}
\end{figure}

In Figure \ref{fig:j0q0}, we show the sum of the jerk and curvature ($j
+ \Omega_k$) parameters as a function of $q_0$.  A constant value of
the jerk is assumed. The grey shading
indicates the region where the luminosity distance expansion is
valid.\footnote{Specifically, the error incurred by not including the
4th order term in the Taylor series expansion results in a distance
modulus error of $\pm$0.3 mag at a redshift of 0.7.}  The blue
contours represent the CSP data using the
\citet{CaldwellKamionkowski2004} parameterization. The black contours,
representing the baryon acoustic oscillation data, were generated
using the transformation equations above. The joint constraints from
the CSP and baryon acoustic oscillation data yield a value of \jo\ and
\qo\ at the 95\% confidence level. Since jerk and curvature are
combined, no assumption of flatness is required.  We note that to within the uncertainties, these
parameters are consistent with the standard dynamical model with $\wsym =
-1$.  With future, larger data sets, the discrimination amongst
competing models will be sharpened and this type of kinematical
prescription will offer a valuable independent test of the current
standard cosmology.

\subsection{Systematic Uncertainties}
\label{sec:cosmo_errors}

To quantify the effects of systematic errors, we model the effects of
known and potential systematic errors for each of the five
cosmological parameters discussed in this paper ($\wsym$, $\Omega_m$,
$\Omega_{DE}$, $q_0$, and $j_0$).  We present in Table
\ref{tab:syserrors} the main sources of uncertainty that could
contribute to a systematic error, including magnitude and color
offsets between the low- and high-redshift samples, errors in the
color terms used to transform the instrumental magnitudes to our
natural system, and biases due to the method of
$\chi^2$-minimization. We have aimed to give a conservative estimate
of the uncertainties.  The first column describes the potential source
of error and the second column provides a bound on its magnitude.  For
each of the cosmological parameters, we tabulate the rate of change of
the parameter with respect to the systematic (columns 3, 5, 7, 9, 11,
13, and 15) and the resulting error on the parameter (columns 4, 6, 8,
10, 12, 14, and 16).  The main sources of potential systematic errors
and how we simulate their effects are discussed in more detail in
Appendix \ref{sec:app_systematics}.

The final systematic errors adopted are obtained by summing in
quadrature the contributions listed in Table \ref{tab:syserrors}.
From this table, we can see that the current systematic total
uncertainty in our measurement of \wsym\ is $\pm$9\%. A key element
for future SNe~Ia studies is improving the absolute calibration of
photometric standards (particularly in the era of the Joint Dark
Energy Mission, JDEM), ensuring that the calibration minimizes color
uncertainties as optical and NIR measurements are
compared. Decreasing the uncertainty due to reddening is another
critical component of minimizing the overall systematic errors.

Finally, we summarize in Table \ref{tab:summarycosmology} the values
of cosmological parameters from this paper calculated under different
sets of assumptions, as described in \S \ref{sec:ibandHubble}, along
with both their statistical and systematic uncertainties.  When the
analysis of our total sample of \NTypeIa\ high-redshift SNe~Ia is
complete, our statistical and systematic uncertainties will be
comparable.

\section{Summary and Future Measurements}

\label{sec:summary}
We have used ground-based NIR measurements of SNe~Ia to
yield an independent Hubble diagram based on {\it rest-frame}
$i$-band data. Reddening effects are a priori lower in the red
than the ultraviolet-blue-visual, and we find that $i$-band
photometry is an effective tool for mimimizing systematic effects for
SNe~Ia.  Our new observations of \Nlowz\ SNe~Ia at $0.01 < z < 0.08$
and \Ngood\ SNe~Ia at $0.12 < z < 0.70$, yield the following results:

\noindent
1) These first CSP data provide independent evidence for an
accelerating universe.  In the context of a cosmological model
including a component of dark energy, $\Omega_{DE} >$ 0 at
signficantly greater than the 99\% confidence level.

\noindent
2) Joint SN~Ia plus baryon acoustic oscillation constraints yield \wo\
and \OmegaMatterFlat. When the analysis of our total sample of
\NTypeIa\ high-redshift SNe~Ia is complete, our statistical and
systematic uncertainties will be comparable, and at the $\pm$10$\%$ level.

\noindent
3) A purely kinematic solution, with no assumptions about the matter
and energy content of the Universe, yields values of \jo\ and \qo\
for the cosmic jerk and the deceleration parameter,
respectively. These results are consistent with an acceleration of the
expansion of the Universe, and with the current standard model of
cosmology. 

\noindent
4) The current sample of SN~Ia photometry is inconsistent with the
application of a standard Milky Way reddening law alone, suggesting
either that intrinsic color effects dominate the standard reddening
corrections, that the SNe~Ia are being reddened by dust with different
properties than that in the Milky Way, or perhaps there is
circumstellar dust about the SN~Ia. Future SNe~Ia studies will need to
disentangle these effect to decrease the systematic errors for SN~Ia
cosmology.

Beyond a redshift of ~0.7, the $i$-band (at restframe ~7600\AA)
~is shifted beyond the 1.2$\mu m$ J-band in the NIR, rendering
$i$-band measurements impossible from the ground. Even before
this limit is reached, observations from space are desirable in order
to eliminate the bright terrestrial sky background. The CSP data
illustrate quantitatively the utility of longer-wavelength data in
minimizing systematic uncertainties for SNe~Ia cosmology, of relevance
for future planned space missions such as the NASA/DOE Joint Dark
Energy Mission (JDEM). We suggest that a combination of ground, HST,
and future space observations are needed to measure accurately the
$i$-band Hubble diagram and constrain the values of $\wsym$ and
$\wsym_a$ to the highest possible accuracy.

Finally, we note that while SNe~Ia currently provide the most
compelling evidence for the acceleration of the Universe, ultimately a
combination of different techniques will be required to measure not
only $\wsym$, but also its time evolution.  Existing degeneracies are
such that accurate constraints can be obtained only in the combination
of several techniques, or explicitly making the assumption that the
Universe is flat ($\Omega_{TOT}$ = 1). Future studies of baryon
acoustic oscillations, weak lensing, clusters of galaxies, and
SNe~Ia will yield further complementary and independent estimates
of $\wsym$ and $\wsym_a$. SNe~Ia will remain a valuable component
of future investigations because they naturally cover the redshift
range (0.1 to 0.6) where dark energy is measured to be dominant, and
cosmic variance is not a major issue for them.

\medskip
\medskip
\medskip

\acknowledgments We thank each of the SNLS, ESSENCE and SDSS-II teams
for all of their dedicated efforts in discovering SNe~Ia,
and for providing coordinates and finder charts in a timely manner,
allowing the CSP to follow-up these objects.  We acknowledge the National Science Foundation (NSF) through
grant AST03-06969 for support of the low-redshift component of the CSP
and the Department of Energy through grant DE-FG02-07ER41512 for
support of the high-redshift CSP. WLF acknowledges the Aspen Center
for Physics for its hospitality in June 2007 during the workshop
``Supernovae as Cosmological Distance Indicators,'' as this paper was
being written for publication. Funding for the SDSS and SDSS-II has
been provided by the Alfred P. Sloan Foundation, the Participating
Institutions, the National Science Foundation, the U.S. Department of
Energy, the National Aeronautics and Space Administration, the
Japanese Monbukagakusho, the Max Planck Society, and the Higher
Education Funding Council for England. The SDSS Web Site is
http://www.sdss.org/.

\appendix

\section{Systematic Error Budget}
\label{sec:app_systematics}

In this appendix, we outline our method for estimating the effects of
systematic errors in the determination of various
parameters.  We have categorized the various sources of error  based on their
functional form.  We then compute the sensitivity of the cosmological parameters to
these functions through simulations.  Finally, we compute the magnitude 
of each effect by summing the systematics for each category
in quadrature.  Table \ref{tab:syserrors} summarizes the sensitivity of
each parameter to each systematic. We also illustrate the effects of these
systematic effects on the calculated parameters  in graphical
form. 

\subsection{Magnitude Offset Between Low- and High-z}

The low- and high-z observations are obtained on different telescopes,
and therefore are calibrated independently, which can potentially
introduce a simple magnitude offset between the zero-points of the
low- and high-z data.  We model this effect by simply adding fixed
values, $\delta_m$, to the high-z peak magnitudes and re- compute the
best-fit cosmology.

\begin{figure}
\centerline{
\includegraphics[width=5in]{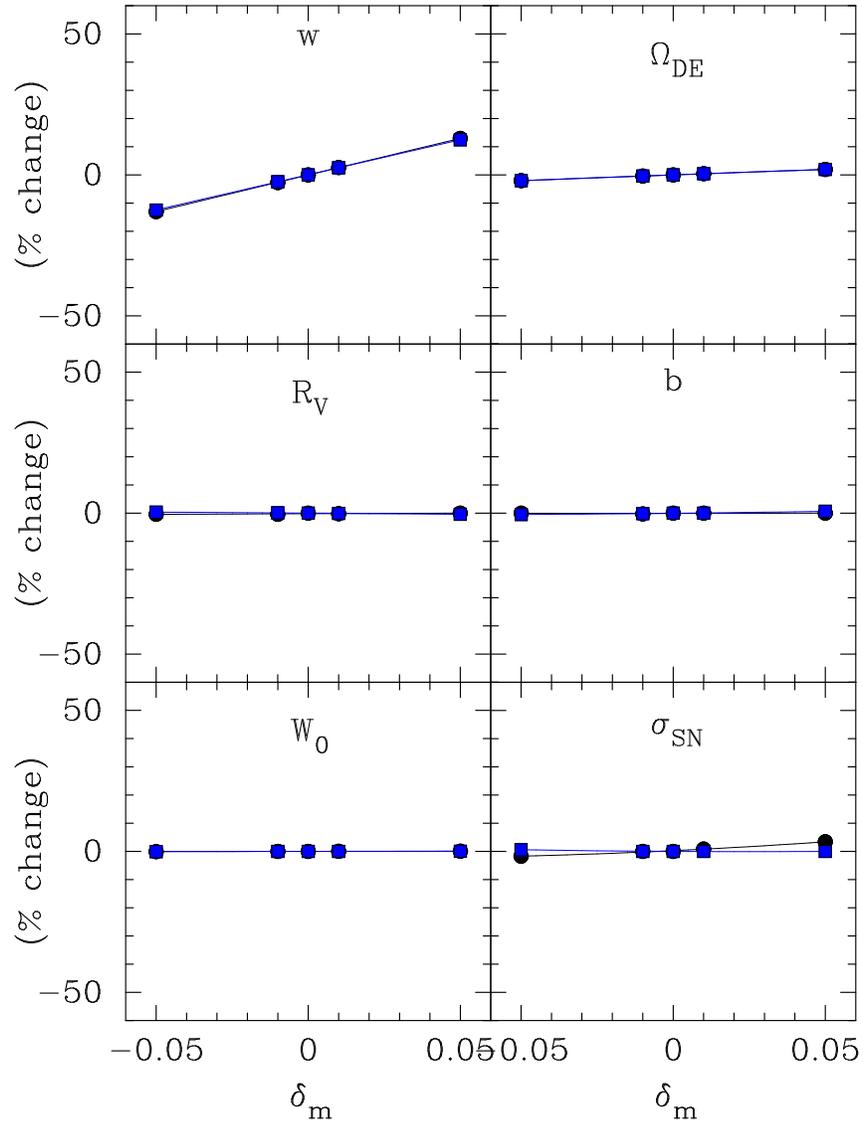}
}
\caption{The effect of a magnitude offset $\delta_m$ on the best-fit
  parameters, \wsym, $\Omega_{DE}$, $R_V$, $B$, $W_0$, and
  $\sigma_{SN}$.  The black circles are for i-band, while the red
  squares are for $B$-band. In almost all cases, the differences in
  the points are negligible. The lines are quadratic fits. The largest
  sensitivity to a magnitude offset is for \wsym. For an offset of
  0.025 magnitudes, the percentage change in \wsym\ amounts to
  7\%.\label{fig:systematic_mag}}
\end{figure}

The results of this simulation are shown in Figure \ref{fig:systematic_mag} and
show that to a good approximation, the systematic effect of a magnitude offset 
on \wsym\ is a linear function of $\delta_m$ with slope $dw/d\delta_m = -2.69$ in 
the $i$-band and $dw/d\delta_m = -2.66$ in the $B$-band.  Furthermore, there
is no significant effect on the 3 nuisance parameters.

To estimate the value of $\delta_m$, we consider the following possible
sources of such an offset:   1) error in the $Y$ and $J$ zero-points, 2) error in
the $i$-band zero-point, and 3) errors
in the extinction coefficients.  We describe each in turn.

The contribution to the systematic uncertainty in $\delta_m$ has a
different origin for the $Y$ and $J$ filters. As described in
\ref{sec:calibration}, we have used Kurucz models to extend the energy
distributions of the \citet{Persson1998} standards from $J$ ($1.25
\mu$) to $Y$ ($1.035 \mu$) \citep{Hamuy2006}. That this was viable was
confirmed by \citet{Contreras2009}, who showed that the zero-color A0
star Feige 16 does in fact have $Y - J = -0.009 \pm 0.016$, confirming
this calibration to an accuracy of $\pm$0.01 mag. Conservatively, we adopt
an error in the zero-point for the $Y$-band of $\pm$0.025.

In the case of $J$, the main problem is the changing amount of water vapor
above the telescope. An increase manifests itself as a systematic narrowing
of the passband. This may or may not show up in standard star measurements,
depending on when during the night they were measured.  We adopt
an error in the zero-point for $J$-band of $\pm$0.015.

The common ancestor to the \citet{Persson1998} and \citet{Smith2002}
systems is Vega. We therefore must determine the zero-point of
the $i$-band relative to Vega instead of $BD+17\arcdeg 4708$
in order to compute consistent zero-points.  We have investigated the
effects of a systematic error in the SED of Vega by comparing synthetic
$i$-band photometry using the \citet{Bohlin2007} and \citet{Bohlin2004} SEDs.
The difference is 0.01 mag.  We further investigated uncertainties in our 
$i$-band filter transmission function.  In constructing all our CSP filters, 
shifts in wavelength needed to be applied in order to have the synthetic 
color terms match the observed color terms \citep{Contreras2009}.  A typical
uncertainty in these shifts is approximately  $10 \AA$, which is equivalent
to an error of $\pm$0.001 mag.  We therefore estimate a zero-point error
of $\pm$0.01 for the $i$-band.

According to \citet{Hamuy2006} and \citet{Contreras2009}, the dispersion in
the extinction coefficent for the  $i$-band is $\pm$0.03.  The median
airmass of our SNe~Ia observations in the $i$-band was 1.315 and so we
estimate a systematic error of $\pm 0.009$ mag due to dispersion in the
extinction coefficient.  Similar errors are expected for the $Y$- and
$J$-band filters, though the median airmass for our high-redshift
observations is typically lower (1.24).  We therefore estimate a total
error of $\pm 0.01$ mag.
    
Given the external check with the standard star Feige 16
\citep{Contreras2009}, we believe our calibration to be robust. Nevertheless,
we allow for a 
conservative systematic magnitude offset between low- and
high-redshift of  $\delta_m = \pm 0.025$ mag.

\subsection{Color Offset Between Low- and High-z}

\begin{figure}
\centerline{
\includegraphics[width=5in]{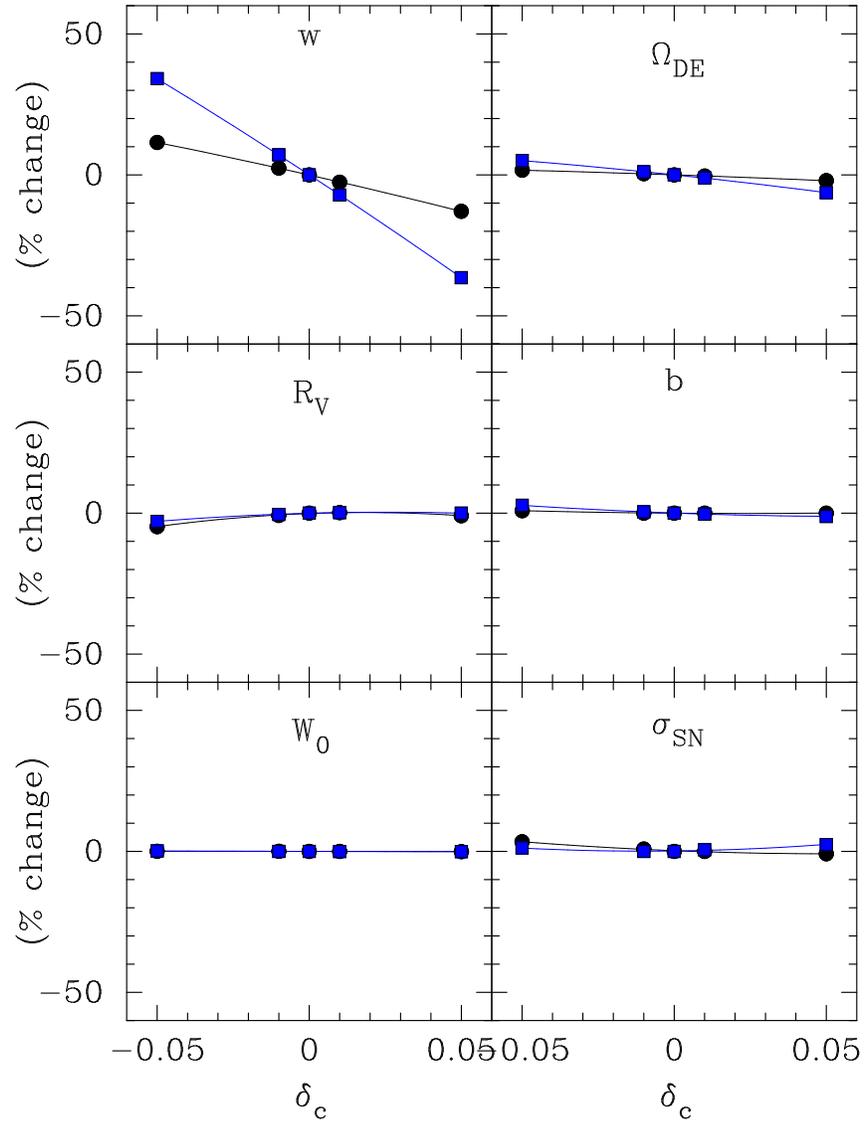}
}
\caption{The effect of a color offset $\delta_c$ on the best-fit
  parameters, as listed in Figure \ref{fig:systematic_mag}.  The 
black circles are for i-band, while the red squares are for $B$-band.  The lines are
quadratic fits. The largest
  sensitivity to a color offset is again for \wsym. For a color offset of 0.02
  magnitudes, the percentage change in \wsym\ amounts to
  5\%.\label{fig:systematic_color}}
\end{figure}

This systematic is analogous to the magnitude offset and can occur for the
same reasons; however, it propagates through the analysis differently as it
is multiplied by the reddening coefficient.  We proceed in the same manner
as before, adding a color offset $\delta_c$ to the $(B-V)$ colors at high-z
and compute the best-fit cosmology.  As in the previous case, the effect on 
the cosmological parameters is linear (see Figure \ref{fig:systematic_color}).
However, the effect on the $B$-band is more
than twice that in the $i$-band ($dw/d\delta_c = 2.53$ and $dw/d\delta_c = 7.53$,
respectively).  This is simply due to the fact that the reddening coefficient
is twice as large at $B$ relative to the $i$-band.  An advantage is
that the colors we use
in our analysis are all constructed from the same optical photometric systems and avoid
the NIR and $i$-band.  We estimate the total possible color offset to be 
$\delta_c = \pm 0.02$.

\subsection{Color Gradient}

Color terms are used to transform the magnitudes of the standard stars
to the natural CSP system.  These color terms are empirically
determined and therefore suffer from measurement uncertainty.  An
error in the color terms could potentially introduce a color gradient
in the data: color errors that correlate with the instrumental color
of the SNe~Ia.  Figure \ref{fig:systematic_color_term} shows the
sensitivity functions.  As expected, the impact in the $B$-band is
larger than in the $i$-band.  From \citet{Hamuy2006} and
\citet{Contreras2009}, we estimate $\delta\left(CT\right) = 0.015$,
which is the typical 1-$\sigma$ error in the color terms.

\subsection{Estimation of Uncertainties}

The method of $\chi^2$-minimization suffers from a well-known bias
when there is measurement error in the independent variables
\citep{Kelly2007}.  The root of the problem lies in the fact that the
numerator of $\chi^2$ (see equation \ref{eq:chi_cosmo}) is sensitive
to the model parameters only, whereas the denominator is sensitive to
the nuisance parameters as well as the variances (see equation
\ref{eq:sigmasq}).  As such, the best-fit solution is a function not
only of the data, but the variances as well, complicating the
computation of the overall systematic uncertainties.

\begin{figure}
\centerline{
\includegraphics[width=5in]{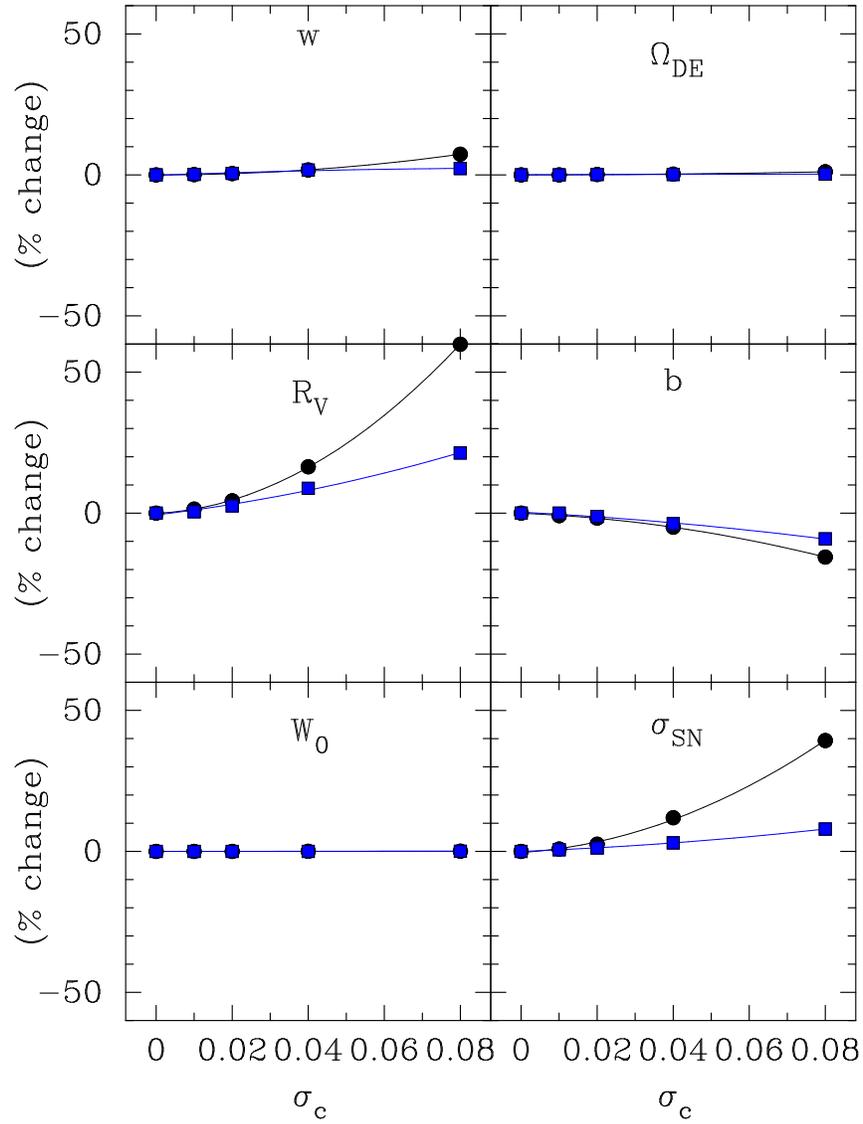}
}
\caption{The effect of extra variance in the color measurements 
$\sigma_c$ on the best-fit parameters, as listed in Figure \ref{fig:systematic_mag}.  The 
black circles are for i-band, while the red squares are for $B$-band.  The lines are
quadratic fits.  A variance in the color measurements impacts the
determination of the reddening law or color term. A variance of $\pm$0.03
  magnitudes results in a  percentage change in $R_V$ of
  10\%. The impact on cosmology is very small.  \label{fig:systematic_color_disp}}
\end{figure}

\begin{figure}
\centerline{
\includegraphics[width=5in]{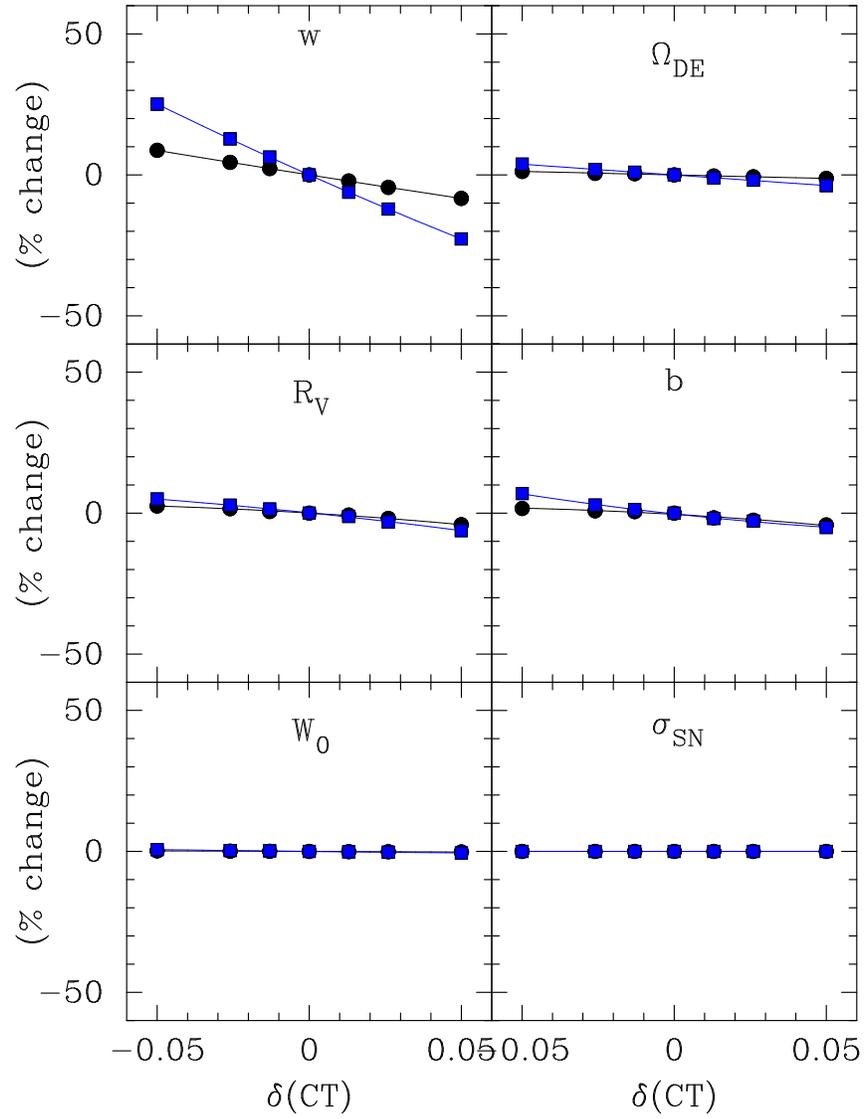}
}
\caption{The effect of an error in the color-term derived from local
sequences of standards, as listed in Figure \ref{fig:systematic_mag}.
The black circles are for $i$-band, while the red squares are for $B$-band.  The lines are
quadratic fits.  An error in the color term has the largest impact on \wsym.  For
an error of 0.015, the percentage change in \wsym\ amounts to
  3\%. \label{fig:systematic_color_term}}
\end{figure}

We have investigated the effect of adding extra variance to both the
colors $\sigma^2_c$ and \dm.  Figure \ref{fig:systematic_color_disp}
shows the strong dependence of $R_V$ on $\sigma_c$.  The $i$-band data
are about a factor of two less sensitive to changes in $R_V$ than the
$B$-band data.  As a result, the bias due to the denominator of $\chi^2$ is
larger in the $i$-band.  If we were to include more heavily reddened objects
(for instance SN~2005A and SN~2006X), then this effect would disappear
entirely.  The effects of extra variance in \dm\ are not significant
and are therefore not shown, though they are included in Table
\ref{tab:syserrors} for completeness.

\clearpage


\clearpage


\clearpage

\clearpage

\clearpage
\newpage
\LongTables
\begin{deluxetable}{lcllc}
\tablecolumns{5}
\tablewidth{0pc}
\tablecaption{Filter nomenclature\label{tab:filters}}
\tablehead{\colhead{Name} & \colhead{Eff. Wavelength} & \colhead{Description} & 
           Zero-point & \colhead{reference}\\
           \colhead{} & \colhead{(Angstroms)} & \colhead{} & \colhead{} & \colhead{}}
\startdata
$Y$ & 10333.6 & $Y$-band on PANIC, Magellan I, LCO & 12.6870 & \citet{Persson1998} \\
$J$ & 12440.3 & $J$-band on PANIC, Magellan I, LCO & 12.8520 & " \\
$B$ & 4905.0 & $B$-band on Swope, LCO & 13.6672 & \citet{Hamuy2006} \\
$V$ & 4905.0 & $V$-band on Swope, LCO & 13.7758 & " \\
$i$ & 4905.0 & $i$-band on Swope, LCO & 13.6873 & " \\
$g_m$ & 4905.0 & $g$-band on Megacam, CFHT & 15.5363 & \citet{Astier2006} \\
$r_m$ & 6282.6 & $r$-band on Megacam, CFHT & 14.8061 & " \\
$i_m$ & 7725.0 & $i$-band on Megacam, CFHT & 14.5543 & " \\
$z_m$ & 8901.6 & $z$-band on Megacam, CFHT & 13.9996 & " \\
$g_s$ & 4718.9 & $g$-band on 2.5m, APO & 14.2013 & \citet{Smith2002} \\
$r_s$ & 6185.2 & $r$-band on 2.5m, APO & 14.2157 & " \\
$i_s$ & 7499.7 & $i$-band on 2.5m, APO & 13.7775 & " \\
$R_{4m}$ & 6660.2& $R$-band on 4m Blanco telescope, CTIO & 15.1820 & \citet{Miknaitis2007} \\
$I_{4m}$ & 7973.8& $I$-band on 4m Blanco telescope, CTIO & 14.4580 & " \\
\enddata
\end{deluxetable}
\clearpage

\begin{deluxetable}{lcccll}
\tablecolumns{6}
\tablewidth{0pc}
\tablecaption{CSP high-z SNe\label{tab:CSPSNe}}
\tablehead{\colhead{Name} & \colhead{RA (2000)} & 
           \colhead{DEC (2000)} & \colhead{Epochs} & \colhead{z} & Cross-ID\\
                 \colhead{} & \colhead{(d:m:s)} &
                 \colhead{(h:m:s)} & 
           \colhead{} & \colhead{} & \colhead{}}
\startdata
SNLS 03D4gl & 22:14:44.16 & -17:31:44.40 & 3 & 0.571 & \nodata \\
SNLS 04D1oh & 02:25:02.38 & -04:14:10.68 & 2 & 0.59 & \nodata \\
SNLS 04D1pg & 02:27:04.15 & -04:10:31.33 & 3 & 0.515 & \nodata \\
SNLS 04D1rh & 02:27:47.16 & -04:15:13.68 & 4 & 0.435 & \nodata \\
SNLS 04D1sk & 02:24:22.56 & -04:21:13.32 & 2 & 0.6634 & \nodata \\
SNLS 04D2an & 10:00:52.32 & 02:02:28.68 & 8 & 0.62 & \nodata \\
SNLS 05D1dn & 02:24:26.64 & -04:59:29.40 & 4 & 0.566 & \nodata \\
SNLS 05D1hk & 02:24:39.17 & -04:38:03.01 & 2 & 0.2631 & \nodata \\
SNLS 05D1hn & 02:24:36.26 & -04:10:54.95 & 1 & 0.1489 & \nodata \\
SNLS 05D1ix & 02:24:19.94 & -04:40:11.75 & 3 & 0.49 & \nodata \\
SNLS 05D1iy & 02:27:39.96 & -04:25:21.36 & 1 & 0.2478 & \nodata \\
SNLS 05D2ah & 10:01:28.80 & 01:51:46.08 & 6 & 0.184 & \nodata \\
SNLS 05D2bt & 10:01:40.32 & 02:33:57.96 & 5 & 0.679 & \nodata \\
SNLS 05D2bv & 10:02:17.04 & 02:14:26.16 & 5 & 0.474 & \nodata \\
SNLS 05D2ck & 10:00:45.12 & 02:34:22.08 & 4 & 0.698 & \nodata \\
SNLS 05D2dw & 09:58:32.16 & 02:01:56.28 & 2 & 0.417 & \nodata \\
SNLS 05D2eb & 10:00:14.64 & 02:24:26.64 & 2 & 0.5344 & \nodata \\
SNLS 05D2mp & 09:59:08.64 & 02:12:14.69 & 3 & 0.3537 & \nodata \\
SNLS 05D4cw & 22:14:50.16 & -17:44:19.32 & 3 & 0.375 & \nodata \\
SNLS 05D4fo & 22:15:20.88 & -17:16:05.16 & 4 & 0.373 & \nodata \\
ESS d149wcc4-11 & 02:10:53.98 & -04:25:49.80 & 3 & 0.342 & SN2003jy \\
ESS e108wdd8-4 & 02:30:09.00 & -09:04:35.76 & 3 & 0.469 & SN2003km \\
SDSS 12855 & 22:01:01.44 & 00:42:58.47 & 2 & 0.165 & \nodata \\
SDSS 13025 & 22:46:16.08 & 00:24:57.21 & 3 & 0.224 & \nodata \\
SDSS 13835 & 00:24:14.31 & -00:14:53.69 & 2 & 0.247 & \nodata \\
SDSS 15287 & 21:35:50.40 & -01:03:26.75 & 1 & 0.235 & \nodata \\
SDSS 16442 & 21:59:47.28 & -00:43:59.59 & 3 & 0.281 & \nodata \\
SDSS 3241 & 20:50:36.24 & -00:21:14.76 & 5 & 0.259 & SN2005gh \\
SDSS 3331 & 02:18:14.74 & 00:47:47.76 & 5 & 0.208 & SN2005ge \\
SDSS 4679 & 01:26:06.79 & 00:40:36.80 & 4 & 0.333 & SN2205gy \\
SDSS 5183 & 03:33:48.96 & 00:42:33.70 & 6 & 0.384 & SN2005gq \\
SDSS 5549 & 00:13:00.13 & 00:14:53.70 & 5 & 0.119 & SN2005hx \\
SDSS 6699 & 21:31:15.60 & -01:03:25.16 & 4 & 0.311 & SN2005ik \\
SDSS 7243 & 21:52:18.96 & 00:28:19.09 & 3 & 0.204 & SN2005jm \\
SDSS 7512 & 03:28:21.67 & -00:19:34.10 & 2 & 0.22 & SN2005jo \\
\enddata
\end{deluxetable}

\clearpage
\begin{deluxetable}{cccccccccc}
\tablecolumns{8}
\tablewidth{0pc}
\tablecaption{CSP high-z Photometry\label{tab:Photo}}
\tablehead{\colhead{Name} & \colhead{MJD\tablenotemark{a}} & \colhead{$Y_c$} & \colhead{$\sigma_Y$} & \colhead{$K_{Y,I}$\tablenotemark{b}} &
          \colhead{$J_c$} & \colhead{$\sigma_J$} & \colhead{$K_{J,I}$\tablenotemark{c}} \\
          \colhead{}& \colhead{(days)} & \colhead{(mag)} & \colhead{(mag)} & 
          \colhead{(mag)} & \colhead{(mag)} & \colhead{(mag)} & \colhead{(mag)} }
\startdata
SNLS 03D4gl & 52947.0 &22.57 & 0.09 & -0.97 &22.84 & 0.15 & -1.37\\
 & 52947.0 &22.54 & 0.05 & -0.91 &23.03 & 0.13 & -1.40\\
 & 52952.0 &22.64 & 0.05 & -0.90 &22.80 & 0.08 & -1.38\\
SNLS 04D1oh & 53302.1 & \ldots & \ldots & \ldots & 22.81 & 0.09 & -1.39 \\
 & 53314.2 & \ldots & \ldots & \ldots & 22.78 & 0.10 & -1.37 \\
SNLS 04D1pg & 53328.2 & \ldots & \ldots & \ldots & 22.72 & 0.08 & -1.46 \\
 & 53338.1 & \ldots & \ldots & \ldots & 22.95 & 0.08 & -1.48 \\
 & 53344.2 & \ldots & \ldots & \ldots & 23.28 & 0.19 & -1.59 \\
SNLS 04D1rh & 53356.1 &22.44 & 0.06 & -1.21 &22.26 & 0.07 & -1.34\\
 & 53356.1 &22.60 & 0.09 & -1.23 &22.53 & 0.11 & -1.46\\
SNLS 04D1sk & 53356.1 & \ldots & \ldots & \ldots & 23.33 & 0.19 & -1.50 \\
 & 53363.0 & \ldots & \ldots & \ldots & 23.78 & 0.33 & -1.51 \\
SNLS 04D2an & 53027.2 &22.75 & 0.05 & -1.11 &22.92 & 0.07 & -1.41\\
 & 53027.2 &22.63 & 0.05 & -1.08 &23.23 & 0.11 & -1.40\\
 & 53034.1 &22.84 & 0.10 & -1.03 &23.07 & 0.14 & -1.39\\
 & 53034.1 &22.67 & 0.05 & -0.95 &23.58 & 0.17 & -1.39\\
 & 53040.2 &23.11 & 0.07 & -0.91 &23.64 & 0.17 & -1.41\\
 & 53040.2 &23.27 & 0.12 & -0.90 &23.52 & 0.22 & -1.41\\
 & 53045.2 &23.32 & 0.07 & -1.03 &23.38 & 0.14 & -1.41\\
 & 53045.2 &23.33 & 0.10 & -1.06 &23.73 & 0.20 & -1.41\\
SNLS 05D1dn & 53641.3 & \ldots & \ldots & \ldots & 22.76 & 0.13 & -1.35 \\
 & 53654.3 & \ldots & \ldots & \ldots & 22.64 & 0.10 & -1.35 \\
 & 53659.2 & \ldots & \ldots & \ldots & 22.94 & 0.13 & -1.36 \\
 & 53666.2 & \ldots & \ldots & \ldots & 23.11 & 0.19 & -1.41 \\
SNLS 05D1hk & 53720.0 & 21.10 & 0.03 & -1.02 & \ldots & \ldots & \ldots \\
 & 53724.1 & 21.13 & 0.03 & -0.98 & \ldots & \ldots & \ldots \\
SNLS 05D1hn & 53721.0 & 20.63 & 0.03 & -0.97 & \ldots & \ldots & \ldots \\
SNLS 05D1ix & 53720.1 & \ldots & \ldots & \ldots & 22.07 & 0.09 & -1.54 \\
 & 53722.1 & \ldots & \ldots & \ldots & 22.27 & 0.09 & -1.54 \\
 & 53724.0 & \ldots & \ldots & \ldots & 22.04 & 0.09 & -1.51 \\
SNLS 05D1iy & 53721.1 & 21.65 & 0.02 & -0.92 & \ldots & \ldots & \ldots \\
SNLS 05D2ah & 53390.2 &20.84 & 0.03 & -0.86 &21.46 & 0.05 & -0.39\\
 & 53390.2 &20.93 & 0.02 & -0.87 &\ldots & \ldots & \ldots\\
 & 53391.2 &\ldots & \ldots & \ldots &21.71 & 0.06 & -0.38\\
 & 53392.2 &20.93 & 0.02 & -0.92 &21.71 & 0.08 & -0.37\\
 & 53393.3 &21.02 & 0.03 & -0.95 &21.77 & 0.07 & -0.37\\
 & 53393.3 &20.93 & 0.02 & -1.16 &21.66 & 0.04 & -0.37\\
SNLS 05D2bt & 53405.1 & \ldots & \ldots & \ldots & 22.98 & 0.09 & -1.54 \\
 & 53413.2 & \ldots & \ldots & \ldots & 23.28 & 0.24 & -1.55 \\
 & 53418.1 & \ldots & \ldots & \ldots & 23.07 & 0.34 & -1.56 \\
 & 53424.1 & \ldots & \ldots & \ldots & 23.47 & 0.28 & -1.54 \\
 & 53430.1 & \ldots & \ldots & \ldots & 23.09 & 0.20 & -1.54 \\
SNLS 05D2bv & 53405.3 & \ldots & \ldots & \ldots & 22.21 & 0.04 & -1.35 \\
 & 53413.2 & \ldots & \ldots & \ldots & 22.64 & 0.13 & -1.36 \\
 & 53418.2 & \ldots & \ldots & \ldots & 22.58 & 0.08 & -1.41 \\
 & 53424.2 & \ldots & \ldots & \ldots & 22.76 & 0.15 & -1.58 \\
 & 53430.1 & \ldots & \ldots & \ldots & 22.90 & 0.14 & -1.60 \\
SNLS 05D2ck & 53418.3 & \ldots & \ldots & \ldots & 23.28 & 0.25 & -1.54 \\
 & 53424.3 & \ldots & \ldots & \ldots & 23.29 & 0.30 & -1.58 \\
 & 53430.2 & \ldots & \ldots & \ldots & 23.51 & 0.23 & -1.59 \\
 & 53444.1 & \ldots & \ldots & \ldots & 23.61 & 0.20 & -1.55 \\
SNLS 05D2dw & 53447.1 &21.99 & 0.07 & -1.16 &22.20 & 0.09 & -1.31\\
 & 53447.1 &22.20 & 0.06 & -1.14 &22.00 & 0.06 & -1.30\\
SNLS 05D2eb & 53447.1 &22.23 & 0.12 & -0.82 &22.40 & 0.15 & -1.51\\
 & 53447.1 &22.49 & 0.04 & -0.78 &\ldots & \ldots & \ldots\\
SNLS 05D2mp & 53720.3 & 22.44 & 0.04 & -0.94 & \ldots & \ldots & \ldots \\
 & 53722.2 & 22.55 & 0.04 & -0.95 & \ldots & \ldots & \ldots \\
 & 53724.3 & 22.54 & 0.04 & -0.96 & \ldots & \ldots & \ldots \\
SNLS 05D4cw & 53591.1 & 22.35 & 0.07 & -1.01 & \ldots & \ldots & \ldots \\
 & 53596.1 & 22.66 & 0.08 & -1.03 & \ldots & \ldots & \ldots \\
 & 53608.1 & 22.55 & 0.08 & -1.05 & \ldots & \ldots & \ldots \\
SNLS 05D4fo & 53648.1 &22.14 & 0.45 & -1.00 &22.07 & 0.15 & -1.35\\
 & 53648.1 &22.32 & 0.07 & -1.00 &21.64 & 0.06 & -1.50\\
 & 53654.1 &22.50 & 0.12 & -1.00 &21.79 & 0.11 & -1.48\\
 & 53654.1 &22.76 & 0.10 & -1.00 &21.96 & 0.11 & -1.56\\
ESS d149wcc4-11 & 52947.3 & 22.01 & 0.04 & -0.91 & \ldots & \ldots & \ldots \\
 & 52952.2 & 21.92 & 0.03 & -0.92 & \ldots & \ldots & \ldots \\
 & 52956.2 & 21.97 & 0.03 & -0.91 & \ldots & \ldots & \ldots \\
ESS e108wdd8-4 & 52980.1 &22.42 & 0.04 & -1.29 &22.25 & 0.08 & -1.38\\
 & 52980.1 &22.24 & 0.03 & -1.30 &22.28 & 0.09 & -1.39\\
 & 52982.1 &22.20 & 0.03 & -1.31 &22.28 & 0.08 & -1.39\\
SDSS 12855 & 54000.1 & 20.61 & 0.02 & -0.89 & \ldots & \ldots & \ldots \\
 & 54008.0 & 20.92 & 0.02 & -1.10 & \ldots & \ldots & \ldots \\
SDSS 13025 & 53996.1 & 21.02 & 0.03 & -1.00 & \ldots & \ldots & \ldots \\
 & 54000.1 & 21.08 & 0.04 & -0.99 & \ldots & \ldots & \ldots \\
 & 54008.1 & 21.58 & 0.04 & -1.20 & \ldots & \ldots & \ldots \\
SDSS 13835 & 54008.1 & 21.14 & 0.01 & -0.90 & \ldots & \ldots & \ldots \\
 & 54021.2 & 21.40 & 0.02 & -1.01 & \ldots & \ldots & \ldots \\
SDSS 15287 & 54029.0 & 21.13 & 0.02 & -0.94 & \ldots & \ldots & \ldots \\
SDSS 16442 & 54060.0 & 21.30 & 0.04 & -0.86 & \ldots & \ldots & \ldots \\
 & 54065.0 & 21.46 & 0.03 & -0.80 & \ldots & \ldots & \ldots \\
 & 54070.0 & 21.79 & 0.04 & -0.80 & \ldots & \ldots & \ldots \\
SDSS 3241 & 53641.0 & 21.31 & 0.02 & -0.86 & \ldots & \ldots & \ldots \\
 & 53648.0 & 21.27 & 0.02 & -0.93 & \ldots & \ldots & \ldots \\
 & 53654.0 & 21.54 & 0.02 & -0.90 & \ldots & \ldots & \ldots \\
 & 53659.0 & 21.79 & 0.02 & -0.95 & \ldots & \ldots & \ldots \\
 & 53666.0 & 21.98 & 0.03 & -1.05 & \ldots & \ldots & \ldots \\
SDSS 3331 & 53641.2 & 20.80 & 0.03 & -0.88 & \ldots & \ldots & \ldots \\
 & 53648.3 & 20.50 & 0.04 & -1.06 & \ldots & \ldots & \ldots \\
 & 53655.2 & 20.77 & 0.03 & -1.03 & \ldots & \ldots & \ldots \\
 & 53659.2 & 20.82 & 0.03 & -1.10 & \ldots & \ldots & \ldots \\
 & 53676.2 & 20.89 & 0.04 & -1.32 & \ldots & \ldots & \ldots \\
SDSS 4679 & 53655.1 & 21.99 & 0.02 & -0.89 & \ldots & \ldots & \ldots \\
 & 53659.1 & 22.02 & 0.03 & -0.88 & \ldots & \ldots & \ldots \\
 & 53666.2 & 22.35 & 0.04 & -0.89 & \ldots & \ldots & \ldots \\
 & 53676.2 & 22.37 & 0.03 & -0.91 & \ldots & \ldots & \ldots \\
SDSS 5183 & 53654.3 & 22.39 & 0.06 & -1.03 & \ldots & \ldots & \ldots \\
 & 53655.3 & 22.46 & 0.05 & -1.04 & \ldots & \ldots & \ldots \\
 & 53666.2 & 23.22 & 0.09 & -1.04 & \ldots & \ldots & \ldots \\
 & 53676.3 & 22.94 & 0.07 & -1.06 & \ldots & \ldots & \ldots \\
 & 53720.2 & 23.68 & 0.14 & -1.09 & \ldots & \ldots & \ldots \\
 & 53721.2 & 23.72 & 0.16 & -1.09 & \ldots & \ldots & \ldots \\
SDSS 5549 & 53666.1 & 19.88 & 0.04 & -0.56 & \ldots & \ldots & \ldots \\
 & 53676.1 & 20.20 & 0.04 & -0.65 & \ldots & \ldots & \ldots \\
 & 53682.1 & 20.03 & 0.11 & -0.93 & \ldots & \ldots & \ldots \\
 & 53684.1 & 20.18 & 0.03 & -0.95 & \ldots & \ldots & \ldots \\
 & 53689.1 & 20.11 & 0.04 & -0.97 & \ldots & \ldots & \ldots \\
SDSS 6699 & 53676.0 & 21.72 & 0.02 & -0.90 & \ldots & \ldots & \ldots \\
 & 53682.0 & 21.87 & 0.05 & -0.85 & \ldots & \ldots & \ldots \\
 & 53684.0 & 22.16 & 0.04 & -0.84 & \ldots & \ldots & \ldots \\
 & 53689.0 & 22.42 & 0.05 & -0.85 & \ldots & \ldots & \ldots \\
SDSS 7243 & 53682.1 & 20.82 & 0.02 & -0.83 & \ldots & \ldots & \ldots \\
 & 53684.1 & 20.87 & 0.02 & -0.88 & \ldots & \ldots & \ldots \\
 & 53689.0 & 20.94 & 0.01 & -0.88 & \ldots & \ldots & \ldots \\
SDSS 7512 & 53682.3 & 21.21 & 0.10 & -0.89 & \ldots & \ldots & \ldots \\
 & 53684.3 & 21.25 & 0.07 & -0.89 & \ldots & \ldots & \ldots \\
\enddata
\tablenotetext{a}{MJD = JD - $2400000.5$.}
\tablenotetext{b}{K-correction from observed filter $Y_c$ to rest-frame filter I.}
\tablenotetext{c}{K-correction from observed filter $J_c$ to rest-frame filter I, except for
                  SNLS05D2ah, for which the K-correction is to rest-frame Y.}
\end{deluxetable}

\clearpage
\begin{turnpage}
\begin{deluxetable}{lcccccccc}
\tablecolumns{9}
\tablewidth{0pc}
\tablecaption{Derived Light-curve Parameters for low-z SNe\label{tab:lcparams_lowz}}
\tablehead{\colhead{Name} & 
           \colhead{z} & 
           \colhead{$\mu$ ($\sigma$)\tablenotemark{a}} & 
           \colhead{$T_{max}$($\sigma$)\tablenotemark{b}} & 
           \colhead{$\Delta m_{15} $($\sigma$)\tablenotemark{c}} & 
           \colhead{$(i_{max})$ ($\sigma$)\tablenotemark{d}} &
           \colhead{$(B_{max}-V_{max})$ ($\sigma$)\tablenotemark{d}} &
           \colhead{$E(B-V)_{host}$ ($\sigma$)\tablenotemark{e}} & 
           \colhead{$A_{TOT}$\tablenotemark{f}} \\
           \colhead{} & \colhead{} & \colhead{(mag)} & 
           \colhead{(days)} & \colhead{(mag)} & 
           \colhead{(mag)} &\colhead{(mag)} & \colhead{(mag)} & \colhead{(mag)}}
\startdata
 SN2004ef & 0.03097 & 35.499(0.139) & 264.8(0.0) & 1.389(0.007) & 17.273(0.010) & 0.123 (0.007) & 0.157 (0.018) & 0.214\\
 SN2004eo & 0.01569 & 33.697(0.189) & 278.9(0.1) & 1.366(0.012) & 15.435(0.016) & 0.095 (0.015) & 0.129 (0.021) & 0.249\\
 SN2004ey & 0.01578 & 34.027(0.190) & 304.6(0.0) & 0.954(0.006) & 15.459(0.011) & -0.070 (0.008) & -0.003 (0.015) & 0.163\\
 SN2004gs & 0.02663 & 35.484(0.142) & 356.3(0.0) & 1.550(0.006) & 17.391(0.009) & 0.202 (0.009) & 0.223 (0.022) & 0.247\\
 SN2004gu & 0.04583 & 36.484(0.129) & 362.1(0.2) & 0.758(0.011) & 18.052(0.020) & 0.145 (0.021) & 0.253 (0.027) & 0.268\\
 SN2005ag & 0.07937 & 37.643(0.121) & 414.1(0.1) & 0.889(0.007) & 19.123(0.011) & 0.004 (0.007) & 0.085 (0.016) & 0.128\\
 SN2005al & 0.01239 & 34.092(0.202) & 430.6(0.1) & 1.243(0.010) & 15.609(0.015) & -0.089 (0.009) & -0.057 (0.016) & 0.012\\
 SN2005el & 0.01490 & 33.967(0.189) & 647.0(0.1) & 1.299(0.014) & 15.538(0.018) & -0.055 (0.014) & -0.025 (0.020) & 0.112\\
 SN2005eq & 0.02896 & 35.489(0.141) & 654.5(0.1) & 0.778(0.008) & 16.962(0.015) & 0.039 (0.010) & 0.135 (0.020) & 0.215\\
 SN2005hc & 0.04591 & 36.582(0.128) & 667.4(0.1) & 0.844(0.008) & 18.050(0.014) & 0.009 (0.008) & 0.095 (0.017) & 0.129\\
 SN2005hj & 0.05797 & 36.976(0.127) & 674.0(0.2) & 0.739(0.017) & 18.462(0.026) & 0.067 (0.014) & 0.170 (0.023) & 0.207\\
 SN2005iq & 0.03402 & 35.891(0.136) & 687.9(0.1) & 1.230(0.016) & 17.457(0.019) & -0.033 (0.011) & 0.005 (0.017) & 0.032\\
 SN2005ir & 0.07631 & 37.589(0.127) & 685.0(0.2) & 0.875(0.025) & 19.086(0.037) & 0.027 (0.016) & 0.112 (0.022) & 0.141\\
 SN2005kc & 0.01511 & 33.871(0.197) & 698.3(0.0) & 1.150(0.016) & 15.632(0.020) & 0.200 (0.014) & 0.267 (0.019) & 0.409\\
 SN2005ki & 0.01919 & 34.615(0.159) & 705.9(0.0) & 1.381(0.007) & 16.237(0.010) & -0.031 (0.010) & -0.010 (0.019) & 0.029\\
 SN2005M & 0.02200 & 35.052(0.151) & 405.9(0.0) & 0.799(0.003) & 16.518(0.006) & 0.024 (0.004) & 0.117 (0.017) & 0.146\\
 SN2005na & 0.02630 & 35.118(0.145) & 740.3(0.2) & 1.005(0.013) & 16.618(0.020) & -0.018 (0.015) & 0.048 (0.019) & 0.137\\
 SN2006ax & 0.01673 & 34.324(0.169) & 827.5(0.1) & 0.949(0.008) & 15.763(0.014) & -0.060 (0.011) & 0.009 (0.017) & 0.068\\
 SN2006bh & 0.01084 & 33.329(0.238) & 833.6(0.0) & 1.387(0.008) & 14.977(0.010) & -0.007 (0.009) & 0.016 (0.018) & 0.047\\
 SN2006gt & 0.04474 & 36.441(0.130) & 1003.0(0.1) & 1.675(0.012) & 18.415(0.020) & 0.225 (0.020) & 0.234 (0.031) & 0.264\\
 SN2006py & 0.05786 & 36.839(0.137) & 1071.0(0.3) & 1.016(0.049) & 18.421(0.060) & 0.070 (0.021) & 0.142 (0.025) & 0.205\\
\enddata
\tablenotetext{a}{Distance modulus ($h=0.72$).}
\tablenotetext{b}{Time of maximum for rest-frame B light-curve (JD - 2400000.5).}
\tablenotetext{c}{Decline rate parameter.}
\tablenotetext{d}{Galactic reddening from \citet{Schlegel1998}.}
\tablenotetext{e}{Host galaxy reddening assuming \citet{Phillips1999} colors.}
\tablenotetext{f}{Total (galactic + host galaxy) absorption in NIR toward SN
                  assuming $R_V = 1.7$.}
\end{deluxetable}
\end{turnpage}

\clearpage
\begin{turnpage}
\begin{deluxetable}{lllllllll}
\tablecolumns{9}
\tablewidth{0pc}
\tablecaption{Derived Light-curve Parameters\label{tab:lcparams}}
\tablehead{\colhead{Name} & 
           \colhead{z} & 
           \colhead{$\mu$ ($\sigma$)\tablenotemark{a}} & 
           \colhead{$T_{max}$($\sigma$)\tablenotemark{b}} & 
           \colhead{$\Delta m_{15} $($\sigma$)\tablenotemark{c}} & 
           \colhead{$(i_{max})$ ($\sigma$)\tablenotemark{d}} &
           \colhead{$(B_{max}-V_{max})$ ($\sigma$)\tablenotemark{d}} &
           \colhead{$E(B-V)_{host}$ ($\sigma$)\tablenotemark{e}} & 
           \colhead{$A_{TOT}$\tablenotemark{f}} \\
           \colhead{} & \colhead{} & \colhead{(mag)} & 
           \colhead{(days)} & \colhead{(mag)} & 
           \colhead{(mag)} &\colhead{(mag)} & \colhead{(mag)} & \colhead{(mag)}}
\startdata
SNLS 03D4gl & 0.571 & 42.765(0.170) & 52954.1(0.6) & 0.719(0.151) & 24.192(0.120) & 0.020 (0.029) & 0.122 (0.038) & 0.136\\
SNLS 04D1oh & 0.59 & 42.680(0.180) & 53306.9(0.4) & 0.946(0.069) & 24.070(0.087) & -0.026 (0.056) & 0.046 (0.058) & 0.067\\
SNLS 04D1pg & 0.515 & 42.353(0.148) & 53325.5(0.3) & 0.813(0.050) & 24.059(0.078) & 0.138 (0.029) & 0.238 (0.034) & 0.248\\
SNLS 04D1rh & 0.435 & 41.942(0.134) & 53349.2(0.4) & 0.877(0.048) & 23.353(0.069) & -0.007 (0.019) & 0.074 (0.024) & 0.100\\
SNLS 04D1sk & 0.6634 & 43.307(0.227) & 53354.6(0.7) & 1.399(0.109) & 24.810(0.153) & -0.032 (0.062) & -0.013 (0.066) & 0.011\\
SNLS 04D2an & 0.62 & 42.957(0.151) & 53031.0(0.8) & 0.872(0.098) & 24.347(0.110) & \ldots & \ldots & \ldots\\
SNLS 05D1dn & 0.566 & 42.417(0.153) & 53647.3(0.6) & 0.742(0.044) & 23.941(0.091) & 0.062 (0.031) & 0.165 (0.036) & 0.181\\
SNLS 05D1hk & 0.2631 & 40.438(0.176) & 53718.1(0.3) & 0.889(0.029) & 21.926(0.136) & 0.027 (0.014) & 0.109 (0.021) & 0.130\\
SNLS 05D1hn & 0.1489 & 39.128(0.244) & 53713.2(0.1) & 0.850(0.038) & 21.231(0.219) & 0.317 (0.020) & 0.428 (0.026) & 0.436\\
SNLS 05D1ix & 0.49 & 42.022(0.131) & 53718.6(0.0) & 0.896(0.017) & 23.526(0.070) & 0.034 (0.006) & 0.116 (0.016) & 0.133\\
SNLS 05D1iy & 0.2478 & 40.509(0.149) & 53715.5(0.1) & 1.191(0.041) & 22.216(0.096) & 0.089 (0.017) & 0.142 (0.021) & 0.165\\
SNLS 05D2ah & 0.184 & 39.656(0.104) & 53382.5(0.0) & 0.995(0.010) & 21.310(0.016) & 0.090 (0.005) & 0.166 (0.014) & 0.178\\
SNLS 05D2bt & 0.679 & 43.169(0.189) & 53400.8(0.3) & 0.941(0.076) & 24.306(0.115) & -0.142 (0.051) & -0.080 (0.054) & -0.056\\
SNLS 05D2bv & 0.474 & 42.194(0.131) & 53403.8(0.1) & 1.031(0.037) & 23.471(0.065) & -0.089 (0.011) & -0.033 (0.017) & -0.016\\
SNLS 05D2ck & 0.698 & 43.088(0.256) & 53415.7(0.8) & 1.644(0.108) & 24.721(0.103) & -0.004 (0.097) & -0.010 (0.101) & 0.006\\
SNLS 05D2dw & 0.417 & 41.933(0.123) & 53453.3(0.1) & 0.879(0.029) & 23.240(0.059) & -0.056 (0.013) & 0.021 (0.020) & 0.043\\
SNLS 05D2eb & 0.5344 & 42.376(0.218) & 53450.4(0.4) & 0.836(0.056) & 23.885(0.182) & 0.043 (0.024) & 0.133 (0.030) & 0.140\\
SNLS 05D2mp & 0.3537 & 41.577(0.115) & 53710.2(0.1) & 0.804(0.014) & 23.018(0.054) & 0.016 (0.010) & 0.108 (0.019) & 0.125\\
SNLS 05D4cw & 0.375 & 41.394(0.174) & 53581.0(0.9) & 1.396(0.126) & 22.920(0.097) & -0.021 (0.048) & -0.000 (0.053) & 0.030\\
SNLS 05D4fo & 0.373 & 41.689(0.120) & 53658.0(0.1) & 1.269(0.025) & 23.232(0.063) & 0.003 (0.011) & 0.041 (0.018) & 0.069\\
ESS d149wcc4-11 & 0.342 & 41.337(0.121) & 52955.3(0.3) & 0.894(0.045) & 22.791(0.038) & 0.011 (0.026) & 0.091 (0.030) & 0.110\\
ESS e108wdd8-4 & 0.469 & 42.391(0.176) & 52980.6(0.8) & 0.516(0.107) & 23.469(0.092) & -0.115 (0.047) & -0.001 (0.055) & 0.028\\
SDSS 12855 & 0.165 & 39.423(0.175) & 53995.5(0.3) & 1.347(0.079) & 21.209(0.024) & 0.105 (0.041) & 0.142 (0.045) & 0.196\\
SDSS 13025 & 0.224 & 40.019(0.186) & 53993.2(0.8) & 1.047(0.089) & 21.783(0.081) & 0.134 (0.047) & 0.208 (0.050) & 0.289\\
SDSS 13835 & 0.247 & 40.635(0.156) & 54011.0(0.6) & 0.816(0.054) & 22.002(0.043) & -0.020 (0.037) & 0.067 (0.041) & 0.093\\
SDSS 15287 & 0.235 & 40.614(0.156) & 54028.6(0.4) & 0.820(0.065) & 21.968(0.056) & -0.026 (0.031) & 0.060 (0.035) & 0.112\\
SDSS 16442 & 0.281 & 40.217(0.271) & 54060.2(1.0) & 1.042(0.143) & 21.983(0.143) & 0.136 (0.092) & 0.210 (0.094) & 0.284\\
SDSS 3241 & 0.259 & 40.827(0.197) & 53647.2(0.4) & 1.005(0.056) & 21.970(0.053) & -0.148 (0.067) & -0.093 (0.069) & 0.079\\
SDSS 3331 & 0.208 & 39.746(0.182) & 53649.5(0.4) & 0.922(0.070) & 21.479(0.064) & 0.136 (0.046) & 0.224 (0.048) & 0.255\\
SDSS 4679 & 0.333 & 41.454(0.235) & 53652.7(0.6) & 0.704(0.052) & 22.734(0.064) & -0.045 (0.093) & 0.052 (0.095) & 0.084\\
SDSS 5183 & 0.384 & 42.069(0.390) & 53649.3(1.3) & 0.946(0.130) & 23.149(0.133) & -0.170 (0.163) & -0.110 (0.164) & 0.042\\
SDSS 5549 & 0.119 & 38.745(0.123) & 53665.5(0.4) & 0.920(0.047) & 20.333(0.054) & 0.069 (0.022) & 0.152 (0.026) & 0.182\\
SDSS 6699 & 0.311 & 41.063(0.194) & 53675.4(0.6) & 1.075(0.081) & 22.468(0.070) & -0.036 (0.067) & 0.021 (0.069) & 0.085\\
SDSS 7243 & 0.204 & 40.148(0.170) & 53685.7(0.2) & 0.929(0.066) & 21.561(0.053) & -0.013 (0.037) & 0.062 (0.040) & 0.178\\
SDSS 7512 & 0.22 & 40.353(0.203) & 53681.8(0.4) & 0.967(0.112) & 21.902(0.114) & 0.045 (0.047) & 0.121 (0.050) & 0.235\\
\enddata
\tablenotetext{a}{Distance modulus ($h=0.72$).}
\tablenotetext{b}{Time of maximum for rest-frame B light-curve (JD - 2400000.5).}
\tablenotetext{c}{Decline rate parameter.}
\tablenotetext{d}{Galactic reddening from \citet{Schlegel1998}.}
\tablenotetext{e}{Host galaxy reddening assuming \citet{Phillips1999} colors.}
\tablenotetext{f}{Total (galactic + host galaxy) absorption in NIR toward SN
                  assuming $R_V = 1.7$.}
\end{deluxetable}
\end{turnpage}
\clearpage

\begin{deluxetable}{llll}
\tablecolumns{4}
\tablewidth{0pc}
\tablecaption{Binned Hubble Diagram\label{tab:binned_hubble}}
\tablehead{\colhead{$<z>$} & \colhead{N} & \colhead{$<\mu>$} &
           \colhead{$\sigma(\mu)$} }
\startdata
0.01 &  6 & 33.906 & 0.018\\
0.02 &  3 & 34.635 & 0.049\\
0.03 &  5 & 35.505 & 0.028\\
0.05 &  5 & 36.619 & 0.032\\
0.08 &  2 & 37.595 & 0.045\\
0.13 &  2 & 38.930 & 0.117\\
0.17 &  2 & 39.627 & 0.099\\
0.24 & 10 & 40.357 & 0.054\\
0.35 &  7 & 41.397 & 0.057\\
0.46 &  5 & 41.984 & 0.065\\
0.56 &  5 & 42.543 & 0.075\\
0.68 &  3 & 43.023 & 0.107\\
\enddata
\end{deluxetable}

\clearpage
\begin{deluxetable}{lrlll}
\tablecolumns{5}
\tablewidth{0pc}
\tablecaption{Cosmological Results\label{tab:cosmologies}}
\tablehead{\colhead{$R_V$} & \colhead{Band} & \colhead{$\Omega_m$} &
           \colhead{$w$} & $rms$ }
\startdata
$1.74$ & $W^{i^\prime}_{BV}$ & $0.27 \pm 0.02$ & $-1.05 \pm 0.13$ & $0.13$\\
      & $W^B_{BV}$          & $0.27 \pm 0.02$ & $-1.08 \pm 0.14$ & $0.15$\\
$3.1$ & $W^{i^\prime}_{BV}$ & $0.26 \pm 0.02$ & $-1.20 \pm 0.13$ & $0.20$\\
      & $W^B_{BV}$          & $0.25 \pm 0.03$ & $-1.24 \pm 0.16$ & $0.24$\\
\enddata
\end{deluxetable}

\clearpage
\begin{turnpage}
\begin{deluxetable}{lr|rr|rr|rrrr|rrrrrr}
\tablecolumns{16}
\tablewidth{0pc}
\tablecaption{Systematic Errors\label{tab:syserrors}}
\tablehead{\colhead{}&\colhead{}&\multicolumn{2}{c}{$\Omega_k=0$}&
           \multicolumn{2}{c}{$w=-1$} & \multicolumn{4}{c}{$j=$constant} &
           \multicolumn{6}{c}{nuisance}\\
           \colhead{Systematic} & \colhead{Error} & 
           \colhead{$\frac{dw}{d\sigma}$} & \colhead{$\Delta w$} &
           \colhead{$\frac{d\Omega_{DE}}{d\sigma}$} &
           \colhead{$\Delta\Omega_{DE}$} &
           \colhead{$\frac{dq_{0}}{d\sigma}$} &
           \colhead{$\Delta q_{0}$} &
           \colhead{$\frac{dj_k}{d\sigma}$} &
           \colhead{$\Delta j_k$} &
           \colhead{$\frac{dR_V}{d\sigma}$} &
           \colhead{$\Delta R_V$} &
           \colhead{$\frac{db}{d\sigma}$} &
           \colhead{$\Delta b$} &
           \colhead{$\frac{dW_0}{d\sigma}$} &
           \colhead{$\Delta W_0$} }
\startdata
Mag. offset ($\delta m$) & 0.025  &
  -2.69  & 0.067 &   
   2.73  & 0.068 &    
  -2.75  & 0.069 &    
   8.10  & 0.203 &    
   0.04  &  0.001  &    
   0.002  &  0.000  &    
   -0.25  & 0.006  \\    
Color offset ($\delta c$) &  0.02  &
    2.53  & 0.051 &   
   -2.58  & 0.052 &   
    2.68  & 0.054 &   
   -7.77  & 0.155  & 
   0.38  &  0.008  &    
   -0.03  &  0.001  &    
    0.21  & 0.004  \\    
Color Term ($\delta(CT)$)  &  0.015 &
      1.8 & 0.027 &   
    -1.73 & 0.026 &   
    1.83  & 0.027 &   
   -5.82  & 0.087 &   
   -0.67  &  0.010  &   
   -0.22  &  0.003  &   
    0.71  & 0.011  \\   
$\Delta m_{15}$ variance ($\sigma_\Delta$) & 0.05 &
     -0.007 & 0.000 &   
     -0.01 & 0.001 &   
     -0.04 & 0.002 &   
      0.14 & 0.007 &    
      0.09  &  0.005  &    
     -0.06  &  0.003  &    
     -0.03  & 0.002  \\    
Color variance ($\sigma_c$) & 0.03 &
     0.03 & 0.001 &   
     0.52  & 0.016 &   
     -0.173& 0.005 &   
  0.53      & 0.016 &    
   \ldots  &  0.100  &    
   \ldots  &  0.011  &    
   \ldots   & 0.004  \\    
   & & & & & & & & & & & & & & & \\ 
Total & &
   & 0.088 &  
   & 0.091 &  
   & 0.092 &  
   & 0.270 &  
   & 0.097 &  
   & 0.012 &  
   & 0.014\\  
\enddata
\end{deluxetable}
\end{turnpage}

\clearpage
\begin{deluxetable}{ll}
\tablecolumns{2}
\tablewidth{0pc}
\tablecaption{Summary of Cosmological Parameters\label{tab:summarycosmology}}
\tablehead{\colhead{Constraints} & \colhead{Cosmological Parameters}}
\startdata
$w = -1$ & \OmegaMatter \\
Combine with BAO & \OmegaLambda \\
  &  \\
\hline
$\Omega_k$ = 0 & \OmegaMatterFlat \\
Combine with BAO & \wo \\
  & \\
\hline
Constant j      &  \qo \\
Combine with BAO      &  \jo \\
\enddata
\end{deluxetable}

\end{document}